\documentclass[AMA,STIX1COL]{WileyNJD-v2}

\articletype{RESEARCH ARTICLE}%

\received{DD MONTH YEAR}
\revised{DD MONTH YEAR}
\accepted{DD MONTH YEAR}

\usepackage{mathtools}
\usepackage{subfigure}
\usepackage{epsfig}
\usepackage{siunitx}
\DeclareSIUnit\Molar{M}     %

\usepackage{textcomp}    %

\usepackage{color}

\usepackage{hyperref}   %

\makeatletter %

\def\input@path{{figures/}}

\newcommand{\defvariable}{:=}%

\newcommand{\secref}[1]{Sec.~\ref{#1}}%
\newcommand{\appref}[1]{Appendix~\ref{#1}}%
\newcommand{\figref}[1]{Fig.~\ref{#1}}%
\renewcommand{\eqref}[1]{Eq.~(\ref{#1})}%

\newcommand\DeclareBoldMathCommand[2]{%
  \protected@edef\@tempb{%
    \noexpand\DeclareRobustCommand{\csname #1\endcsname}{\boldsymbol{\ensuremath{#2}}}}
  \@tempb}
\@tfor\AM@letter:=abcdefghijklmnopqrstuvwxyz%
\do{\DeclareBoldMathCommand{v\AM@letter}{\AM@letter}}
\@tfor\AM@letter:=ABCDEFGHIJKLMNOPQRSTUVWXYZ%
\do{\DeclareBoldMathCommand{v\AM@letter}{\AM@letter}}
\DeclareBoldMathCommand{vnull}{0}
\DeclareBoldMathCommand{vone}{1}
\DeclareBoldMathCommand{valpha}{\alpha}
\DeclareBoldMathCommand{vbeta}{\beta}
\DeclareBoldMathCommand{vgamma}{\gamma}
\DeclareBoldMathCommand{vdelta}{\delta}
\DeclareBoldMathCommand{vepsilon}{\epsilon}
\DeclareBoldMathCommand{vvarepsilon}{\varepsilon}
\DeclareBoldMathCommand{vzeta}{\zeta}
\DeclareBoldMathCommand{veta}{\eta}
\DeclareBoldMathCommand{vtheta}{\theta}
\DeclareBoldMathCommand{vvartheta}{\vartheta}
\DeclareBoldMathCommand{viota}{\iota}
\DeclareBoldMathCommand{vkappa}{\kappa}
\DeclareBoldMathCommand{vlambda}{\lambda}
\DeclareBoldMathCommand{vmu}{\mu}
\DeclareBoldMathCommand{vnu}{\nu}
\DeclareBoldMathCommand{vpi}{\pi}
\DeclareBoldMathCommand{vxi}{\xi}
\DeclareBoldMathCommand{vrho}{\varrho}
\DeclareBoldMathCommand{vsigma}{\sigma}
\DeclareBoldMathCommand{vtau}{\tau}
\DeclareBoldMathCommand{vupsilon}{\upsilon}
\DeclareBoldMathCommand{vphi}{\varphi}
\DeclareBoldMathCommand{vchi}{\chi}
\DeclareBoldMathCommand{vpsi}{\psi}
\DeclareBoldMathCommand{vomega}{\omega}
\DeclareBoldMathCommand{vGamma}{\Gamma}
\DeclareBoldMathCommand{vDelta}{\Delta}
\DeclareBoldMathCommand{vTheta}{\Theta}
\DeclareBoldMathCommand{vLambda}{\Lambda}
\DeclareBoldMathCommand{vXi}{\Xi}
\DeclareBoldMathCommand{vPi}{\Pi}
\DeclareBoldMathCommand{vSigma}{\Sigma}
\DeclareBoldMathCommand{vUpsilon}{\Upsilon}
\DeclareBoldMathCommand{vPhi}{\Phi}
\DeclareBoldMathCommand{vPsi}{\Psi}
\DeclareBoldMathCommand{vOmega}{\Omega}

\newcommand\DeclareDiscreteBoldMathCommand[2]{%
  \protected@edef\@tempc{%
    \noexpand\DeclareRobustCommand{\csname #1\endcsname}{\boldsymbol{\mathrm{#2}}}}
  \@tempc}
\@tfor\AM@letter:=abcdefghijklmnopqrstuvwxyz%
\do{\DeclareDiscreteBoldMathCommand{vd\AM@letter}{\AM@letter}}
\@tfor\AM@letter:=ABCDEFGHIJKLMNOPQRSTUVWXYZ%
\do{\DeclareDiscreteBoldMathCommand{vd\AM@letter}{\AM@letter}}
\DeclareDiscreteBoldMathCommand{vdnull}{0}
\DeclareDiscreteBoldMathCommand{vdone}{1}
\DeclareDiscreteBoldMathCommand{vdalpha}{\alpha}
\DeclareDiscreteBoldMathCommand{vdbeta}{\beta}
\DeclareDiscreteBoldMathCommand{vdgamma}{\gamma}
\DeclareDiscreteBoldMathCommand{vddelta}{\delta}
\DeclareDiscreteBoldMathCommand{vdepsilon}{\epsilon}
\DeclareDiscreteBoldMathCommand{vdvarepsilon}{\varepsilon}
\DeclareDiscreteBoldMathCommand{vdzeta}{\zeta}
\DeclareDiscreteBoldMathCommand{vdeta}{\eta}
\DeclareDiscreteBoldMathCommand{vdtheta}{\theta}
\DeclareDiscreteBoldMathCommand{vdvartheta}{\vartheta}
\DeclareDiscreteBoldMathCommand{vdiota}{\iota}
\DeclareDiscreteBoldMathCommand{vdkappa}{\kappa}
\DeclareDiscreteBoldMathCommand{vdlambda}{\lambda}
\DeclareDiscreteBoldMathCommand{vdmu}{\mu}
\DeclareDiscreteBoldMathCommand{vdnu}{\nu}
\DeclareDiscreteBoldMathCommand{vdpi}{\pi}
\DeclareDiscreteBoldMathCommand{vdxi}{\xi}
\DeclareDiscreteBoldMathCommand{vdrho}{\varrho}
\DeclareDiscreteBoldMathCommand{vdsigma}{\sigma}
\DeclareDiscreteBoldMathCommand{vdtau}{\tau}
\DeclareDiscreteBoldMathCommand{vdupsilon}{\upsilon}
\DeclareDiscreteBoldMathCommand{vdphi}{\varphi}
\DeclareDiscreteBoldMathCommand{vdchi}{\chi}
\DeclareDiscreteBoldMathCommand{vdpsi}{\psi}
\DeclareDiscreteBoldMathCommand{vdomega}{\omega}
\DeclareDiscreteBoldMathCommand{vdGamma}{\Gamma}
\DeclareDiscreteBoldMathCommand{vdDelta}{\Delta}
\DeclareDiscreteBoldMathCommand{vdTheta}{\Theta}
\DeclareDiscreteBoldMathCommand{vdLambda}{\Lambda}
\DeclareDiscreteBoldMathCommand{vdXi}{\Xi}
\DeclareDiscreteBoldMathCommand{vdPi}{\Pi}
\DeclareDiscreteBoldMathCommand{vdSigma}{\Sigma}
\DeclareDiscreteBoldMathCommand{vdUpsilon}{\Upsilon}
\DeclareDiscreteBoldMathCommand{vdPhi}{\Phi}
\DeclareDiscreteBoldMathCommand{vdPsi}{\Psi}
\DeclareDiscreteBoldMathCommand{vdOmega}{\Omega}

\providecommand*{\dd}{%
  \@ifnextchar^{\@dd}{\@dd^{}}}
\def\@dd^#1{%
  \mathop{\mathrm{\mathstrut d}}%
  \nolimits^{#1}\dd@gobblespace}
\def\dd@gobblespace{%
  \futurelet\diffarg\dd@opspace}
\def\dd@opspace{%
  \let\dd@space\!%
  \ifx\diffarg(%
\let\dd@space\relax%
\else%
\ifx\diffarg[%
\let\dd@space\relax%
\else%
\ifx\diffarg\{%
\let\dd@space\relax%
\fi%
\fi%
\fi%
\dd@space}

\newcommand{\Frac}{%
  \@ifnextchar[%
  {\Frac@i}
  {\Frac@ii}}
\newcommand{\Frac@i}{}
\def\Frac@i[#1]#2#3{%
  \genfrac{}{}{#1}{}{\displaystyle{#2}}{\displaystyle{#3}}}
\newcommand{\Frac@ii}[2]{\frac{\displaystyle{#1}}{\displaystyle{#2}}}
      \newcommand{\diff@diffspace}{\,}
\newcommand{\diff@mathfrac}[2]{\frac{#1}{#2}}
\newcommand{\diff@mathFrac}[2]{\Frac{#1}{#2}}
\newcommand{\diff@textfrac}[2]{%
  \bgroup #1\egroup\mkern-1mu/\mkern-1mu\bgroup #2\egroup}
\newcommand{\diff}{%
  \global\let\diff@diffop\dd
  \global\let\diff@frac\diff@mathfrac
  \@ifnextchar[%
  {\diff@i}
  {\diff@ii}}
\newcommand{\Diff}{%
  \global\let\diff@diffop\dd
  \global\let\diff@frac\diff@mathFrac
  \@ifnextchar[%
  {\diff@i}
  {\diff@ii}}
\newcommand{\tdiff}{%
  \global\let\diff@diffop\dd
  \global\let\diff@frac\diff@textfrac
  \@ifnextchar[%
  {\diff@i}
  {\diff@ii}}
\newcommand{\pdiff}{%
  \global\let\diff@diffop\partial
  \global\let\diff@frac\diff@mathfrac
  \@ifnextchar[%
  {\diff@i}
  {\diff@ii}}
\newcommand{\Pdiff}{%
  \global\let\diff@diffop\partial
  \global\let\diff@frac\diff@mathFrac
  \@ifnextchar[%
  {\diff@i}
  {\diff@ii}}
\newcommand{\tpdiff}{%
  \global\let\diff@diffop\partial
  \global\let\diff@frac\diff@textfrac
  \@ifnextchar[%
  {\diff@i}
  {\diff@ii}}

\newcommand*{\diff@i}{}
\def\diff@i[#1]#2#3{\eval{\diff@ii{#2}{#3}}_{#1}}

\newcommand*{\diff@ii}[2]{%
  \begingroup
  \toks0={}\count0=0
  \diff@degree #2\diff@degree
  \diff@frac{\diff@diffop\ifnum\count0>1^{\the\count0}\fi\diff@diffspace#1}%
  {\the\toks0}%
  \endgroup}

\newcommand*{\diff@degree}[1]{%
  \ifx #1\diff@degree \expandafter\diff@stopd
  \else \expandafter\diff@addd \fi #1^1$#1\diff@addd}
\newcommand{\diff@stopd}{}
\def\diff@stopd #1\diff@addd{}
\newcommand*{\diff@addd}{}
\def\diff@addd #1^#2#3$#4\diff@addd{%
  \advance\count0 #2
  \toks0=\expandafter{\the\toks0%
    {\diff@diffop\diff@diffspace #4}%
    \diff@diffspace}\diff@degree}

\def\rs#1{\@ifnextchar[%
  {\@rs{#1}}{\@@rs{#1}}}
\def\@rs#1[#2]#3{\mathinner{%
    \setbox\@ne\hbox{$\displaystyle{\vphantom{#3}}#1{#3}\m@th$}%
    \setbox\tw@\hbox{$\displaystyle{#3}#2\m@th$}%
    \hskip\wd\@ne\hskip-\wd\tw@\mathord{\hskip\wd\tw@\hskip-\wd\@ne%
      {\vphantom{#3}}#1{#3}#2}}}
\def\@@rs#1#2{\mathinner{%
    \setbox\@ne\hbox{$\displaystyle{\vphantom{#2}}#1{#2}\m@th$}%
    \hskip\wd\@ne\mathord{\hskip-\wd\@ne%
      {\vphantom{#2}}#1{#2}}}}

\newcommand{\MR}{\mathalpha{\mathbb{R}}}

\newcommand*{\abs}[1]{\mathinner{\vert#1\vert}}

\newcommand*{\norm}[1]{\mathinner{\Vert#1\Vert}}

\definecolor{notecolor}{cmyk}{0,1,1,.2}
\newcommand*\AM@notesname{Notes}

\makeatother

\graphicspath{{figures/}}

\begin{document}

\title{Analytical disk-cylinder interaction potential laws for the computational modeling of adhesive, deformable (nano)fibers}

\author{Maximilian J.~Grill*}
\author{Wolfgang A.~Wall}
\author{Christoph Meier}

\corres{* \email{maximilian.grill@tum.de}}

\address{Institute for Computational Mechanics, Technical University of Munich, Boltzmannstr.~15, 85748 Garching b.~M\"unchen, Germany}

\abstract[Summary]{%
The analysis of complex fibrous systems or materials on the micro- and nanoscale, which have a high practical relevance for many technical or biological systems, requires accurate analytical descriptions of the adhesive and repulsive forces acting on the fiber surfaces.
While such analytical expressions are generally needed both for theoretical studies and for computer-based simulations, the latter motivates us here to derive disk-cylinder interaction potential laws that are valid for arbitrary mutual orientations in the decisive regime of small surface separations.
The chosen type of fundamental point-pair interaction follows the simple Lennard-Jones model with inverse power laws for both the adhesive van der Waals part and the steric, repulsive part.
We present three different solutions, ranging from highest accuracy to the best trade-off between simplicity of the expression and sufficient accuracy for our intended use.
The validity of simplifying approximations and the accuracy of the derived potential laws is thoroughly analyzed, using both numerical and analytical reference solutions for specific interaction cases.
Most importantly, the correct asymptotic scaling behavior in the decisive regime of small separations is achieved, and also the theoretically predicted $(1\!/\!\sin\!\alpha)$-angle dependence (for non-parallel cylinders) is obtained by the proposed analytical solutions.
As we show in the outlook to our current research, the derived analytical disk-cylinder interaction potential laws may be used to formulate highly efficient computational models for the interaction of arbitrarily curved fibers, such that the disk represents the cross-section of the first and the cylinder a local approximation to the shape of the second fiber.
}


\maketitle

\section{Introduction}
\label{sec::introduction}
Filamentous actin, collagen, and DNA are just a few popular examples of the many different fiber-like, deformable structures that can be found on the nano to microscale in biological systems.
The interactions between such slender, elastic fibers are crucial to the complex, hierarchical assemblies they form.
Typical examples of assemblies include networks (e.g. the cytoskeleton or extracellular matrix) and bundles (e.g. muscle or tendon) and play a key role in numerous functions of the human body.
The rise of computational modeling and simulation of these complex biophysical systems nourishes the hope to shed light on some of the yet poorly understood aspects, e.g., the basic working principles and their impact on human physiology and pathophysiology.
In addition, such interaction effects can also play a crucial role in the design of novel materials in different technical applications.

By deriving analytical interaction potential laws, this work lays the foundation for an accurate and efficient computational model for short-ranged molecular interactions between curved slender fibers undergoing large 3D deformations like the one presented in our own recent contribution~\cite{GrillSBIP}.
To further explain the motivation for the analytical work in the present article, the key ideas of this novel computational model shall be summarized here.
One important aspect is the dimensionally reduced description of the fibers based on the geometrically exact beam theory~\cite{Reissner1972, Simo1986a, Simo1986b, Simo1988, Cardona1988, Crisfield1999, betsch2002frame, leyendecker2006objective, Romero2008, cesarek2012, Bauchau2014, sonneville2014, Meier2014, Meier2019}, relying on the fundamental kinematic assumption of undeformable fiber cross-sections.
The novel approach therefore belongs to the class of beam-beam interaction formulations, which have mainly focused on modeling macroscopic contact phenomena~\cite{wriggers1997,litewka2005,durville2010,Kulachenko2012,Chamekh2014,GayNeto2016a,Konyukhov2016,Weeger2017,meier2016,Meier2017a,bosten2022mortar}, but were recently also extended to other types of interactions such as electrostatic or vdW forces~\cite{GrillSSIP}.
Due to this versatility with respect to the type of interaction, but also due to the modularity and simple integration in (nonlinear) finite element solver frameworks for structural mechanics, such beam interaction formulations are widely used and its large number of applications ranges from biological to industrial materials and from nano to macroscale~\cite{durville2010,Kulachenko2012,Mueller2014rheology,Mueller2015interpolatedcrosslinks,Weeger2016,Negi2018,meier2018geometrically,pattinson2019additive,steinbrecher2020mortar,GrillParticleMobilityHydrogels,Eichinger2021,BundlesPNAS,Khristenko2021,steinbrecher2021}.

\begin{figure}[htpb]%
  \centering
  \def\svgwidth{0.65\textwidth}
\begingroup%
  \makeatletter%
  \providecommand\color[2][]{%
    \errmessage{(Inkscape) Color is used for the text in Inkscape, but the package 'color.sty' is not loaded}%
    \renewcommand\color[2][]{}%
  }%
  \providecommand\transparent[1]{%
    \errmessage{(Inkscape) Transparency is used (non-zero) for the text in Inkscape, but the package 'transparent.sty' is not loaded}%
    \renewcommand\transparent[1]{}%
  }%
  \providecommand\rotatebox[2]{#2}%
  \newcommand*\fsize{\dimexpr\f@size pt\relax}%
  \newcommand*\lineheight[1]{\fontsize{\fsize}{#1\fsize}\selectfont}%
  \ifx\svgwidth\undefined%
    \setlength{\unitlength}{186.34990815bp}%
    \ifx\svgscale\undefined%
      \relax%
    \else%
      \setlength{\unitlength}{\unitlength * \real{\svgscale}}%
    \fi%
  \else%
    \setlength{\unitlength}{\svgwidth}%
  \fi%
  \global\let\svgwidth\undefined%
  \global\let\svgscale\undefined%
  \makeatother%
  \begin{picture}(1,0.62726745)%
    \lineheight{1}%
    \setlength\tabcolsep{0pt}%
    \put(0,0.23163747){\color[rgb]{0,0,0}\makebox(0,0)[lt]{\lineheight{0}\smash{\begin{tabular}[t]{l}\fbox{master}\end{tabular}}}}%
    \put(0.30480308,0.00840154){\color[rgb]{0,0,0}\makebox(0,0)[lt]{\lineheight{0}\smash{\begin{tabular}[t]{l}\fbox{slave}\end{tabular}}}}%
    \put(0,0){\includegraphics[width=\unitlength,page=1]{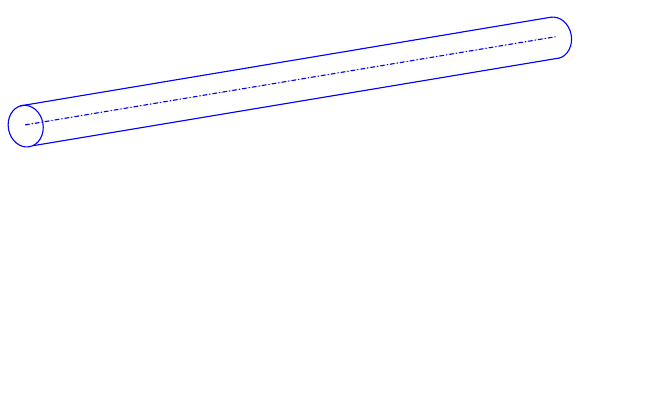}}%
    \put(0.00940326,0.53846098){\color[rgb]{0,0,1}\makebox(0,0)[lt]{\lineheight{0}\smash{\begin{tabular}[t]{l}\fbox{master surrogate}\end{tabular}}}}%
    \put(0.4640014,0.40431491){\color[rgb]{0,0,0}\makebox(0,0)[lt]{\lineheight{0}\smash{\begin{tabular}[t]{l}$\vpsi_{1-2\text{c}}$\end{tabular}}}}%
    \put(0,0){\includegraphics[width=\unitlength,page=2]{master_slave_approach_strategy.pdf}}%
    \put(0.34258386,0.35790422){\color[rgb]{0,0,0}\makebox(0,0)[lt]{\lineheight{0}\smash{\begin{tabular}[t]{l}$\vr_{1-2\text{c}}$\end{tabular}}}}%
    \put(0,0){\includegraphics[width=\unitlength,page=3]{master_slave_approach_strategy.pdf}}%
  \end{picture}%
\endgroup%

  \caption{Illustration of the section-beam interaction potential (SBIP) approach from our recent contribution~\cite{GrillSBIP} and the underlying disk-cylinder interaction resulting from this modeling approach.
            It allows for arbitrary mutual configurations described by distance vector~$\vr_{1-2\text{c}}$ and relative rotation vector~$\vpsi_{1-2\text{c}}$.}
  \label{fig::master_slave_approach_strategy}
\end{figure}

Generally, such a beam-beam interaction formulation needs to evaluate the interaction forces (and moments) acting on two slender fibers for arbitrarily deformed configurations and mutual orientations. A direct evaluation of intermolecular interaction potential and forces between two general bodies in 3D space~\cite{Argento1997,Sauer2007a,Sauer2009a,Sauer2013,Fan2015,Du2019,Mergel2019} requires to integrate molecule densities over their volumes, leading to a sixfold integral (two nested 3D integrals) that can often only be solved by means of numerical integration. However, such a procedure would lead to a forbiddingly high computational effort if large systems of slender fibers shall be modeled. The novelty of our recently developed approach~\cite{GrillSBIP} is that it only requires one single integration step along the centerline of the first fiber (``slave'' beam) to be performed numerically.
This can be achieved by exploiting the short-range nature of the considered class of interaction potentials as well as the fundamental kinematic assumption of undeformable cross-sections, which is justified for fibers that are sufficiently slender.
More specifically, a closed-form analytical solution is applied for the interaction potential between a given section of the first fiber (``slave'' beam) and the entire second fiber (``master'' beam), whose geometry is linearly expanded at the point with smallest distance to the given slave beam section.
Exactly this analytical section-beam interaction potential (SBIP) law shall be derived in the present article.
Here, we focus on circular cross-sections, which leads to the scenario of a disk interacting with a cylinder.
This scenario is illustrated in \figref{fig::master_slave_approach_strategy}.

In the present article, we apply the simple approach of pairwise summation (Hamaker).
Acknowledging the limitations with respect to retardation and other effects~\cite{parsegian2005}, the use of this simple approach is a deliberate decision to enable the derivation of analytical, closed-form expressions that can be used as a good first-order approximation to investigate the role of adhesive contact in complex systems of practical relevance, e.g. in engineering and biology as outlined above.
In the considered context, pairwise summation requires the analytical integration of a point-pair potential~$\Phi$ over all point pairs in the disk-cylinder system.
This strategy will be demonstrated for a generic inverse power law~$\Phi_\text{m}(r)=k_\text{m} \, r^{-m}$ with exponent~$m \geq 6$.
Due to this generality, the resulting reduced interaction law~$\tilde \pi$ can be used to model both the adhesive van der Waals (vdW) part ($m=6$) and the repulsive part ($m=12$) of the Lennard-Jones (LJ) potential.
Moreover, we consider the practically relevant case of circular, undeformable cross-sections and homogeneous densities of the fundamental interacting points in both fibers.

To the best of the authors' knowledge, no such disk-cylinder interaction potential law~$\tilde \pi_\text{m,disk-cyl}$ based on a generic point-pair potential with exponent~$m$ and valid for all mutual orientations can be found in the literature.
The pairwise summation strategy has initially been applied to a large number of traditional geometries of the interacting bodies, such as infinite half spaces, spheres, and infinitely long cylinders or thin wires~\cite{parsegian2005,israel2011}.
Some closely related scenarios including disk-cylinder interaction were studied by Ref.~\cite{montgomery2000} in an attempt to deeper understand particle adherence to surfaces, however, the derived analytical results remain limited to the special cases of parallel and perpendicular mutual orientation of disk and cylinder.
It nevertheless proves to be a valuable source for intermediate expressions such as the interaction energy of a point and a cylinder.
Likewise, the well-known analytical solutions for cylinder-cylinder interactions shall prove useful in the subsequent validation of our derived expressions.
Most notably in this context are the $g^{-3/2}$ law for the vdW interaction energy per unit length of two infinitely long, parallel cylinders and the $g^{-1}$ law for the vdW interaction energy of perpendicular cylinders, both in the limit of surface separations $g$ being much smaller than the cylinder radii~\cite[p.255]{israel2011}.
These scaling laws agree with the following, more general relationship valid for all mutual angles~$\alpha \in \, ]0,\pi/2]$ between two straight cylinders, as stated e.g.~in the textbook~\cite[p.~173]{parsegian2005}:
\begin{equation}\label{eq::pot_ia_vdW_cyl_cyl_skewed_smallseparation}
  \Pi_\text{vdW,cyl-cyl} =  - \frac{A_\text{Ham}}{6} \sqrt{R_1 R_2} \, g^{-1} / \sin\alpha
\end{equation}
Here, $R_1$ and $R_2$ denote the radii of the two interacting cylinders and $A_\text{Ham}$ represents the Hamaker constant.
It is one of the main objectives of this work to correctly capture this angle dependency in the sought-after analytical expression for the disk-cylinder interaction potential~$\tilde \pi_\text{m,disk-cyl}(g,\alpha)$.

Based on different approximations, we present three different analytical solutions for the sought-after disk-cylinder interaction potential. As will be shown in the verification part of this work, the important requirement to capture the correct asymptotic distance scaling, i.e. $\propto \! g^{-3/2}$ for parallel and $\propto  \! g^{-1}$ for perpendicular cylinders, and the theoretically predicted $(1\!/\!\sin\!\alpha)$-angle dependence in the decisive regime of small separations is met by all three solutions. This result is irrespective of the slight differences in the simplifying assumptions being made in their derivation, which leads to different levels of complexity in the final expression on the one hand and different levels of accuracy on the other hand. We chose to present the three most promising expressions, ranging from highest accuracy to the best trade-off between simplicity and accuracy for our intended use within the aforementioned simulation model for fiber-fiber interactions.

The remainder of this article is structured as follows:
\secref{sec::disk-cyl-pot_m_derivation} presents the required steps of the analytical integration and likewise serves as an example for the future derivation of other section-beam interaction potential laws, e.g.~for other types of interactions or cross-section shapes.
In the subsequent \secref{sec::verif_approx_singlelengthspec}, the accuracy of the derived closed-form expression for the disk-cylinder interaction potential~$\tilde \pi_\text{m,disk-cyl}$ will be verified. Eventually,~\secref{sec::conclusion_outlook} summarizes the main findings and conclusions and provides an outlook to promising future extensions and applications of this work.

\section{5D analytical integration of the point-pair interaction potential}
\label{sec::disk-cyl-pot_m_derivation}
As outlined already in the introduction, the beam-beam interaction potential~$\Pi_\text{ia}$ is obtained from the two nested 3D integrals over the two interacting bodies.
The integrand is a product of the molecule densities~$\rho_i$ of the two bodies~$i=1,2$ and the point-pair interaction potential~$\Phi(r)$, where $r=\norm{\vx_1 - \vx_2}$ denotes the distance between two points~$\vx_1$ and $\vx_2$.
\begin{equation}\label{eq::pot_3variants}
 \Pi_\text{ia} = \iint_{V_1,V_2} \rho_1(\vx_1) \rho_2(\vx_2) \Phi(r) \dd V_2 \dd V_1  =
 \int_{l_1}  \underbrace{\int_{l_2} \, \iint_{A_1,A_2} \rho_1(\vx_1) \rho_2(\vx_2) \Phi(r) \dd A_2 \dd A_1 \dd s_2}_{=:\tilde{\pi}(\vr_{1-2\text{c}},\vpsi_{1-2\text{c}}) \,\, \rightarrow \,\, \text{SBIP}} \dd s_1.
\end{equation}
The section-beam interaction potential (SBIP) approach~\cite{GrillSBIP} splits the evaluation into the numerical integration along the centerline length~$l_1$ of the slave beam and the analytical evaluation of the SBIP law~$\tilde{\pi}$.
Assuming circular, undeformable cross-sections of the fibers leads to the scenario of disk-cylinder interaction with a relative distance vector~$\vr_{1-2\text{c}}$ and relative rotation vector~$\vpsi_{1-2\text{c}}$ illustrated in \figref{fig::master_slave_approach_strategy}.
Moreover, we use a generic inverse power law~$\Phi(r)=\Phi_\text{m}(r)=k_\text{m} \, r^{-m}$ with exponent~$m \geq 6$.
After this brief introduction of nomenclature, the complete definition of the resulting mathematical problem to be solved in this work will be stated in the following section.

\subsection{Problem statement}
\begin{figure}[htb]%
  \def\svgwidth{0.95\textwidth}
  \input{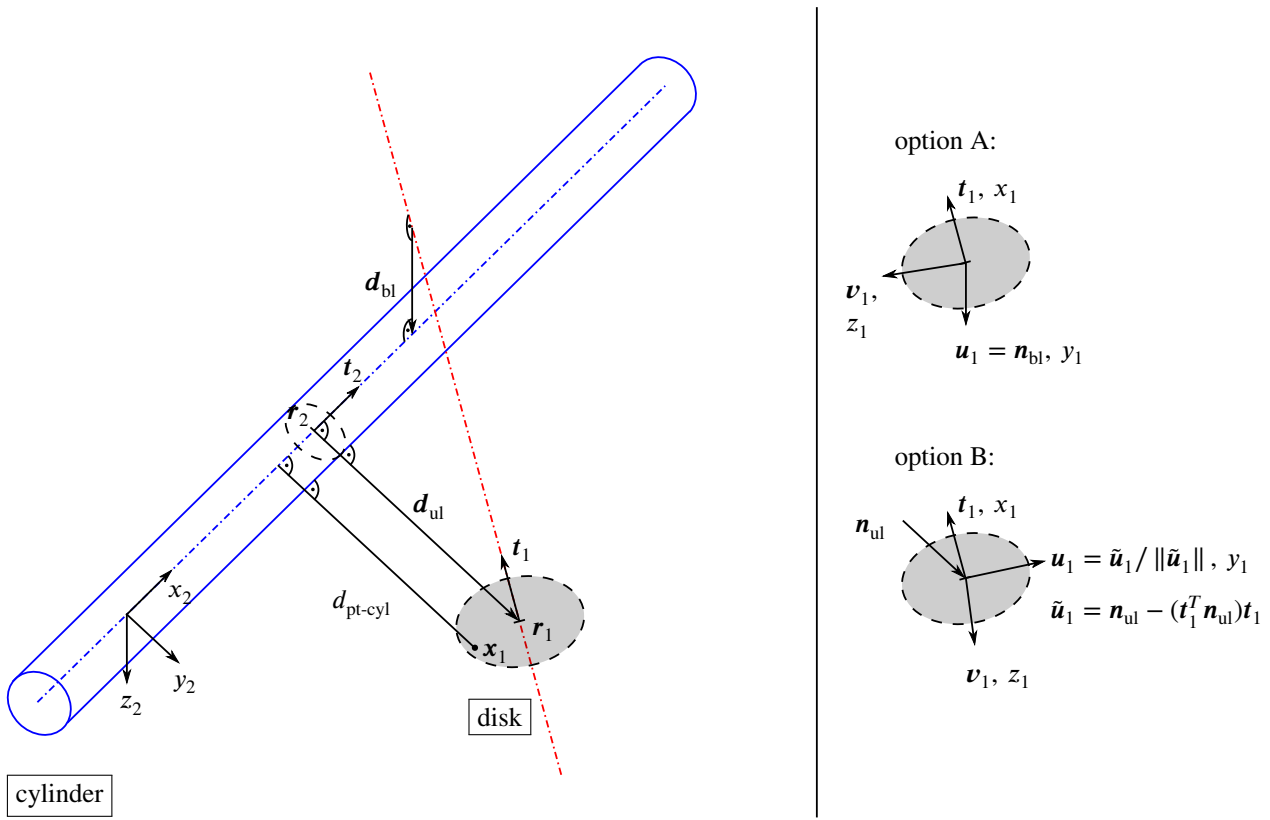}
  \caption{Illustration of the geometrical quantities used to describe the disk-cylinder interaction (left side). In addition, the two different options for the choice of the Cartesian coordinate frame~$(\vt_1,\vu_1,\vv_1)$ used for the analytical integration over the disk-shaped slave cross-section are shown on the right side.}
  \label{fig::disk_cylinder_potential_coord_frames}
\end{figure}

We aim to find the analytical solution for the disk-cylinder interaction potential
\begin{align}\label{eq::disk_cyl_pot_m_5Dint}
  \tilde \pi_{\text{m,disk-cyl}} &\defvariable \iint \limits_{A_\text{disk}} \rho_1 \overbrace{\iiint \limits_{V_\text{cyl}} \rho_2 \, \Phi_\text{m}(r) \dd V}^{=: \, \Pi_\text{m,pt-cyl}} \dd A\\
  &\text{with} \quad r=\norm{\vx_1 - \vx_2} \quad \text{and} \quad \vx_1 \in A_\text{disk}, \, \vx_2 \in V_\text{cyl}.  %
\end{align}
Here,~$\vx_1 \in A_\text{disk}$ denotes any point in the disk, i.e., circular slave cross-section area~$A_1=A_\text{disk}:= \{\vx \in \MR^3 \, | \, \vr_1 + y_1 \vu_1 + z_1 \vv_1, \, y_1^2 + z_1^2 \leq R_1^2\}$.
This disk area is parameterized via two coordinates~$y_1,z_1$ and its corresponding (for now not further specified) in-plane coordinate vectors~$\vu_1, \vv_1$.
The latter complete the (normalized) slave centerline tangent vector~$\vt_1=\vr^\shortmid_1 / \norm{\vr^\shortmid_1}$ to form an orthonormal triad~$(\vt_1,\vu_1,\vv_1)$ and its definition will be discussed later.
Once again, the short prime denotes a differentiation with respect to the element parameter coordinate, i.e.,~$\vr^{\shortmid}_{i}(\xi_i) = \tdiff{\vr_i(\xi_i)}{{\xi_i}}$.
On the master side,~$\vx_2 \in V_\text{cyl}$ with $V_\text{cyl} := \{\vx \in \MR^3 \, | \, \vr_2 + x_2 \vt_2 + y_2 \vu_2 + z_2 \vv_2, \, y_2^2 + z_2^2 \leq R_2^2, \, x_2 \in ] - \infty, \infty [ \}$ denotes any point in the infinitely long auxiliary cylinder oriented along the (normalized) tangent vector~$\vt_2=\vr^\shortmid_2 / \norm{\vr^\shortmid_2}$.
Again, a set of coordinates~$x_2,y_2,z_2$ together with an orthonormal frame~$(\vt_2,\vu_2,\vv_2)$ is chosen for parameterizing the geometry.
Regarding this second basis~$(\vt_2,\vu_2,\vv_2)$, it will turn out that the exact definition does in fact not play a role and is thus left unspecified.
\figref{fig::disk_cylinder_potential_coord_frames} (left side) illustrates the introduced geometrical quantities.
\subsection{General strategy}
The general strategy follows the one generally known as point-pairwise summation (see e.g.~\cite{parsegian2005,israel2011} for details and a discussion) as e.g.~applied in~\cite{montgomery2000} for the analytical calculation of vdW forces for certain geometric configurations, e.g., a cylinder and a perpendicular disk.
Since already for such specific scenarios, no exact analytical solution can be found for the integrals, also the following derivation will make use of the common approach of series expansions in order to find an analytical, closed-form expression for the integral of the leading term(s) of the series.
Due to the rapid decay of the inverse power laws, this approach is known to yield good approximations for the true solution of the integral. In \secref{sec::verif_approx_singlelengthspec} we will verify the resulting accuracy for the specific expressions derived here.

\subsection{Derivation}
Fortunately, the point-cylinder scenario for the case~$m=6$ of vdW interaction has already been studied in~\cite{montgomery2000}.
It can be generalized for exponents~$m \geq 6$ as outlined in \appref{sec::point-cyl-pot_Montgomery_generalization} and at this point we can make use of the final result that the point-cylinder interaction potential follows the proportionality
\begin{align}
  \Pi_\text{m,pt-cyl} \propto g_\text{pt-cyl}^{-m+3}
\end{align}
where~$g_\text{pt-cyl}$ denotes the norm of the \textit{smallest} distance vector $\vg_\text{pt-cyl}$ between the point and the cylinder \textit{surface}.
For now, we do not need to specify the precise expression for the point-cylinder interaction potential~$\Pi_\text{m,pt-cyl}$ and in fact we will later investigate two different variants/approximations presented in Appendices~\ref{sec::point-cyl-pot_Montgomery_generalization} and \ref{sec::point-halfspace-pot_m_derivation}, which deviate by a scalar factor, and discuss its effect on the accuracy of the disk-cylinder interaction potential~$\tilde \pi_{\text{m,disk-cyl}}$ to be derived.

Due to the orthogonality condition~$\vg_\text{pt-cyl} \perp \vt_2$ for the smallest distance from the master centerline curve~$\vr_2$, $g_\text{pt-cyl}$ can be expressed via the smallest distance between a point and the cylinder \textit{axis}~$d_\text{pt-cyl}$ as
\begin{align}\label{eq::gap_pt-cyl}
  g_\text{pt-cyl} \defvariable d_\text{pt-cyl} - R_2,
\end{align}
which in turn can be written as
\begin{align}\label{eq::distance_pt-cyl}
  d_\text{pt-cyl} := \norm{\vd_\text{pt-cyl}} = \norm{\vx_1(s_1,y_1,z_1) - \vr_2(s_{2\text{c}}(s_1,y_1,z_1))}.
\end{align}
Again refer to~\figref{fig::disk_cylinder_potential_coord_frames} for a sketch.
Note that $\vr_2(s_{2\text{c}}(s_1,y_1,z_1))$ is the master centerline position at~$s_{\text{2c}}$, which results from the unilateral closest-point (``c'') projection of the point~$\vx_1$ onto the master centerline curve~$\vr_2$.
Therefore, it depends also on the exact coordinates~$y_1,z_1$ of the point within the slave cross-section and not only on the arc-length parameter~$s_1$, i.e., the position along the slave centerline curve~$\vr_1$.
This fact can be used to express the decisive distance~$d_\text{pt-cyl}$ in terms of the primary centerline fields~$\vr_{1/2}$ as well as the slave arc-length parameter~$s_1$ to be used as integration variable in numerical integration (cf.~\eqref{eq::pot_3variants}) and~$y_1,z_1$ to be used as integration variables for the analytical integration over the disk area.
Since $\vg_\text{pt-cyl}$ is perpendicular to $\vt_2$, \eqref{eq::distance_pt-cyl} can equivalently be written as
\begin{align}
  d_\text{pt-cyl} &= \norm{ \left( \vx_1(s_1,y_1,z_1) - \vr_2(s_{2\text{c}}(s_1,y_1,z_1) \right) \times \vt_2 }\\
                    &= \norm{ \left( \vr_1(s_1) + y_1 \vu_1(s_1) + z_1 \vv_1(s_1) - \vr_2(s_{2\text{c}}(s_1,0,0)) \right) \times \vt_2}\\
                    &= \norm{ \left( d_\text{ul} \vn_\text{ul} + y_1 \vu_1 + z_1 \vv_1 \right) \times \vt_2}.
\end{align}
Here, we have introduced the so-called unilateral inter-axis separation
\begin{align}
  d_\text{ul}:=\norm{ \vr_1(s_1) - \vr_2(s_{2\text{c}}(s_1,0,0)) }
\end{align}
and the corresponding unilateral normal vector~$\vn_\text{ul}$, which result from the unilateral closest-point projection and are known from macroscopic (line) contact formulations (e.g.~\cite{meier2016}).
After a few steps of basic vector algebra and making use of~$\vn_\text{ul}\perp \vt_2$, $\vu_1 \perp \vv_1$, $\vv_1=\vt_1\times \vu_1$, and $\vt_2^T \vt_2 =1$, we end up with
\begin{align}\label{eq::point-cyl-sep_generic_coeff}
  d_\text{pt-cyl}^2 = &\underbrace{ \left( 1- \left(\vu_1^T \vt_2 \right)^2 \right) }_{=:a_y} \, y_1^2 \,
                     + \, \underbrace{ \left( 1- \left(\vt_1^T \left( \vu_1 \times \vt_2 \right) \right)^2 \right) }_{=:a_z} \, z_1^2 \,
                     \underbrace{ - \, 2 \left( \vu_1^T \vt_2 \right) \left( \vt_1^T \left( \vu_1 \times \vt_2 \right) \right) }_{=:a_{yz}} \, y_1 z_1\\
                     &+ \, \underbrace{2 d_\text{ul} \left( \vn_\text{ul}^T \vu_1 \right)}_{=:b_y} \, y_1 \,
                     + \, \underbrace{2 d_\text{ul} \left( \vt_1^T \left( \vu_1 \times \vn_\text{ul}\right) \right)}_{=:b_z} \, z_1 \,
                     + \, \underbrace{d_\text{ul}^2}_{=:c}\nonumber
\end{align}
which aims to express the integrand of \eqref{eq::disk_cyl_pot_m_5Dint} as polynomial in the integration variables~$y_1,z_1$ to be used for the analytical integration over the disk-shaped cross-section area~$A_1$ on the slave side.

At this point, we return to the pending definition of the coordinate vector~$\vu_1$, which shall serve as an in-plane direction within the slave cross-section and thereby complete the unique definition of the coordinate frame~$(\vt_1,\vu_1,\vv_1)$.
To briefly outline the procedure, this direction~$\vu_1$ will firstly be used as the first direction of integration over the cross-section area and secondly the associated coordinate~$y_1$ also defines the point of series expansion~$(y_1=-R_1, z_1=0)$ to be applied later on.
We basically see two reasonable options for the definition of~$\vu_1$ (see~\figref{fig::disk_cylinder_potential_coord_frames} (right side) for an illustration):
\begin{itemize}
  \item option A: $\vu_1 \defvariable \vn_\text{bl} \hspace{0.92cm} \quad \text{with} \quad \vn_\text{bl} \defvariable \left( \vt_1 \times \vt_2 \right) / \norm{ \vt_1 \times \vt_2 }$
  \item option B: $\vu_1 \defvariable \tilde{\vu}_1 / \norm{\tilde{\vu}_1} \quad \text{with} \quad \tilde{\vu}_1 \defvariable \left( \vI - \vt_1 \otimes \vt_1^T \right) \vn_\text{ul}$
\end{itemize}

Option A is a natural choice in the sense that the \textit{bilateral} normal vector~$\vn_\text{bl}$ will always be perpendicular to~$\vt_1$ and thus lie within the cross-section.
Note that its definition in the context of this work deviates from the one known from macroscopic (point) contact models (e.g.~\cite{wriggers1997}).
Here, it is defined via the cross product of the normalized tangent vector~$\vt_1$ at the integration point on the slave side, i.e., the vector perpendicular to the disk plane, and the normalized tangent vector~$\vt_2$ at the unilateral closest point on the master side, i.e., the vector defining the axis of the auxiliary cylinder serving as a surrogate for the actual beam geometry on the master side.
In contrast to this, point contact formulations for beams define the bilateral normal vector as the result of a bilateral closest-point projection, i.e., the minimization of the mutual distance of both beam centerline curves.
This difference is crucial because it carries over to the important topic of non-uniqueness and singularities of the definition (see e.g.~\cite{meier2016,Meier2017a}).
Here, the only critical geometric configuration is the case of parallel disk and cylinder~$\vt_1 \parallel \vt_2$.
In the use case considered here, one could overcome this issue with an alternative definition for this special case~$\vt_1 \parallel \vt_2$ (similar to the so-called all-angle beam contact (ABC)~\cite{Meier2017a}), however at the cost of an additionally required, smooth transition between both definitions.

In option B, $\vu_1$ is defined as the projection of the unilateral normal vector~$\vn_\text{ul}$ into the slave cross-section plane, which can equivalently be regarded as the construction of an orthonormal frame based on~$\vt_1$ and~$\vn_\text{ul}$.
This already reveals the singularity in this second possible definition of~$\vu_1$, which now occurs for~$\vt_1 \parallel \vn_\text{ul}$.
Fortunately, this scenario is by far less critical than~$\vt_1 \parallel \vt_2$ and will be discussed in further detail later on.

To sum up, we consider both options A and B as viable choices for~$\vu_1$ and want to further investigate both of them.
In fact, a third option C finally turned out to be the best compromise between accuracy and simplicity of the expression for our purposes.
\begin{itemize}
  \item option C: $\vu_1 \defvariable \vn_\text{ul}$
\end{itemize}
The main difference to options A and B is that $\vu_1$ does in general not lie exactly within the disk area, which makes it a less accurate and therefore less obvious choice than options A and B.
Depending on the use case, however, either of the options may be favorable, such that we present all three in the following.
For this reason, the introduction of generic scalar coefficients~$a_y,a_z,b_y,b_z,c \in \MR$ in \eqref{eq::point-cyl-sep_generic_coeff} conveniently allow us to directly obtain and compare the three different final expressions for the disk-cylinder potential~$\tilde \pi_{\text{m,disk-cyl}}$ once the analytical 2D integral has been solved as presented in the following steps.
For later reference, let us thus state the resulting expressions for these coefficients in both cases.
By inserting the definition of $\vu_1$ according to either option A or B into \eqref{eq::point-cyl-sep_generic_coeff}, the following expressions are obtained for the polynomial coefficients:
\begin{align}
  \text{option A:} \quad &a_y=1, \, a_z=\cos^2\alpha, \, a_{yz}=0, \, b_y=2 \, d_\text{ul} \sqrt{1- \frac{\cos^2\theta}{\sin^2\alpha}}, \, b_z =2 \, d_\text{ul} \frac{\cos\theta}{\tan\alpha}, c= d_\text{ul}^2\\
  \text{option B:} \quad &a_y=1-\frac{\cos^2\alpha}{\tan^2\theta}, \, a_z=\frac{\cos^2\alpha}{\sin^2\theta}, \, a_{yz}=-2\, \frac{\cos\alpha \, \cos\theta}{\sin^2\theta} \sqrt{\sin^2\alpha - \cos^2\theta},\\&b_y=2\, d_\text{ul} \, \sin\theta, \, b_z=0, \, c=d^2_\text{ul}\nonumber
\end{align}
For the sake of both brevity and clarity, here the dot products occurring in the expressions have been replaced by using the corresponding scalar angles enclosed by the vectors as follows.
\begin{align}
  \abs{\vt^T_1 \vt_2} =: \cos\alpha \quad \text{and} \quad \vt^T_1 \vn_\text{ul} =: \cos\theta
\end{align}\\

\noindent\textit{Remark on the minimal set of degrees of freedom.}
Note that the chosen set~$(d_\text{ul}, \alpha, \theta)$ is just one of the many different ways to uniquely describe the \textit{mutual} configuration of the disk-cylinder system.
Other choices include e.g.~any three of $d_\text{bl}$, $\vartheta$ with $\cos\vartheta \defvariable \vn_\text{bl}^T \vn_\text{ul}$, and the three aforementioned ones.
Our choice however avoids the non-uniqueness of the bilateral normal vector~$\vn_\text{bl}$ for~$\vt_1 \parallel \vt_2$ mentioned above and moreover appears to be most natural in the sense of yielding both compact and illustrative expressions.\\

\noindent\textit{Remark on the interpretation of the angle~$\theta$.}
Whereas the smallest distance between disk midpoint and cylinder axis~$d_\text{ul}$ and the angle included by the cylinder axis and the disk axis (i.e.~the normal to the disk surface)~$\alpha$ are straightforward to interpret and visualize, this seems harder for the angle~$\theta$.
It helps to think of a disk and a cylinder at fixed distance~$d_\text{ul}$ and inter-axis angle~$\alpha \neq 0$ (e.g.~perpendicular), and then begin to move the disk midpoint on the circle with radius~$d_\text{ul}$ around the closest point on the cylinder axis, while keeping~$\alpha$ fixed.
This is the interpretation of the third degree of freedom~$\theta$.
Now consider for instance the cases~$\alpha=\pi/2$ and~$\theta=0$ to see that this configuration will not occur for small~$d_\text{ul}$ if we consider the interaction of two (arbitrarily) curved fibers with bounded curvature, because the fibers would penetrate each other.
However, only these small values of~$d_\text{ul}$ are decisive for the two-fiber interaction potential, such that we can conclude that certain disk-cylinder configurations are less important for the two-fiber interaction potential.
This will be the motivation to later use a further simplified version of the disk-cylinder potential law~$\tilde \pi_\text{m,disk-cyl}$ in the context of the SBIP approach.\\

Coming back to the problem statement in \eqref{eq::disk_cyl_pot_m_5Dint}, we can now reformulate the initial problem to solve
\begin{align}\label{eq::2Dint_gap_inv_m_over_disk}
  \iint \limits_{A_\text{disk}} g_\text{pt-cyl}^{-m+3} \dd A \qquad \text{with} \quad A_\text{disk} = \{ \, (y_1,z_1) \, | \, y_1^2 + z_1^2 \leq R_1^2 \, \}
\end{align}
using $g_\text{pt-cyl}$ from \eqref{eq::gap_pt-cyl} and substituting $d_\text{pt-cyl}$ from \eqref{eq::point-cyl-sep_generic_coeff} to end up with the general expression for the smallest separation of the point and the cylinder \textit{surface}
\begin{align}\label{eq::gap_point-cyl_generic_coeff}
  g_\text{pt-cyl} = \sqrt{ a_y y_1^2 + a_z z_1^2 + a_{yz} y_1 z_1 + b_y y_1 + b_z z_1 + c } - R_2.
\end{align}
Note that up to this point the two options A and B from above are just two different yet equivalent ways of stating the identical problem.
If we could solve the problem defined in either of the two ways in an exact manner, the solution would of course be identical.
However, as outlined in the beginning of this section, our strategy to find an approximative solution includes the two steps to first express~$g_\text{pt-cyl}$ as a multivariate Taylor series expansion and second solve the 2D integral with this simplified integrand analytically.
The point of expansion hereby is of crucial importance and due to the nature of the inverse power interaction law, it should be located at the disk point with smallest disk-cylinder surface separation, where the by far largest contributions come from.
This is where the two options A and B again come into play, because we choose the point of expansion to be at~$(y_1=-R_1, z_1=0)$ in both cases, which means that the point is located on the disk contour and lies either on the bilateral normal direction vector for option A or on the projected unilateral normal direction vector for option B.
Both choices will be the optimal choice in terms of being located at the point of smallest surface separation for some mutual configurations, but not for all of them.
This motivates the investigation of both of them and a final judgment will later be made based on the resulting accuracy of the disk-cylinder potential for all mutual configurations.\\

\noindent\textit{Remark on alternative solution attempts.}
Note that several other approaches to solve the 2D integral over the circular cross-section area have been investigated, yet did not lead to any exact analytical solutions and thus superior accuracy and simplicity of the final disk-cylinder potential expression.
These unsuccessful other attempts include e.g.~coordinate transformations in polar coordinates and the description of the projected rotated disk as an ellipse.\\

The required multivariate series expansion of \eqref{eq::gap_point-cyl_generic_coeff} finally reads
\begin{align}\label{eq::gap_point-cyl_generic_coeff_multivariate_taylor}
  \text{Lin} \left[ g_\text{pt-cyl} \right]_{y_1=-R_1, z_1=0} \, = &\,
          \overbrace{\beta - R_2}^{=:\tilde c}
          \, + \, \overbrace{\frac{1}{2\beta} \left( b_y - 2 a_y R_1 \right)}^{=:\tilde b_y} (y_1+R_1)
          \, + \, \overbrace{\frac{1}{2\beta} \left( b_z - a_{yz} R_1 \right)}^{=:\tilde b_z} z_1\\
          &+ \frac{1}{2} \left( \frac{a_z}{\beta} - \frac{(b_z-a_{yz}R_1)^2}{4\beta^3} \right) z_1^2 \, + \, \text{H.O.T.},\nonumber
\end{align}
where the auxiliary variable~$\beta\defvariable \sqrt{c-b_y R_1 + a_y R_1^2}$ and further abbreviations~$\tilde c$, $\tilde b_y$, $\tilde b_z$ have been introduced for the later use.
Note also that we have already used the knowledge from the subsequent accuracy analysis here that the second order term in~$z_1$ is indeed decisive for the overall accuracy whereas neglecting the other second and higher order terms still gives us good results.
Continuing from \eqref{eq::2Dint_gap_inv_m_over_disk}, the 2D integral can thus be further simplified by
\begin{align}\label{eq::2Dint_gap_inv_m_approx_series_exp}
  \int \limits_{-R_1}^{R_1} \int \limits_{-\sqrt{R_1^2 - z_1^2}}^{\sqrt{R_1^2 - z_1^2}} g_\text{pt-cyl}^{-m+3} \dd y_1 \dd z_1
  \approx \int \limits_{-R_1}^{R_1} \int \limits_{-\sqrt{R_1^2 - z_1^2}}^{\sqrt{R_1^2 - z_1^2}} \left( \text{Lin} \left[ g_\text{pt-cyl} \right]_{y_1=-R_1, z_1=0} \right)^{-m+3} \dd y_1 \dd z_1,
\end{align}
for which a closed-form antiderivative exists for the inner integral in $y_1$ (see e.g.~\cite[p.1017]{Bronshtein2003}):
\begin{align}
  \int ( \tilde b_y (y_1+R_1) + \ldots )^{-m+3} \dd y_1 = \frac{{\tilde b_y}^{-1}}{(-m+4)} (\tilde b_y (y_1+R_1) + \ldots)^{-m+4}, \quad m \neq 4
\end{align}
To keep the expressions simple and enable the subsequent analytical integration in~$z_1$, we once again exploit the fact that the contributions from point-pairs decay rapidly with increasing distance and set the upper integration limit in~$y_1$ to infinity.
Following the same reasoning, the lower integration limit is replaced by its second-order Maclaurin series expansion at~$z_1=0$
\begin{align}
  -\sqrt{R_1^2 - z_1^2} \approx -R_1+z_1^2/(2R_1).
\end{align}
The error introduced by this approximation is expected to be small because the point of expansion and its immediate vicinity include the most important closest point pair.
For a more detailed analysis of the approximation quality, again refer to \secref{sec::verif_approx_singlelengthspec}.
Finally, the integral in~$y_1$ can be solved as follows:
\begin{align}\label{eq::int_y_gap_inv_m_approx_int_limits}
  \int \limits_{-\sqrt{R_1^2 - z_1^2}}^{\sqrt{R_1^2 - z_1^2}} g_\text{pt-cyl}^{-m+3} \dd y_1
  &\approx \lim \limits_{y_\text{max} \rightarrow \infty } \int  \limits_{-R_1+z_1^2/(2R_1)}^{ y_\text{max} } \left( \text{Lin} \left[ g_\text{pt-cyl} \right]_{y_1=-R_1, z_1=0} \right)^{-m+3} \dd y_1\\
  &= \frac{{\tilde b_y}^{-1}}{(-m+4)} ( \tilde c + \tilde b_z z_1 + \underbrace{\left( \frac{\tilde b_y}{2R_1} + \frac{a_z}{2\beta} - \frac{\tilde b_z^2}{2\beta} \right)}_{=:\tilde a_z} z_1^2 )^{-m+4}\label{eq::disk_cyl_pot_m_integral_y_solved}
\end{align}
The remaining fifth and final integral to be evaluated analytically is the one in transversal direction~$z_1$ within the disk-shaped cross-section area.
Naturally, this last step turns out to be the critical point and many of the mentioned other attempts to find an analytical solution for the disk-cylinder interaction potential~$\tilde \pi_\text{m,disk-cyl}$ failed here.
For this specific formulation of the problem described above and the simplifications based on the previously discussed assumptions, an analytical antiderivative exists and is stated in a recursive manner for a generic exponent~$m$ (see e.g.~\cite[p.1019]{Bronshtein2003}):
\begin{align}
  \int ( \underbrace{\tilde a_z z_1^2 + \tilde b_z z_1 + \tilde c}_{=:Z} )^{-m+4} \dd z_1 &= \frac{2 \tilde a_z z_1 +\tilde b_z}{(m-5) \, \Delta} Z^{-m+5} + \frac{2 (2m-11) \, \tilde a_z}{(m-5) \, \Delta} \int Z^{-m+5} \dd z_1\nonumber\\
  &\text{with} \quad (m-4)>1\label{eq::disk_cyl_pot_m_integral_z_antideriv}\\
  \int Z^{-1} \dd z_1 &= \frac{2}{\sqrt{\Delta}} \arctan \left( \frac{2 \tilde a_z z_1 + \tilde b_z }{\sqrt{\Delta}} \right) \quad \text{for} \quad \Delta > 0 \nonumber
\end{align}
Here, the introduced dimensionless quantity~$\Delta \defvariable 4 \tilde a_z \tilde c - \tilde b_z^2$ represents the negative discriminant of the quadratic expression~$Z(z_1)$, which can be identified as~$\text{Lin}[g_{\text{pt-cyl}}]$ evaluated at~$y_1=-R_1+z_1^2/(2R_1)$, i.e., the (approximated) distance between the points on the disk's boundary and the corresponding closest point on the cylinder surface.
From this interpretation, we can follow that~$\Delta>0$ will hold true for all physically sensible scenarios, because the distance will always be a positive value, i.e.,~have no roots and the corresponding discriminant~$\tilde b_z^2 - 4 \tilde a_z \tilde c$ will be negative.

\subsection{Solutions}
Substituting \eqref{eq::disk_cyl_pot_m_integral_y_solved} into \eqref{eq::2Dint_gap_inv_m_approx_series_exp} and making use of the analytical antiderivative (\eqref{eq::disk_cyl_pot_m_integral_z_antideriv}) finally allows us to find a closed-form analytical expression for~$\tilde \pi_\text{m,disk-cyl}$ for any given exponent~$m\geq6$.
The expression will however be lengthy, such that we make a final approximation and replace the exact integration domain~$z_1\in[-R_1,R_1]$ by~$z_1\in]-\infty,\infty[$, which significantly simplifies the expression, because all the recursive terms~$Z^{-m+5}$ from \eqref{eq::disk_cyl_pot_m_integral_z_antideriv} vanish and the~$\arctan$-function evaluates to~$\pm \pi/2$, respectively.
Once again, this is expected to be a good approximation, since only the point pairs in the vicinity of the closest point yield significant contributions to the value of the integral.

\paragraph{Options A and B.}
Finally, the sought-after analytical solution for the disk-cylinder interaction potential with generic exponent thus reads
\begin{align}\label{eq::disk-cyl-pot_m_general}
  \tilde \pi_\text{m,disk-cyl} = \rho_1 K_\text{m} \, \tilde b_y^{-1} \, \tilde a_z^{m-5} \, \Delta^{-m+9/2},
\end{align}
where all the constants have been collected in a newly introduced prefactor~$K_\text{m}$.
For the two parts of the LJ potential and using the point-cylinder potential law%
\footnote{
See \appref{sec::point-cyl-pot_m_comparison} for a comparison of the two alternative expressions for the point-cylinder interaction potential~$\Pi_\text{m,pt-cyl}$ and the reason for using the one presented in \appref{sec::point-halfspace-pot_m_derivation}.
}
from \appref{sec::point-halfspace-pot_m_derivation}, these prefactors are given as
\begin{align}
  K_6 \defvariable \frac{1}{3} \pi^2 \, k_6 \, \rho_2 \quad \text{and} \quad K_{12} \defvariable \frac{286}{15} \pi^2 \, k_{12} \, \rho_2
\end{align}
for the adhesive $m=6$ and repulsive $m=12$ part, respectively.
At this point, we have found an analytical expression for (both parts of) the LJ interaction potential~$\tilde \pi_\text{LJ,disk-cyl}$ of a disk and a cylinder of infinite length valid for arbitrary mutual orientations~$\alpha,\theta$ in the decisive regime of small separations~$g_\text{ul} \ll R_{1/2}$.
Now recall the discussion of the two options A and B for the definition of~$\vu_1$ and thus the associated direction of integration in~$y_1$ as well as the point of expansion at~$y_1=-R_1,z_1=0$.
Re-substitution of the auxiliary variables as follows -- together with the general solution of \eqref{eq::disk-cyl-pot_m_general} -- gives a first impression of the complexity of the different expressions for option A and B.
\begin{align}
  \text{option A:} \quad &\beta = \sqrt{d_\text{ul}^2 -2 R_1 d_\text{ul} \sqrt{1-\cos^2\theta/\sin^2\alpha} + R_1^2 }\nonumber\\
                         &\tilde b_y = \frac{1}{\beta} \left( d_\text{ul} \sqrt{1-\cos^2\theta/\sin^2\alpha} - R_1 \right)\nonumber\\
                         &\tilde b_z = \frac{1}{\beta} d_\text{ul} \cos\theta \cos\alpha / \sin \alpha \label{eq::disk-cyl-pot_m_auxvariables_optA}\\
                         &\tilde a_z = \frac{\tilde b_y}{2 R_1} + \frac{\cos^2\alpha}{2\beta} - \frac{\tilde b_z^2}{2\beta}\nonumber\\
                         &\tilde c = \beta - R_2\nonumber\\
                         &\Delta = 4 \tilde a_z \tilde c - \tilde b_z^2\nonumber
\end{align}
\begin{align}
  \text{option B:} \quad &\beta = \sqrt{ d_\text{ul}^2 - 2 R_1 d_\text{ul} \sin \theta + \left( 1 - \cos^2\alpha \cos^2\theta / \sin^2\theta \right) R_1^2 }\nonumber\\
                         &\tilde b_y = \frac{1}{\beta} \left( d_\text{ul} \sin \theta - ( 1 - \cos^2\alpha \cos^2\theta / \sin^2\theta ) \, R_1 \right)\nonumber\\
                         &\tilde b_z = \frac{1}{\beta} \, \cos\alpha \, \cos\theta / \sin^2\theta \, \sqrt{ \sin^2\alpha - \cos^2\theta } \, R_1 \label{eq::disk-cyl-pot_m_auxvariables_optB}\\
                         &\tilde a_z = \frac{\tilde b_y}{2R_1} + \frac{\cos^2\alpha}{2\beta \, \sin^2\theta} - \frac{\tilde b_z^2}{2\beta}\nonumber\\
                         &\tilde c = \beta - R_2\nonumber\\
                         &\Delta = 4 \tilde a_z \tilde c - \tilde b_z^2\nonumber
\end{align}
Looking at these expressions, we can state that both length and complexity of the terms is similar for both options A and B.
As mentioned before, the resulting accuracy will be analyzed in~\secref{sec::verif_approx_singlelengthspec} and thus complete the assessment of these two options.

\paragraph{Option C -- Final solution to be used in the simulation framework.}
At this point, however, let us turn to a possibility to simplify the resulting expressions without significant loss of accuracy.
In this respect, the following third option C turned out to be a good compromise between accuracy and simplicity of the expression:
\begin{itemize}
  \item option C: $\vu_1 \defvariable \vn_\text{ul}$
\end{itemize}
The main difference to options A and B is that $\vu_1$ does in general not lie exactly within the disk area, which makes it a less obvious choice than options A and B.
However, it will turn out in~\secref{sec::verif_approx_singlelengthspec} that the influence of this approximation on the accuracy is rather insignificant.
In analogy to the derivation for the options A and B, the auxiliary variables follow as
\begin{align}
  \text{option C:} \quad &\beta=d_\text{ul}-R_1\nonumber\\
                         &\tilde b_y = 1\nonumber\\
                         &\tilde b_z = 0 \label{eq::disk-cyl-pot_m_auxvariables_optC}\\
                         &\tilde a_z = \frac{1}{2R_1} + \frac{\cos^2\alpha}{2(d_\text{ul}-R_1)}\nonumber\\
                         &\tilde c = d_\text{ul}-R_1-R_2 =: g_\text{ul}\nonumber\\
                         &\Delta = 4 \tilde a_z \, g_\text{ul},\nonumber
\end{align}
and substitution into the general form of the disk-cylinder interaction potential in \eqref{eq::disk-cyl-pot_m_general} results in the pleasantly simple expression
\begin{align}
  \tilde \pi_\text{m,disk-cyl} = \rho_1 K_\text{m} \, \tilde a_z^{-\frac{1}{2}} \, (4 \, g_\text{ul})^{-m+\frac{9}{2}}.
\end{align}
Note that this option C is equivalent to both options A and B for the special case of~$\theta=\pi/2$, i.e., $\cos\theta=\vt_1^T\vn_\text{ul}=0$ and thus~$\vn_\text{ul} \perp \vt_1$ and~$\vn_\text{ul}\equiv \vn_\text{bl}$.
This means that the scalar angle~$\theta$ has been eliminated and only two degrees of freedom remain.
Now recall from the interpretation of the angle~$\theta$ given above that certain configurations of the disk-cylinder system are more important -- and in fact decisive -- for the two-fiber interaction potential and others are rather irrelevant.
This is the motivation for the special value~$\theta=\pi/2$.
Consider the fact that $\cos\theta=\sin\alpha \sin\vartheta$ and thus $\theta=\pi/2$ if either~$\vartheta=0$, i.e.~$\vn_\text{ul}\parallel \vn_\text{bl}$, which in turn means that we are at the bilateral closest-point pair of the two fibers, or if~$\alpha=0$, i.e., we have parallel beam axes and again a bilateral closest-point pair, which is non-unique in this case.
To conclude, all the disk-cylinder configurations that are decisive for the two-fiber interaction will have~$\theta\approx \pi/2$.
Therefore, the option C disk-cylinder potential law will yield a very high accuracy if applied together with the general SBIP approach on the level of fiber-fiber interactions.
In this way, we have eliminated the least relevant degree of freedom in order to obtain a reduced and thus simpler SBIP law.%
\footnote{
Note the analogy to the discussion of reduced SSIP laws in our previous contribution~\cite{GrillSSIP}.
}

Note that such a simple SBIP expression is especially desirable for the later use in the resulting virtual work contribution and its linearization in the context of an implicit, nonlinear finite element framework for structural dynamics.
In this use case, the required two-fold differentiation of the potential law with respect to the discrete set of primary degrees of freedom can become tedious and at some point unfeasible if the potential law is too complex.
The same reasoning applies to the replacement of the exact integration domain in~$z_1$ by~$z_1\in]-\infty,\infty[$.
In other use cases however, the above presented options A and B of the disk-cylinder interaction potential expressions may still be considered reasonably simple and due to the increased accuracy (especially for those configurations with~$\theta$ far from~$\pi/2$, as discussed above) they may be of great value.
For these reasons, they are included here despite the fact that solely the option C will be used in the final beam interaction formulation to be applied in the numerical examples of this work.

In this light, one small further simplification can be achieved by recalling the initial restriction to the dominating regime of small separations~$g_\text{ul} \ll R_{1/2}$ and thus applying
\begin{align}
  d_\text{ul}-R_1 = g_\text{ul} + R_2 \approx R_2
\end{align}
to the coefficients in \eqref{eq::disk-cyl-pot_m_auxvariables_optC}, which finally leads to
\begin{align}\label{eq::disk-cyl-pot_m}
  \tilde \pi_\text{m,disk-cyl} = \hat K_\text{m} \, \rho_1 \, \sqrt{\frac{2 R_1 R_2}{R_1 \, \cos^2 \alpha + R_2}} \, g_\text{ul}^{-m+\frac{9}{2}} \quad \text{with } \hat K_\text{m} \defvariable 4^{-m+\frac{9}{2}} \, K_\text{m}, \quad m \geq 6.
\end{align}
For convenience in later reference, we explicitly state the most common prefactors for the vdW part~$m=6$ and the repulsive part~$m=12$ of the LJ potential as follows:
\begin{align}
  \hat K_6 = \frac{1}{24} \pi^2 \, k_6 \, \rho_2 \quad \text{and} \quad \hat K_{12} = \frac{143}{15 \cdot 2^{14} } \pi^2 \, k_{12} \, \rho_2
\end{align}
At this point, we have arrived at the final form (\eqref{eq::disk-cyl-pot_m}) of the disk-cylinder interaction potential to be used as reduced interaction law in the context of the SBIP approach\cite{GrillSBIP}.

An immediate verification of these expressions for the special case~$\alpha=0$ confirms that both~$\tilde \pi_\text{6,disk-cyl}$ and~$\tilde \pi_\text{12,disk-cyl}$ are identical to the independently derived analytical solutions for the interaction potential per unit length~$\tilde \pi_\text{6,cyl$\parallel$cyl}$ and~$\tilde \pi_\text{12,cyl$\parallel$cyl}$ of two infinitely long, parallel cylinders (cf.~Equations~(A23) and~(A24) in our previous contribution~\cite{GrillSSIP}).
This is an important finding, as it shows the consistency of the more general expression (\eqref{eq::disk-cyl-pot_m}) valid for all mutual angles~$\alpha$ with previously derived expressions for the important special case~$\alpha=0$.
A much broader and deeper analysis of the accuracy of \eqref{eq::disk-cyl-pot_m} as well as a comparison to the expressions obtained for the other options A and B above will be the content of the following~\secref{sec::verif_approx_singlelengthspec}.

\section{Verification}
\label{sec::verif_approx_singlelengthspec}
This section aims to verify the specific SBIP law~$\tilde \pi_\text{m,disk-cyl}$ derived in the preceding \secref{sec::disk-cyl-pot_m_derivation} in the context of the general SBIP approach\cite{GrillSBIP}.
Let us first recall the underlying assumptions of the general evaluation strategy:
\begin{itemize}
  \item pair-wise summation
  \item very short range of interactions and focus on small separation regime~$g \ll R_\text{1/2}$
  \item linear Taylor expansion of the master beam volume
  \item choice of master and slave
\end{itemize}
Moreover, the following assumptions and approximations have been made in order to obtain a closed-form analytical solution for the disk-cylinder interaction potential at arbitrary mutual orientations in the previous section.
\begin{itemize}
  \item multivariate Taylor expansion of point-cylinder distance, see \eqref{eq::2Dint_gap_inv_m_approx_series_exp}
  \item Taylor expansion of lower integration limit~$y_\text{min}(z)$ and set upper integration limit to $\infty$, see \eqref{eq::int_y_gap_inv_m_approx_int_limits}
  \item set integration limits~$z_\text{min/max}$ to~$\pm \infty$, see comment above \eqref{eq::disk-cyl-pot_m_general}
  \item use of the point-half space expression~$\Pi_\text{m,pt-hs}$ as point-cylinder interaction potential~$\Pi_\text{m,pt-cyl}$, see \eqref{eq::pot_ia_vdW_point-cylinder}
  \item further simplification by using option C:~$\vu_1=\vn_\text{ul}$ and finally~$g_\text{ul}+R_2 \approx R_2$, see steps above \eqref{eq::disk-cyl-pot_m}
\end{itemize}
To allow for a clear analysis of the resulting accuracy, the minimal example of two cylinders, which represents the special case of two straight beams with circular cross-section, is considered in the following.
In this case, analytical reference solutions are known for the limit of small separations and specifically the influence of the second list of assumptions can be assessed, because the first list is exactly fulfilled.

\subsection{Van der Waals interaction potential of two cylinders for all separations and all mutual angles}
\label{sec::verif_disk-cyl-pot_twocylinders}
\begin{figure}[htpb]
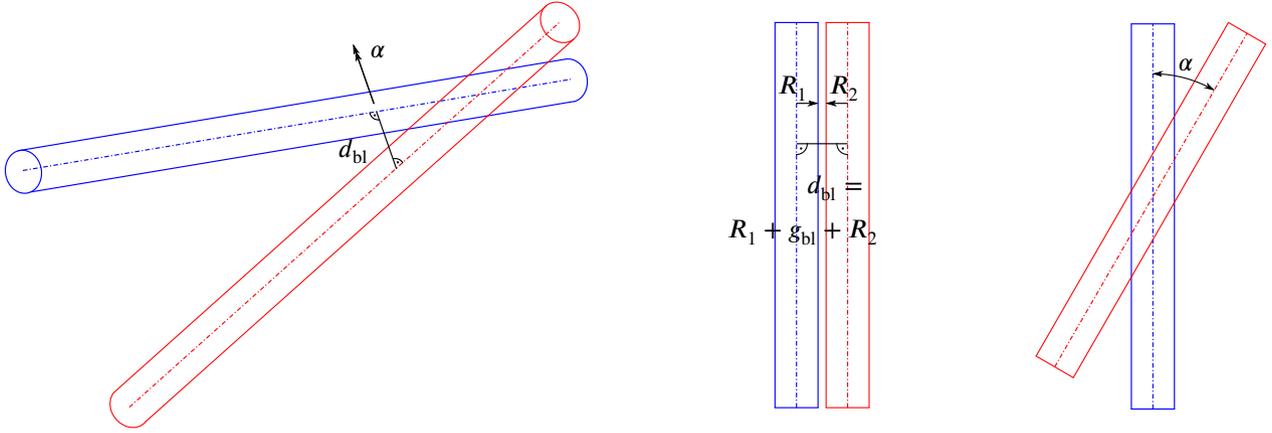
%
  \centering
  \vspace{-1cm}
  \def\svgwidth{0.45\textwidth}
  \input{two_cylinders_bilateral_gap_mutal_angle_perspective_view.pdf_tex}
  \def\svgwidth{0.45\textwidth}
  \hspace{-4.5cm}
  \input{two_cylinders_bilateral_gap_mutal_angle_side_view.pdf_tex}
  \def\svgwidth{0.45\textwidth}
  \hspace{-3cm}
  \input{two_cylinders_bilateral_gap_mutal_angle_top_view.pdf_tex}
  \caption{Illustration of the two-cylinder interaction scenario used for verification purposes.
           Perspective view (left), side view (center), and top view (right) of the cylinders including their smallest inter-axis distance~$d_\text{bl}$ and mutual angle~$\alpha$.}
  \label{fig::two_cylinders_bilateral_gap_mutal_angle}
\end{figure}

The relative configuration of two cylinders is uniquely described by their (bilateral) smallest inter-axis separation~$d_\text{bl}$ and their mutual angle~$\alpha$, as depicted in~\figref{fig::two_cylinders_bilateral_gap_mutal_angle}.
Note that the surface separation also known as gap~$g_\text{bl}=d_\text{bl}-R_1-R_2$ will often be used instead of the inter-axis separation~$d_\text{bl}$ in the following discussion.
Exemplarily, vdW interaction with exponent~$m=6$ of the point pair potential will be considered, however the results are expected to be analogous for the repulsive part of LJ or any other short-ranged interaction.
As mentioned before, analytical reference solutions obtained via 6D analytical integration of the point-pair vdW potential~$\Phi_\text{vdW}$ for mutual angles~$\alpha \in \, ]0,\pi/2]$ (see \eqref{eq::pot_ia_vdW_cyl_cyl_skewed_smallseparation} and e.g.~\cite[p.~173]{parsegian2005}) as well as the special case of parallel cylinders~$\alpha=0$ (see e.g.~\cite[p.~172]{parsegian2005} and the quick reference in Table 1 of our previous contribution~\cite{GrillSSIP}) are available from literature.
However, keep in mind that these reference solutions are derived for the limit of small separations~$g_\text{bl} \ll R_{1/2}$ and \textit{infinite} length of the cylinders such that the solution for parallel cylinders is given as interaction potential per unit length~$\tilde \pi_\text{vdW,cyl$\parallel$cyl}$ instead of the total two-body interaction potential~$\Pi_\text{vdW,cyl-cyl}$, which would be infinite.
For simplicity, a fixed length~$L=20$ and radius~$R=1$ is chosen exemplarily for the numerically evaluated solutions, which turned out to have no noticeable influence as long as the slenderness~$L/R$ is sufficiently large.

\begin{figure}[htpb]%
  \centering
   \subfigure[Mutual angle $\alpha=0^\circ$]{
    \includegraphics[width=0.45\textwidth]{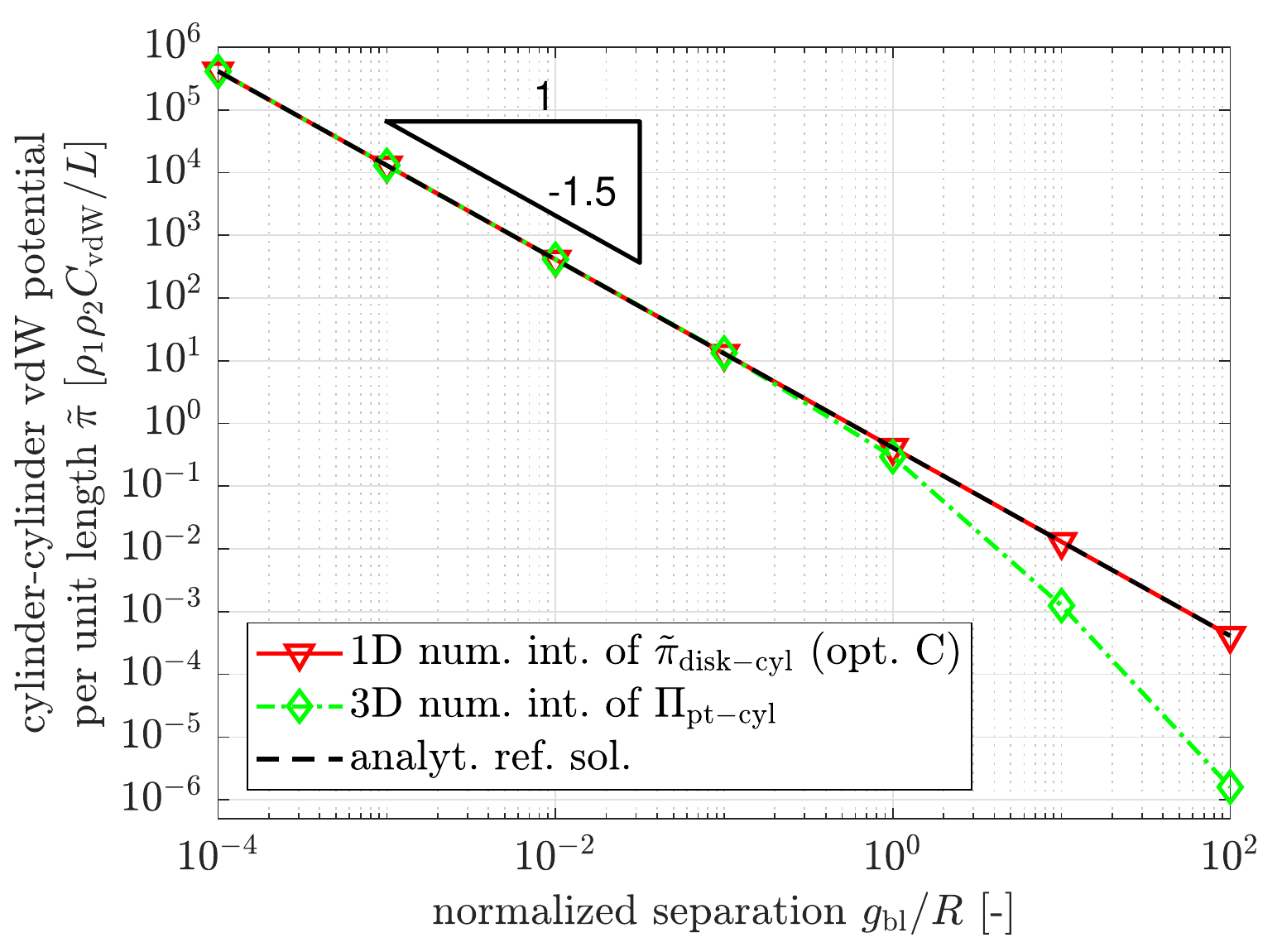}
    \label{fig::cyl-cyl_ia_pot_SBIP_optC_over_sep_angle0}
   }
   \subfigure[Mutual angle $\alpha=2.8125^\circ$]{
    \includegraphics[width=0.45\textwidth]{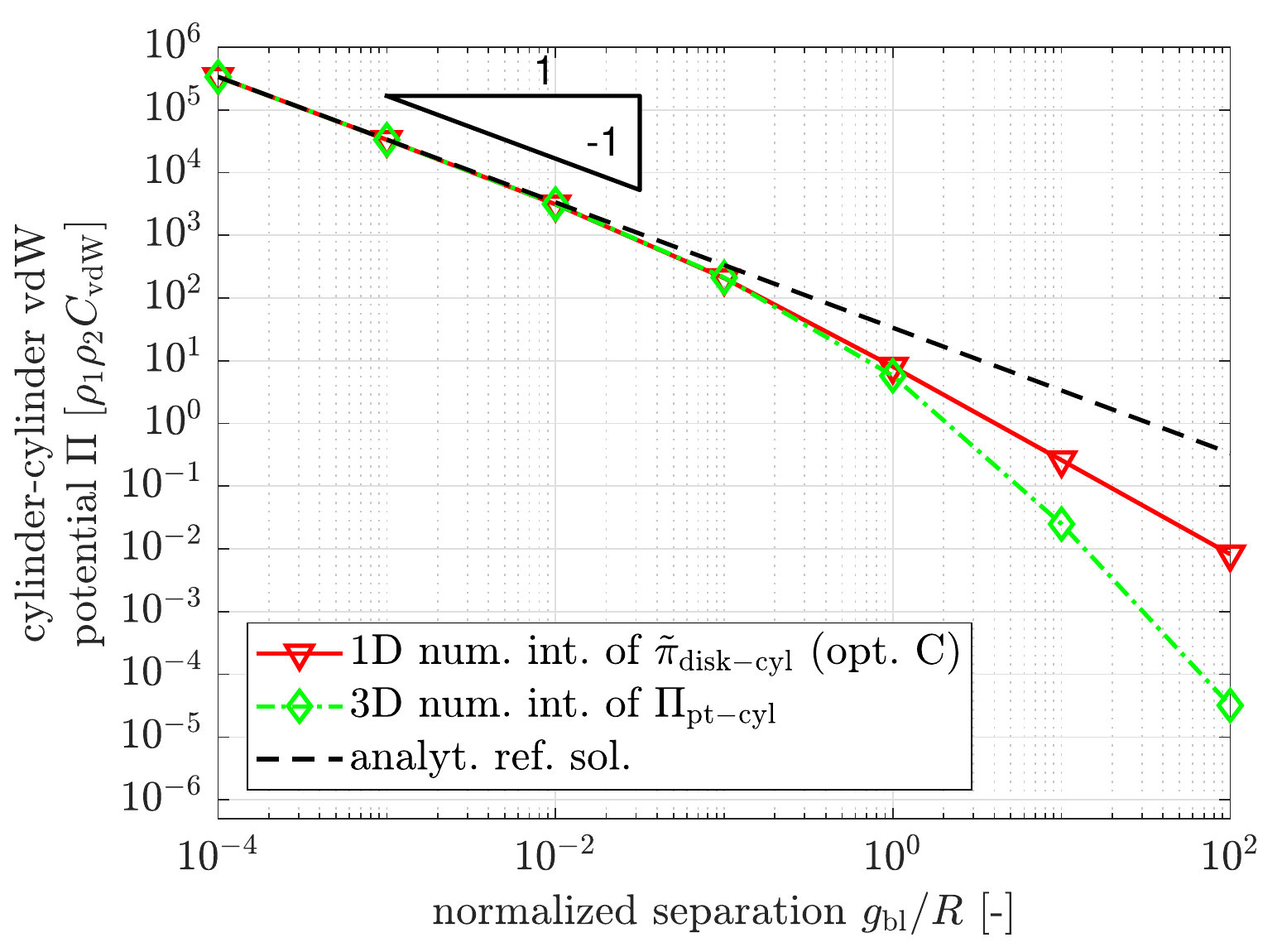}
    \label{fig::cyl-cyl_ia_pot_SBIP_optC_over_sep_angle2_8125}
   }
   \subfigure[Mutual angle $\alpha=11.25^\circ$]{
    \includegraphics[width=0.45\textwidth]{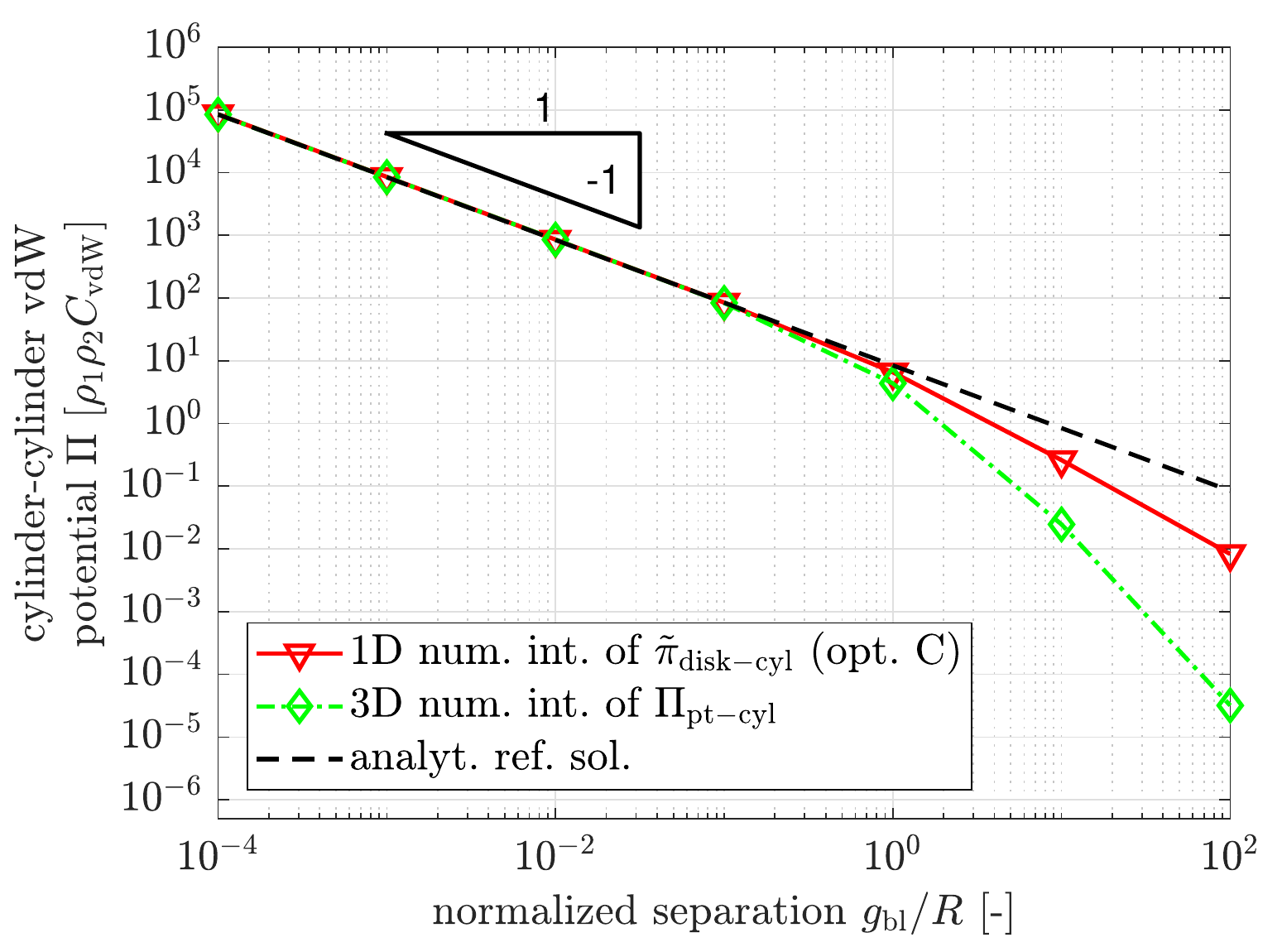}
    \label{fig::cyl-cyl_ia_pot_SBIP_optC_over_sep_angle11_25}
   }
   \subfigure[Mutual angle $\alpha=22.5^\circ$]{
    \includegraphics[width=0.45\textwidth]{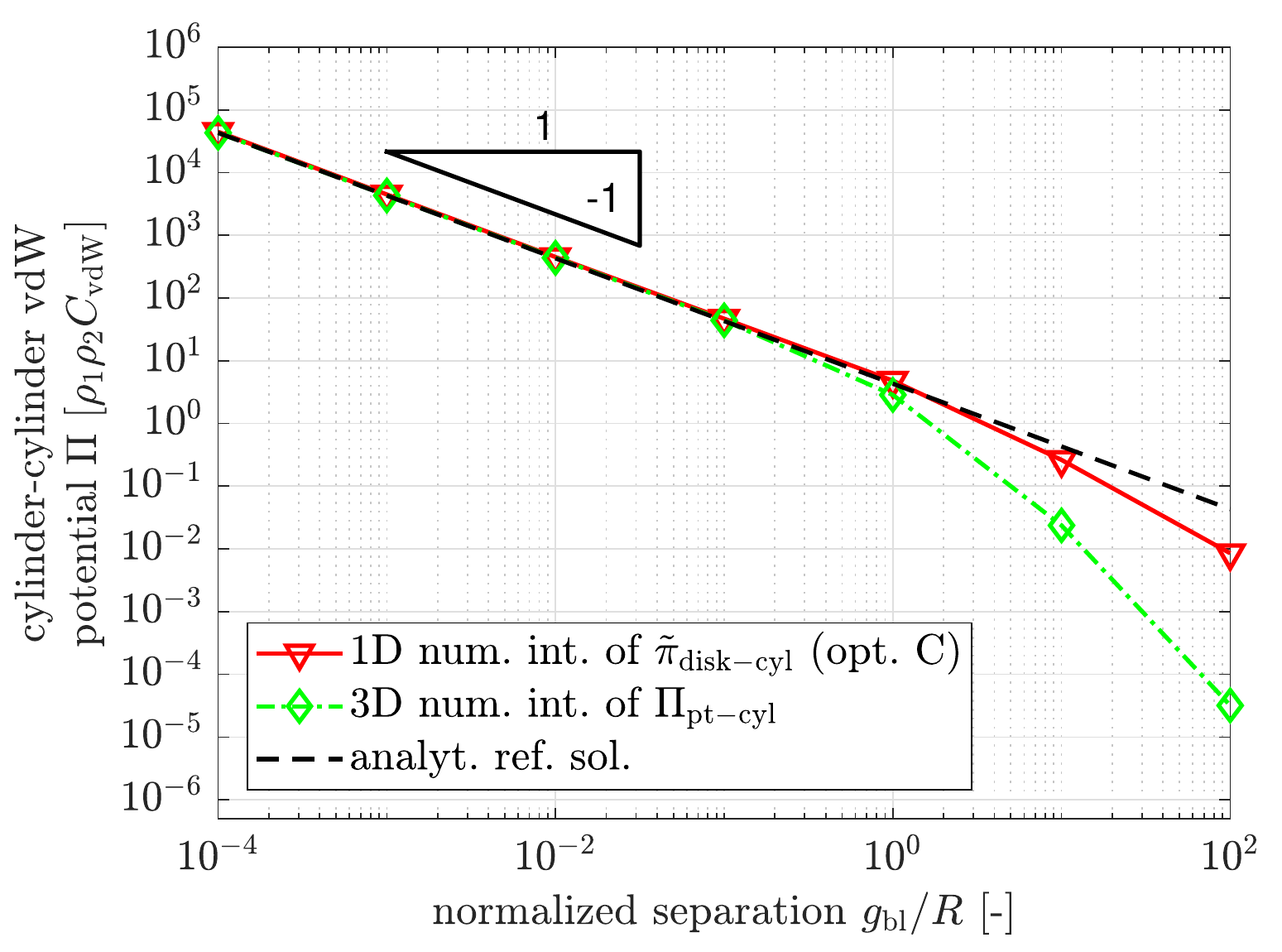}
    \label{fig::cyl-cyl_ia_pot_SBIP_optC_over_sep_angle22_5}
   }
   \subfigure[Mutual angle $\alpha=45^\circ$]{
    \includegraphics[width=0.45\textwidth]{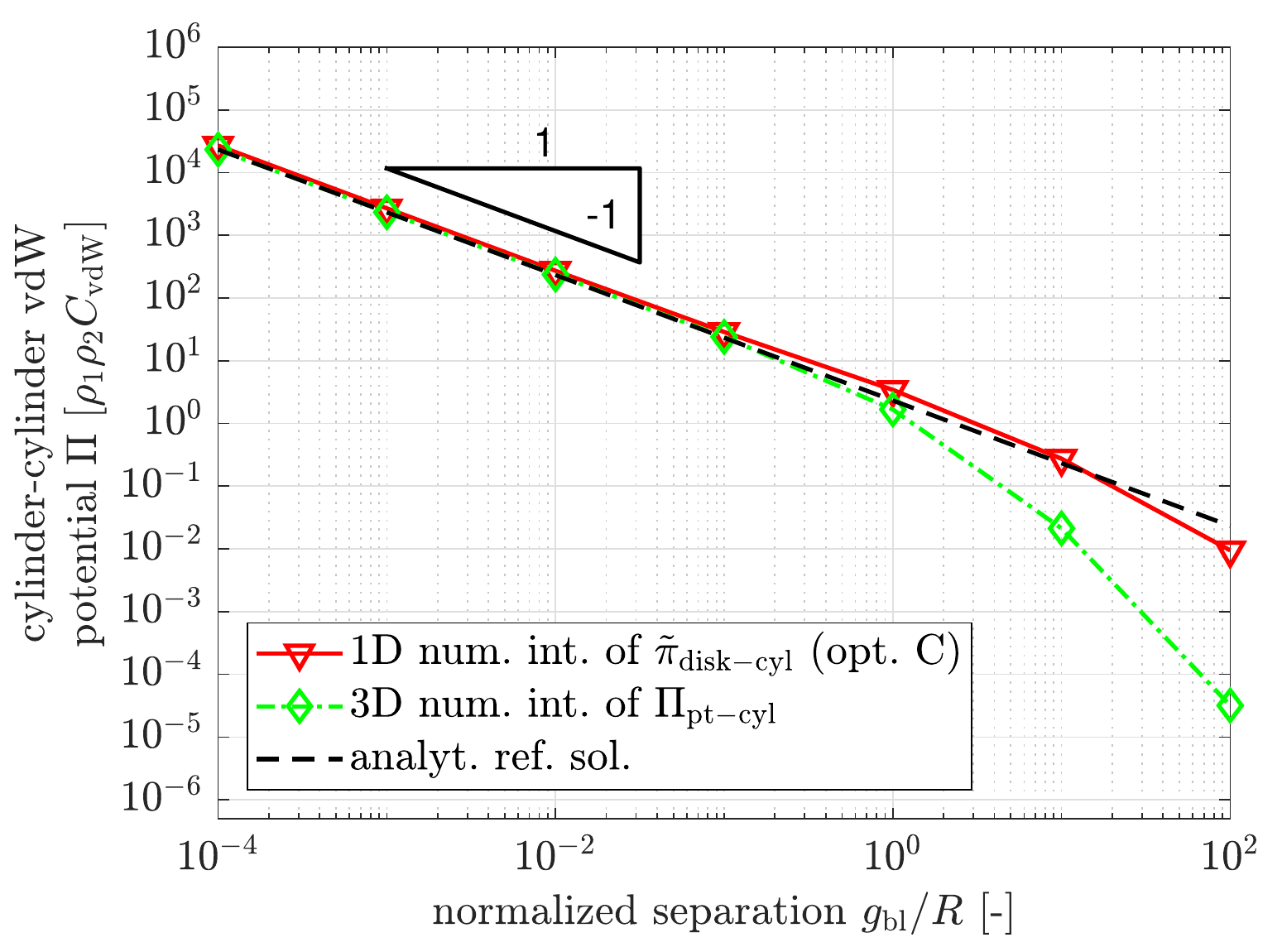}
    \label{fig::cyl-cyl_ia_pot_SBIP_optC_over_sep_angle45}
   }
   \subfigure[Mutual angle $\alpha=90^\circ$]{
    \includegraphics[width=0.45\textwidth]{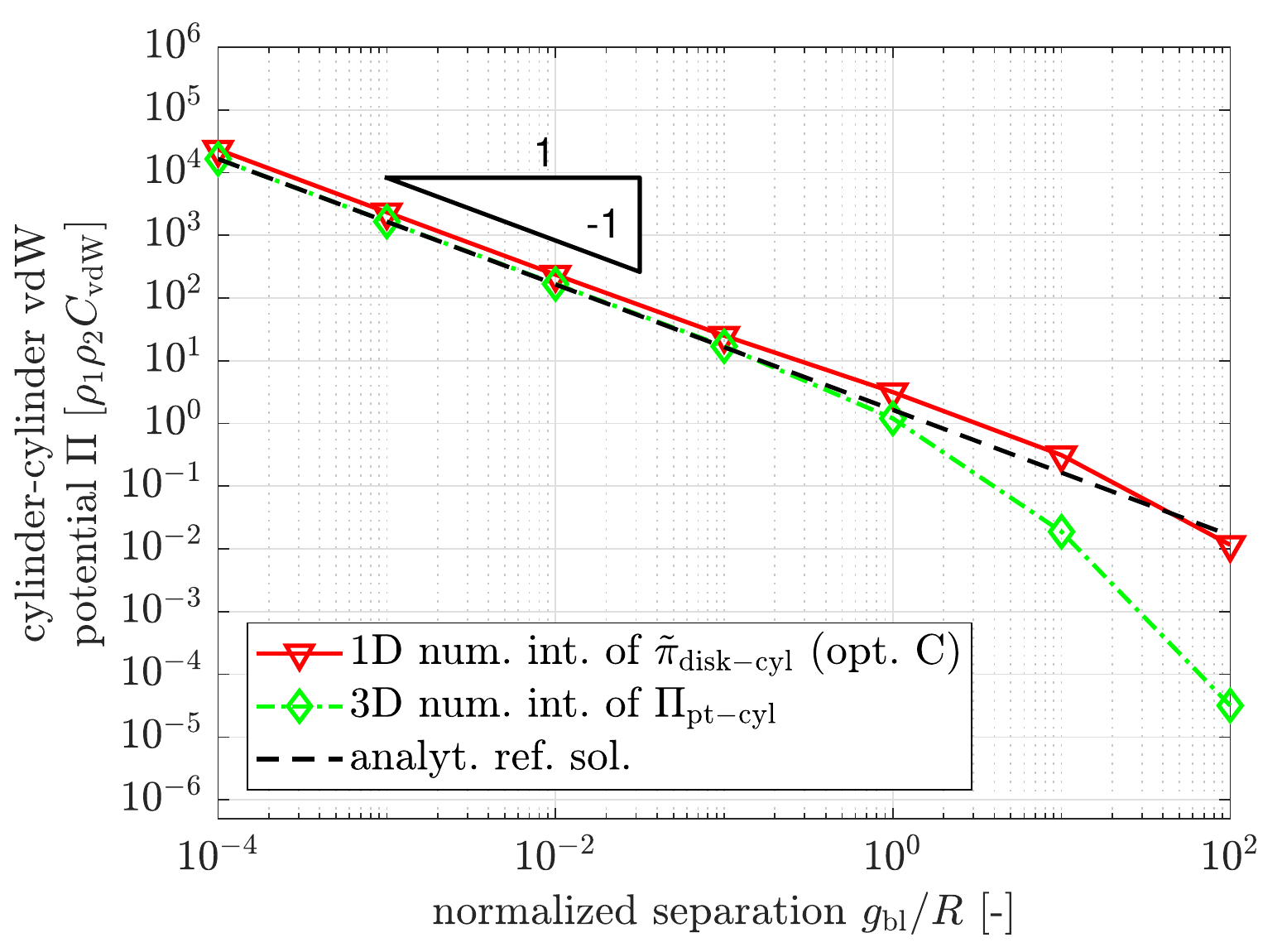}
    \label{fig::cyl-cyl_ia_pot_SBIP_optC_over_sep_angle90}
   }
   \caption{Interaction potential of two cylinders as a function of the dimensionless minimal surface separation~$g_\text{bl}/R$ at different mutual angles~$\alpha$. Verification of the analytical expression for the disk-cylinder potential~$\tilde \pi_\text{6,disk-cyl}$ from \eqref{eq::disk-cyl-pot_m} (used together with the SBIP approach\cite{GrillSBIP}; red line with triangles) by means of a numerical reference solution obtained via 3D Gaussian quadrature of the point-half space potential~$\Pi_\text{6,pt-hs}$ from \eqref{eq::pot_ia_vdW_point-cylinder} (green line with diamonds) and by means of analytical reference solutions summarized in~\secref{sec::introduction} (black dashed line).}
  \label{fig::cyl-cyl_ia_pot_SBIP_optC_over_sep}
\end{figure}

\figref{fig::cyl-cyl_ia_pot_SBIP_optC_over_sep} shows a double-logarithmic plot of the dimensionless vdW interaction potential~$\Pi_\text{vdW,cyl-cyl}$ as a function of the dimensionless surface separation~$g_\text{bl}/R$ at different mutual angles~$\alpha$.
The derived analytical expression for the disk-cylinder interaction potential~$\tilde \pi_\text{6,disk-cyl}$ (option C) from \eqref{eq::disk-cyl-pot_m} is numerically integrated along the second cylinder axis (1D) as proposed in the general SBIP approach\cite{GrillSBIP} and depicted as red line with triangles.%
\footnote{
Gaussian quadrature with 5 integration points has been applied in each of the 40 integration segments, which were chosen increasingly fine around the bilateral closest point of the cylinders, and it has been verified that a further increase of the number of integration points has no noticeable influence on the results.%
}
For verification, the analytical reference solutions for parallel and skew cylinders as summarized in~\secref{sec::introduction} are plotted as black dashed line.
In addition, the point-half space potential~$\Pi_\text{6,pt-hs}$ from \eqref{eq::pot_ia_vdW_point-cylinder}, which has been used as point-cylinder potential already in the analytical 5D integration to end up with the disk-cylinder potential~$\tilde \pi_\text{6,disk-cyl}$ is used for 3D numerical integration over the entire volume of the second cylinder and shown as green line with diamonds.%
\footnote{
In addition to the 1D integration scheme above, we applied Gaussian quadrature with 12 integration segments in $z_1$- and 16 segments in $y_1$-direction (see \figref{fig::disk_cylinder_potential_coord_frames} for the definition of the coordinates), once again with adaptive fineness and 5 integration points each, and verified that a further increase of the number of integration points has no noticeable influence on the results.%
}
This option serves two purposes at the same time.
First, \figref{fig::cyl-cyl_ia_pot_SBIP_optC_over_sep} shows that the two-cylinder interaction potential obtained in this way perfectly matches the analytical reference solutions derived for the limit of small separations.
This verifies that the point-half space potential is the consistent approximation for the point-cylinder potential under the assumptions of short-ranged interactions at small separations as discussed and motivated in~\ref{sec::point-cyl-pot_m_comparison} and is thus an important step of verification also of the disk-cylinder interaction potential~$\tilde \pi_\text{m,disk-cyl}$.
And second, it is an important numerical reference solution, because it allows to judge the accuracy of the assumptions and simplifications made in the steps of analytically integrating the point-cylinder potential over the disk-shaped slave cross-section (see list of assumptions above).
Note also in this respect that obtaining a fully numerical reference solution via 6D numerical integration of the point-pair potential~$\Phi_\text{vdW}$ once again failed due to its infeasible computational cost, especially in the decisive regime of small separations.
Refer to our previous contribution~\cite{GrillSSIP} for a more detailed discussion of this topic.
The 3D numerical integration of the point-cylinder potential~$\Pi_\text{pt-cyl}$ therefore is a valuable reference solution for the two-cylinder potential~$\Pi_\text{vdW,cyl-cyl}$ in the regime of intermediate separations, where no analytical reference solution is known.
For the sake of completeness, note that the regime of large separations is covered in our previous contribution~\cite{GrillSSIP} as well.
However, it is of minor practical interest in the case of short-ranged interactions considered here.

Let us now have a detailed look at the most important topic of the accuracy of the analytical disk-cylinder interaction potential~$\tilde \pi_\text{6,disk-cyl}$ (option C) from \eqref{eq::disk-cyl-pot_m}.
First and foremost, the accuracy is excellent in the case of parallel cylinders shown in~\figref{fig::cyl-cyl_ia_pot_SBIP_optC_over_sep_angle0}, which is no surprise as it has already been stated in an immediate assessment at the end of~\secref{sec::disk-cyl-pot_m_derivation} that~$\tilde \pi_\text{m,disk-cyl}$ coincides with the analytical reference solution for the interaction potential per unit length of parallel cylinders~$\tilde \pi_\text{m,cyl$\parallel$cyl}$.
As a consequence, the asymptotic scaling behavior being an inverse power-law with exponent 1.5 is correctly reproduced.
Taking into account also the reference from 3D numerical integration, the accuracy is found to be excellent even for~$g_\text{bl}/R \lesssim 1$, which is already well into the region of intermediate separations and the interaction potential values have dropped by several orders of magnitude.
To give a number, the relative error is approx.~2.3\% for~$g_\text{bl}/R = 0.1$, increases to approx.~39\% for~$g_\text{bl}/R = 1$, and decreases as expected with decreasing separation.
Considering the next plot~\ref{fig::cyl-cyl_ia_pot_SBIP_optC_over_sep_angle2_8125} for~$\alpha=\pi/64$, it is striking to see that the asymptotic scaling behavior now follows the~$g_\text{bl}^{-1}$ law as theoretically predicted for skew cylinders~$\alpha\neq 0$ in \eqref{eq::pot_ia_vdW_cyl_cyl_skewed_smallseparation}.
Despite this sharp transition between both cases~$\alpha=0$ and~$\alpha \neq 0$, the SBIP approach with the ``option C'' disk-cylinder law~$\tilde \pi_\text{m,disk-cyl}$ again shows the correct asymptotic scaling behavior and agrees very well with the 3D numerical reference solution up to separations of~$g_\text{bl}/R \lesssim 1$.
The same statements hold for all other mutual angles shown in~\figref{fig::cyl-cyl_ia_pot_SBIP_optC_over_sep_angle11_25} - \ref{fig::cyl-cyl_ia_pot_SBIP_optC_over_sep_angle90}, although one important point requires some more discussion.
It is clearly visible for~$\alpha=\pi/2$ and noticeable above~$\alpha \approx \pi/4$ that the ``option C'' disk-cylinder law~$\tilde \pi_\text{m,disk-cyl}$ no longer approaches the correct level of the asymptotic solution for small separations, however still shows the correct~$g_\text{bl}^{-1}$ scaling behavior.
This deviation from the analytical as well as numerical reference solution by an almost constant factor of e.g.~approx.~$1.5$ for the worst case~$\alpha=\pi/2$ can be attributed to the additional simplifications made for option C of~$\tilde \pi_\text{m,disk-cyl}$ and is not observable for the more accurate yet more complex option A and B expressions as will be shown in a subsequent analysis further down.
At this point, the most important conclusion to take away from analyzing the accuracy of the ``option C'' disk-cylinder potential~$\tilde \pi_\text{m,disk-cyl}$ as a reduced interaction law within the general SBIP approach therefore is that the accuracy is best for the regime of small separations~$g_\text{bl}/R \lesssim 1$ and particularly for small angles, which is by far the most important one, because the interaction potential values are by far the highest.
Note also in this respect that the only stable equilibrium configuration of two adhesive fibers is the one of straight parallel fibers and it seems especially important to achieve the highest accuracy at and around this special configuration.
Finally, and maybe even most important is the finding that the correct asymptotic scaling behavior~$g_\text{bl}^{-1.5}$ and $g_\text{bl}^{-1}$ is met for both distinctive cases~$\alpha=0$ and $\alpha\neq0$, respectively.

\subsubsection{Specific investigation of the scaling behavior with respect to the mutual angle}
\begin{figure}[htpb]%
  \centering
   \vspace{-5pt}
   \subfigure[Surface separation $g_\text{bl}/R=10^{-3}$]{
    \includegraphics[width=0.45\textwidth]{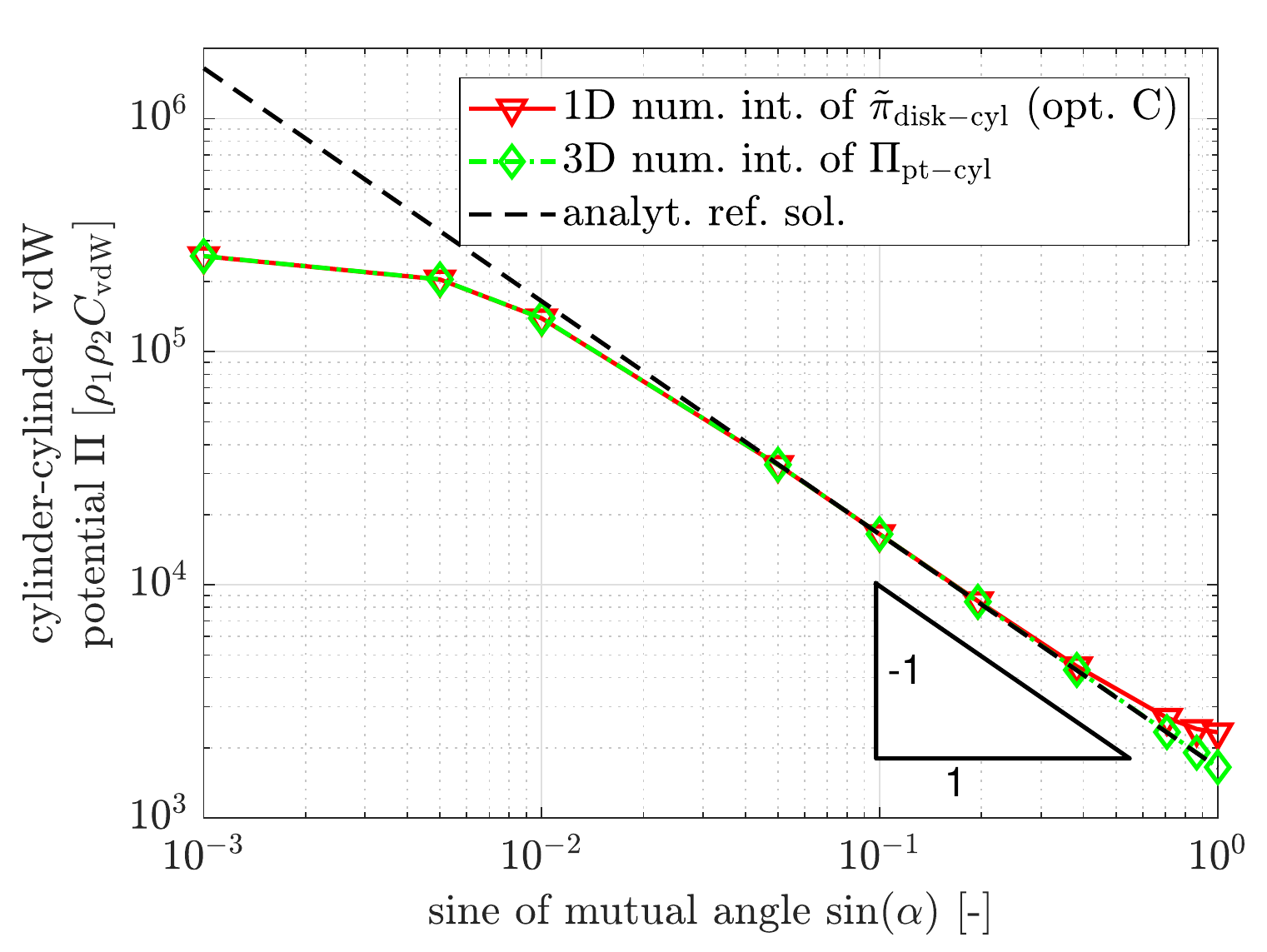}
    \label{fig::cyl-cyl_ia_pot_SBIP_optC_over_angle_sep1e-3}
   }
   \vspace{-5pt}
   \subfigure[Surface separation $g_\text{bl}/R=10^{-2}$]{
    \includegraphics[width=0.45\textwidth]{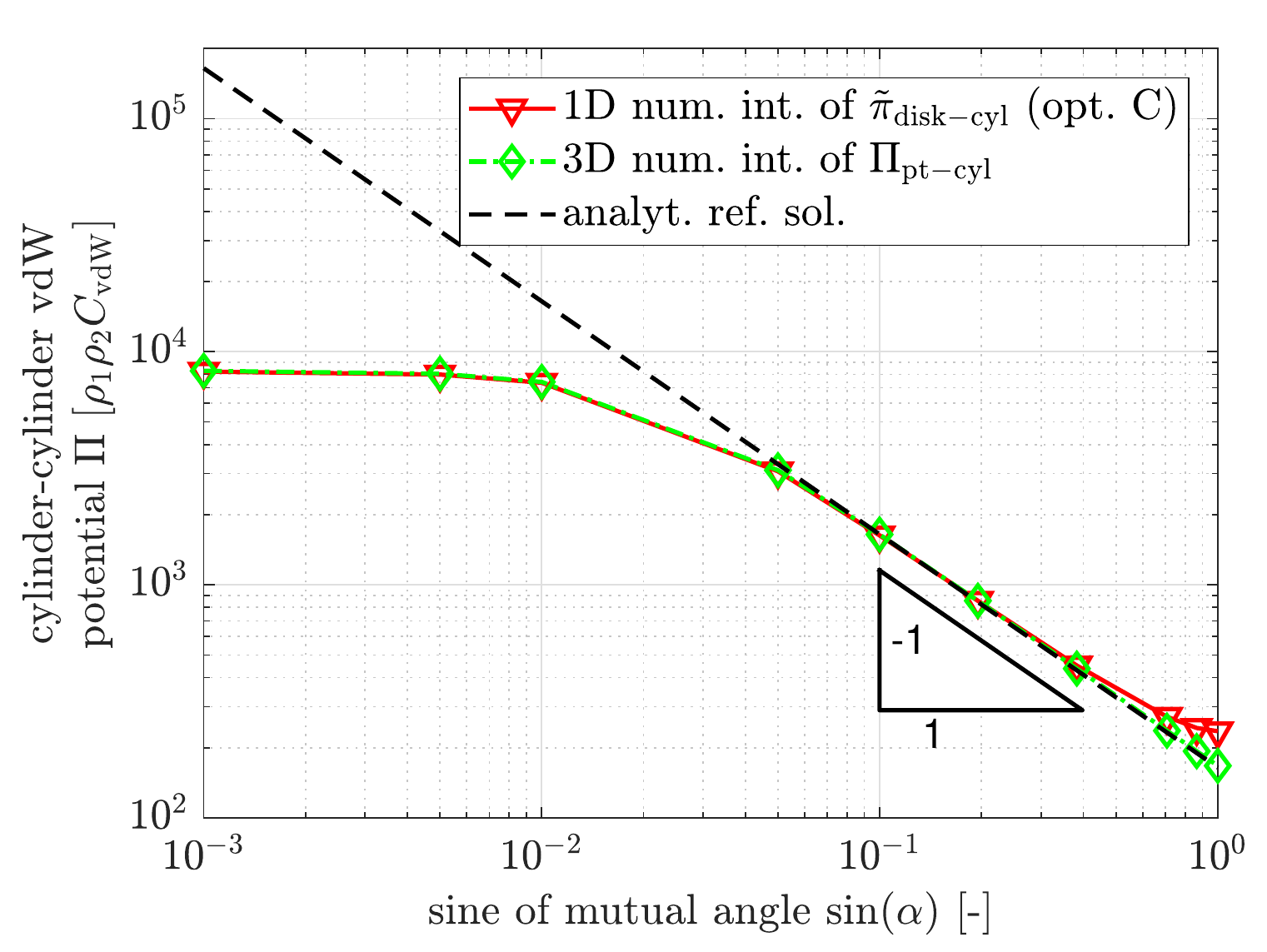}
    \label{fig::cyl-cyl_ia_pot_SBIP_optC_over_angle_sep1e-2}
   }
   \vspace{-5pt}
   \subfigure[Surface separation $g_\text{bl}/R=10^{-1}$]{
    \includegraphics[width=0.45\textwidth]{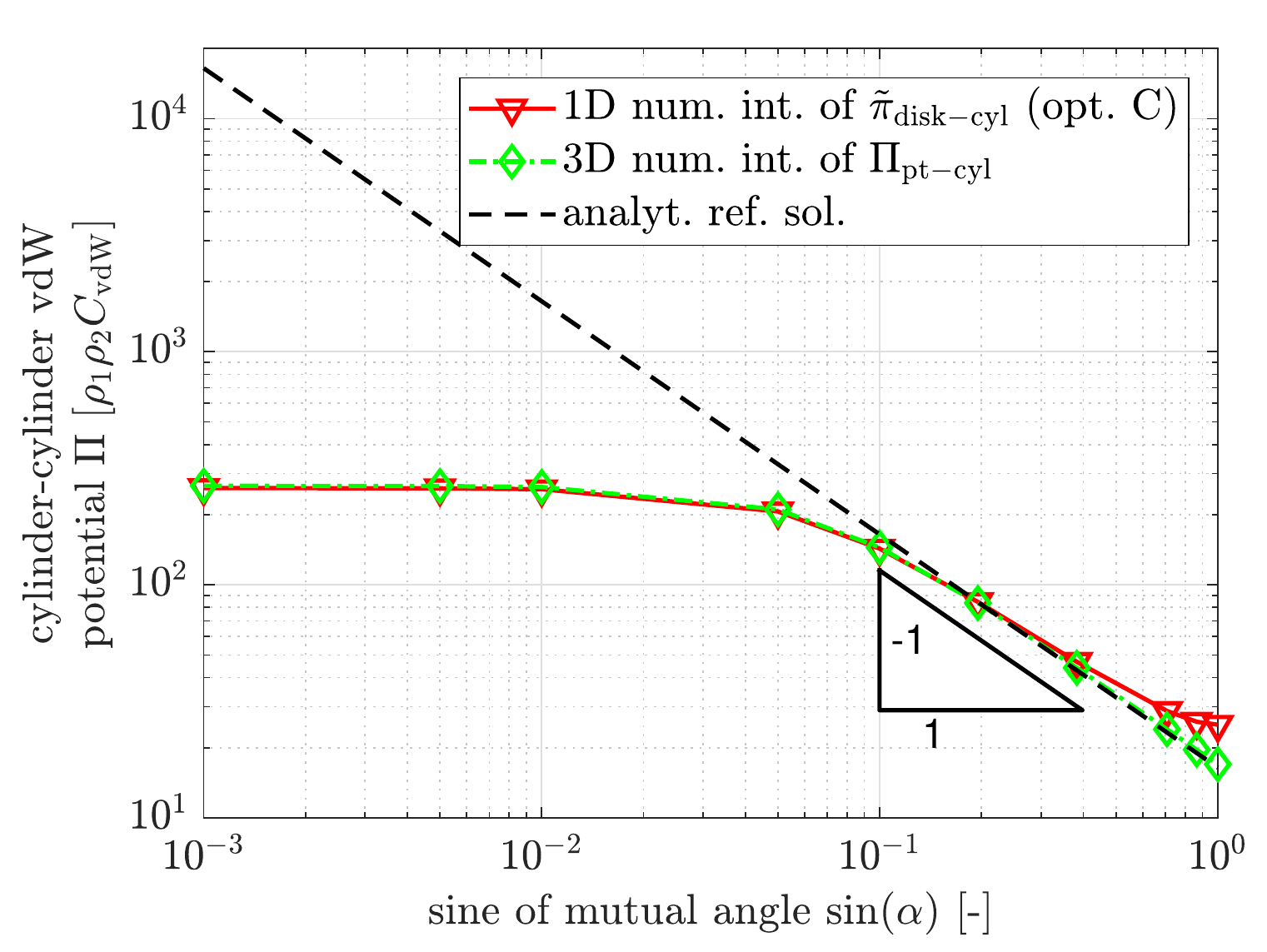}
    \label{fig::cyl-cyl_ia_pot_SBIP_optC_over_angle_sep1e-1}
   }
   \vspace{-5pt}
   \subfigure[Surface separation $g_\text{bl}/R=1$]{
    \includegraphics[width=0.45\textwidth]{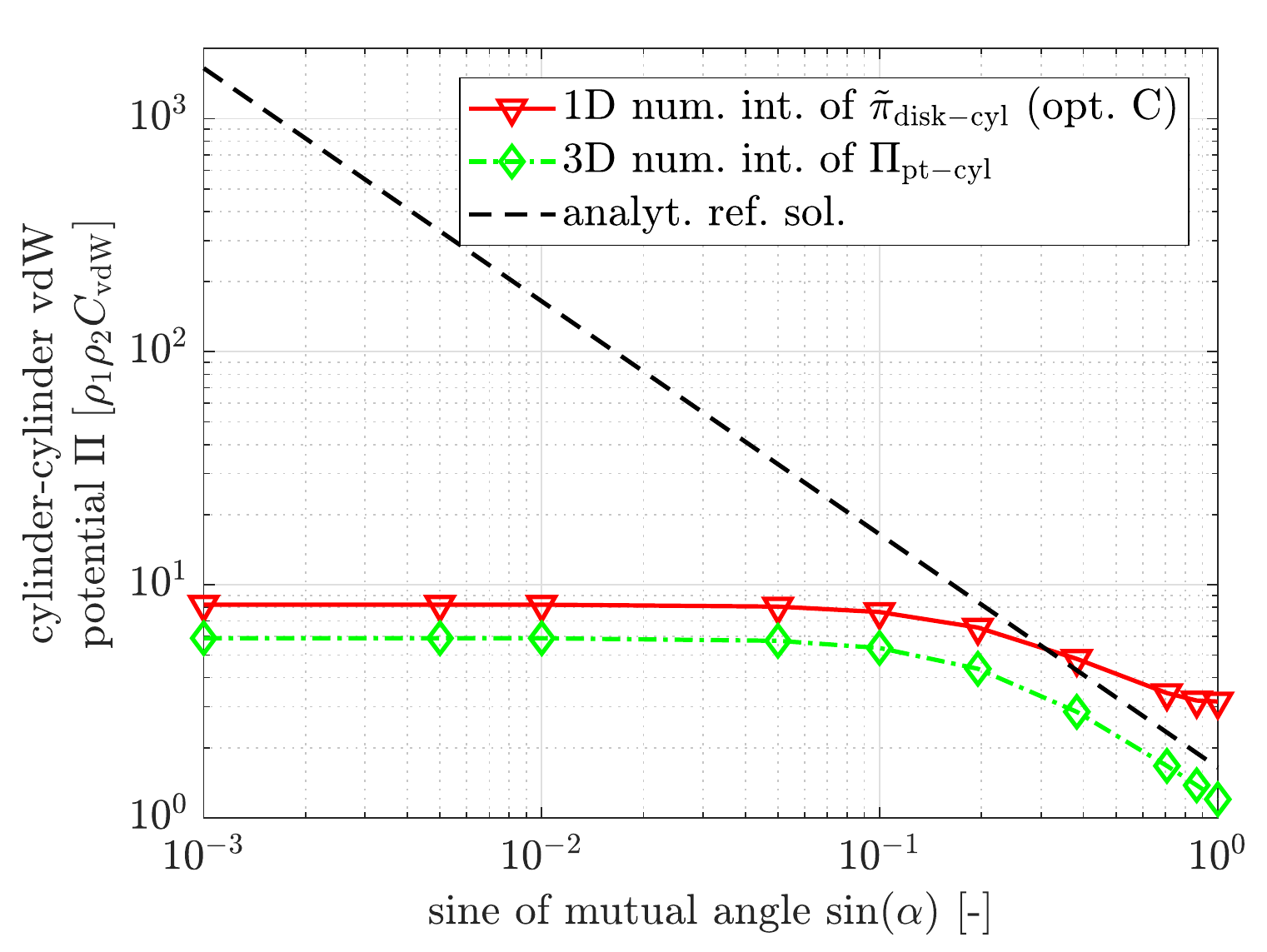}
    \label{fig::cyl-cyl_ia_pot_SBIP_optC_over_angle_sep1e0}
   }
   \caption{Interaction potential of two cylinders as a function of the sine of the mutual angle at different smallest surface separations~$g_\text{bl}/R$. Verification of the analytical expression for the disk-cylinder potential~$\tilde \pi_\text{6,disk-cyl}$ from \eqref{eq::disk-cyl-pot_m} (used together with the SBIP approach\cite{GrillSBIP}; red line with triangles) by means of a numerical reference solution obtained via 3D Gaussian quadrature of the point-half space potential~$\Pi_\text{6,pt-hs}$ from \eqref{eq::pot_ia_vdW_point-cylinder} (green line with diamonds) and by means of analytical reference solutions summarized in~\secref{sec::introduction} (black dashed line).}
  \label{fig::cyl-cyl_ia_pot_SBIP_optC_over_angle}
\end{figure}

$\,$\\
Having observed the sharp transition between the cases~$\alpha=0$ and~$\alpha \neq 0$ above it seems worth to have a more specific look at the angle dependency and especially the~$\sin\alpha^{-1}$ scaling behavior as theoretically predicted by \eqref{eq::pot_ia_vdW_cyl_cyl_skewed_smallseparation} for the limit of small separations.
The double-logarithmic plots in \figref{fig::cyl-cyl_ia_pot_SBIP_optC_over_angle} thus complement the analysis above by showing the dimensionless interaction potential as a function of the sine of the mutual angle~$\alpha$ for various separations.
Note the different scales on the vertical axes, which once again underline the importance of the small separation regime.
The considered scenario of two cylinders and the three different solutions for the two-cylinder interaction potential are identical to the previous \figref{fig::cyl-cyl_ia_pot_SBIP_optC_over_sep}.
Most importantly, the theoretically predicted scaling behavior is confirmed by the numerical reference solution and reproduced by the disk-cylinder potential law~$\tilde \pi_\text{6,disk-cyl}$ (option C) from \eqref{eq::disk-cyl-pot_m}.
Moreover, one can clearly observe the limits of validity of the analytical reference solution (\eqref{eq::pot_ia_vdW_cyl_cyl_skewed_smallseparation}) and particularly the
predicted $1/\sin\alpha$ scaling due to the underlying assumptions of small separations and~$\alpha \neq 0$.
In order to include also the special value~$\alpha=0$ of parallel cylinders and to get a more intuitive impression of the change of the interaction potential over the angle~$\alpha \in \, [0,\pi/2]$, the equivalent plots as function of the angle~$\alpha$ and in semi-logarithmic style are provided in \appref{sec::SBIP_verification_additional_plots} (cf.~\figref{fig::cyl-cyl_ia_pot_SBIP_optC_over_angle_semilog}).

\subsubsection{Comparison of options A, B and C}
\begin{figure}[htpb]%
  \centering
   \subfigure[Mutual angle $\alpha=0^\circ$]{
    \includegraphics[width=0.45\textwidth]{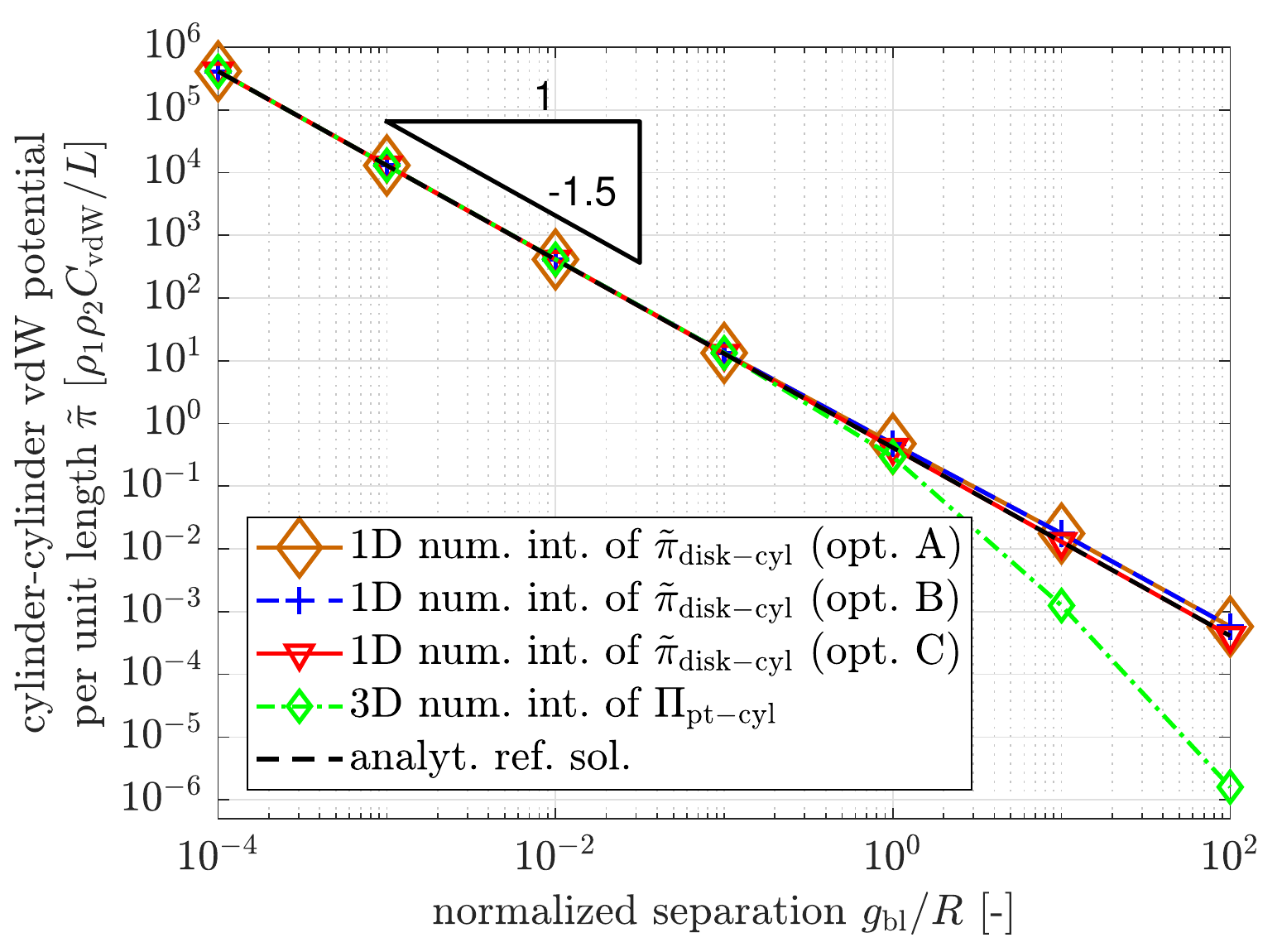}
    \label{fig::cyl-cyl_ia_pot_SBIP_optABC_over_sep_angle0}
   }
   \subfigure[Mutual angle $\alpha=2.8125^\circ$]{
    \includegraphics[width=0.45\textwidth]{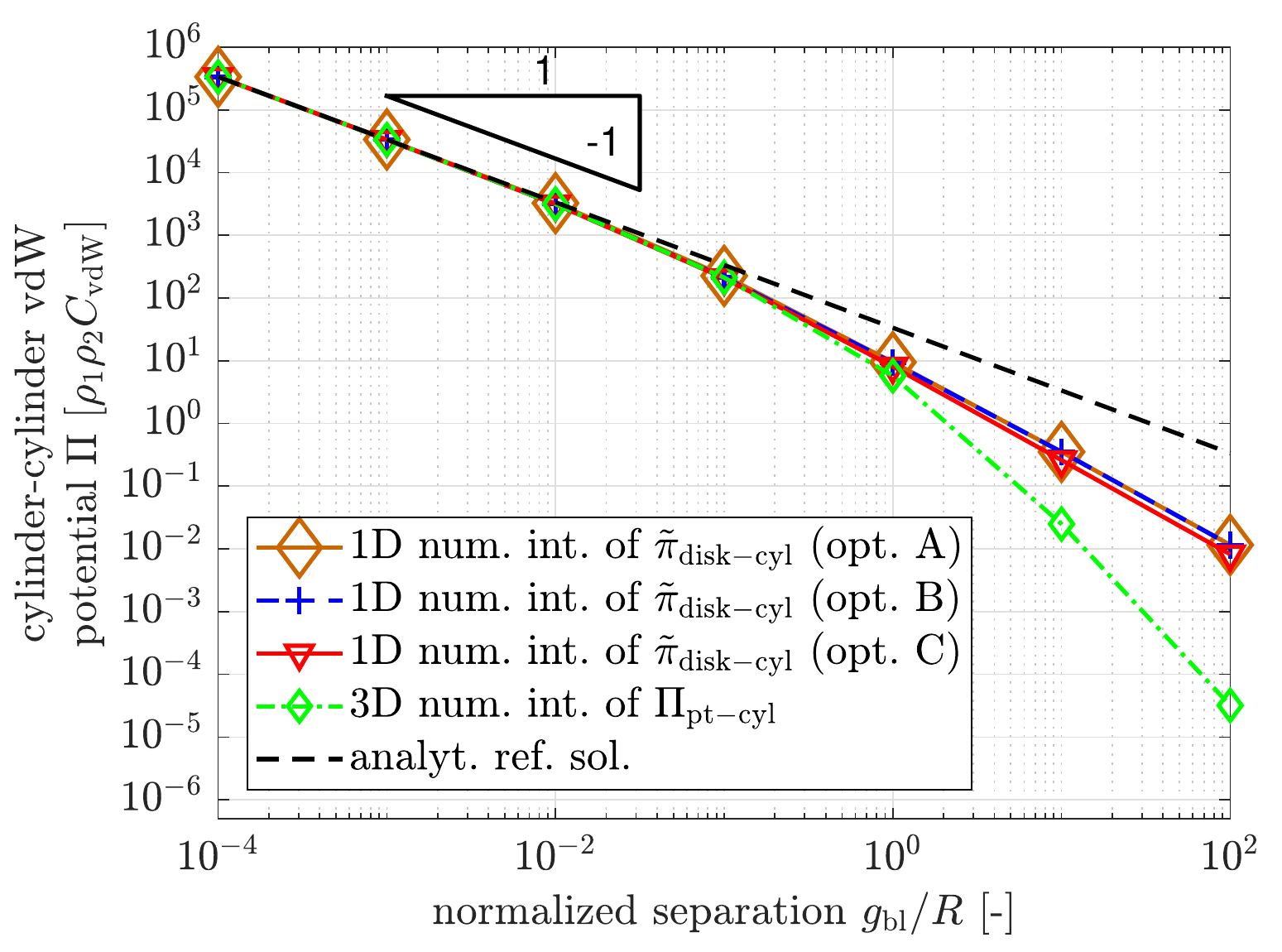}
    \label{fig::cyl-cyl_ia_pot_SBIP_optABC_over_sep_angle2_8125}
   }
   \subfigure[Mutual angle $\alpha=45^\circ$]{
    \includegraphics[width=0.45\textwidth]{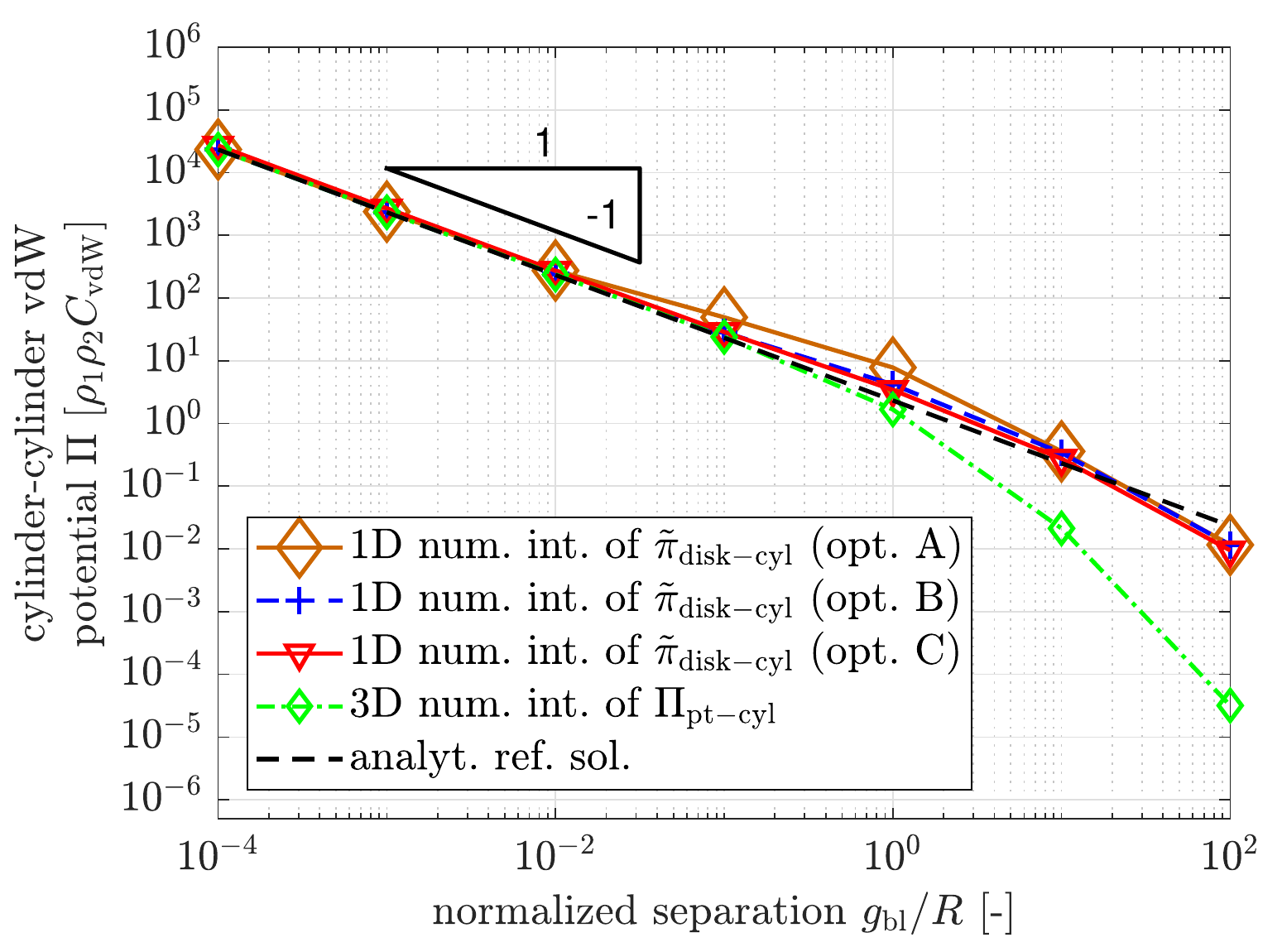}
    \label{fig::cyl-cyl_ia_pot_SBIP_optABC_over_sep_angle45}
   }
   \subfigure[Mutual angle $\alpha=90^\circ$]{
    \includegraphics[width=0.45\textwidth]{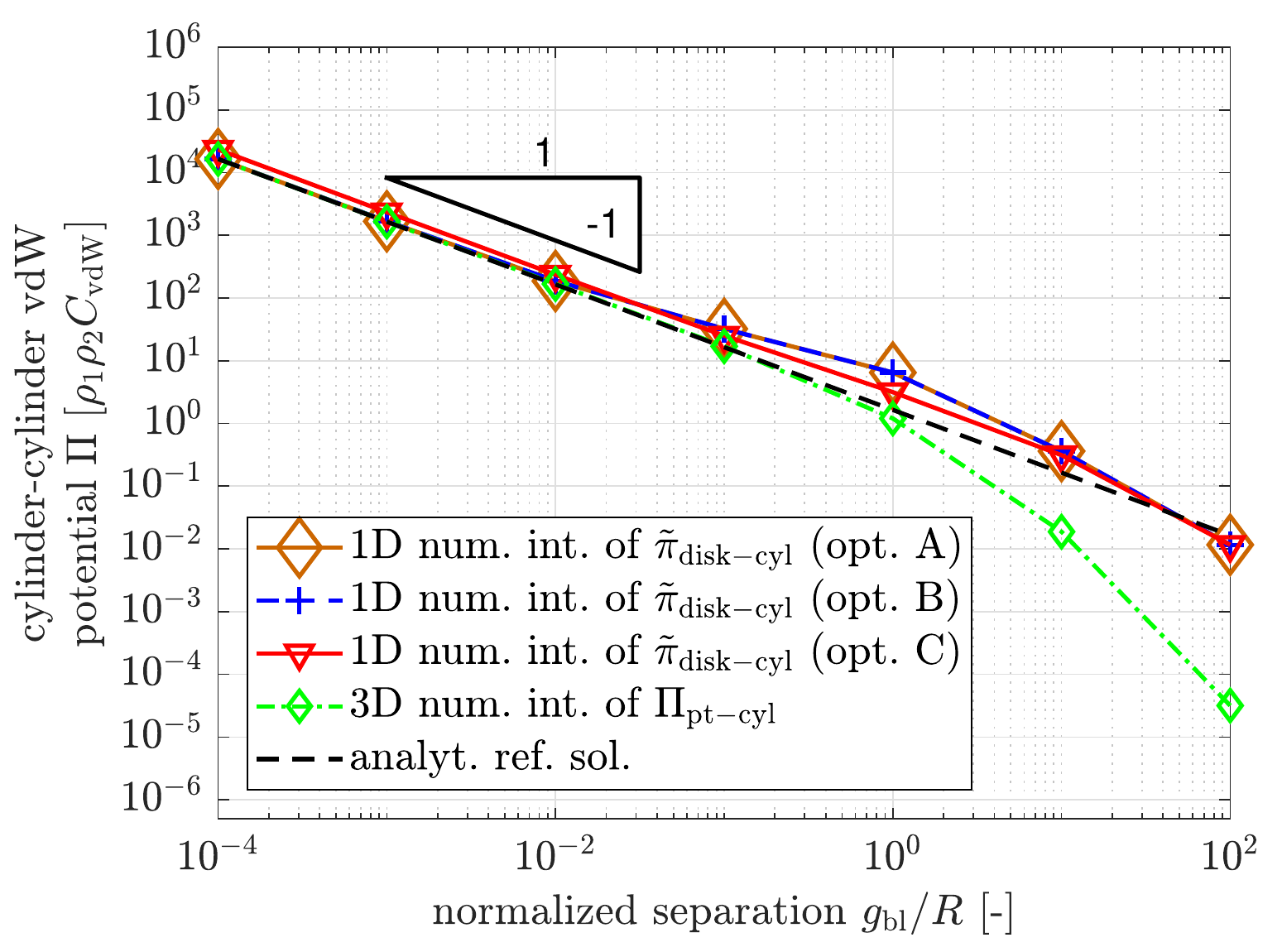}
    \label{fig::cyl-cyl_ia_pot_SBIP_optABC_over_sep_angle90}
   }
  \vspace{-5pt}
  \subfigure[Surface separation $g_\text{bl}/R=10^{-3}$]{
    \includegraphics[width=0.45\textwidth]{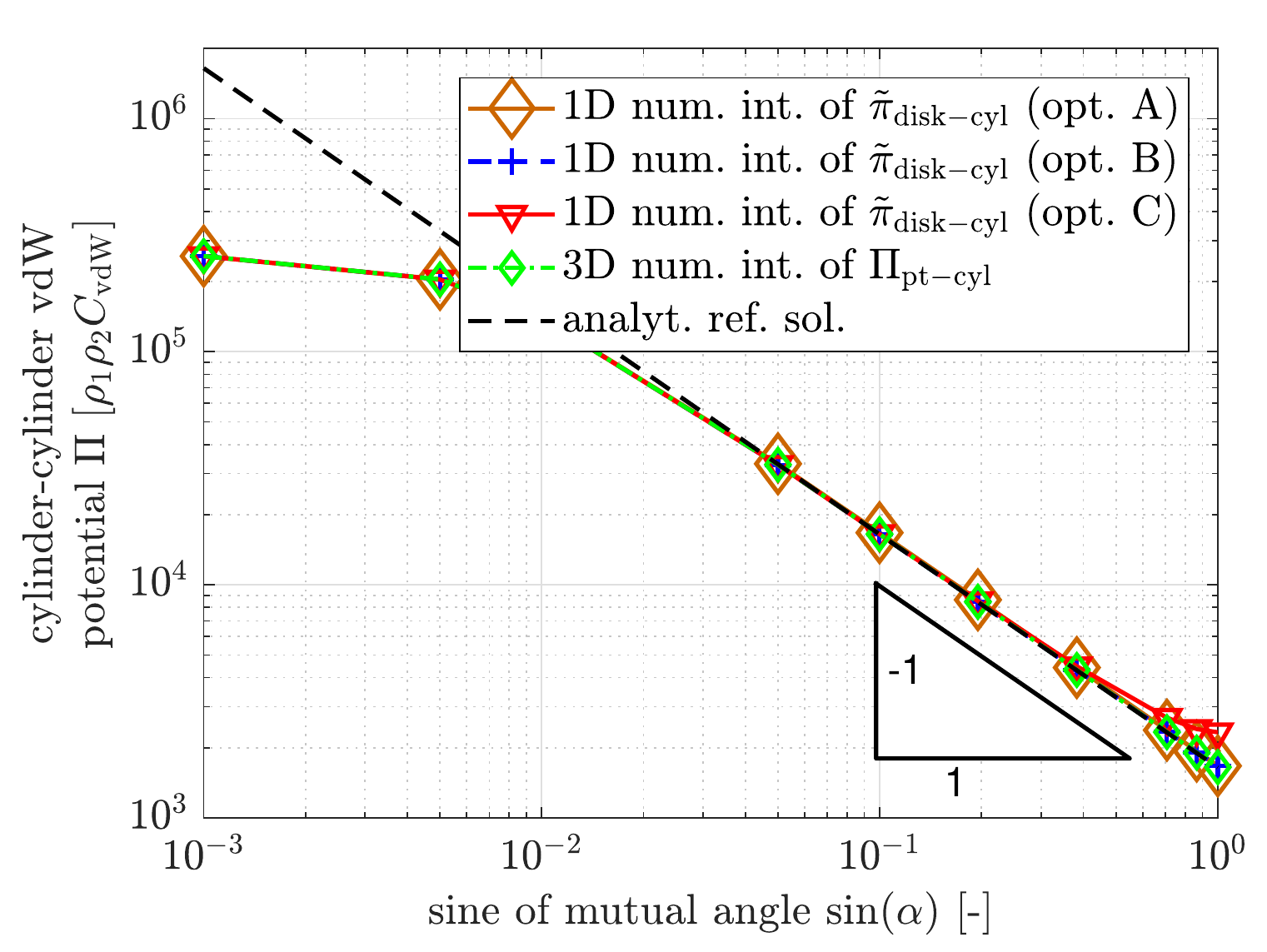}
    \label{fig::cyl-cyl_ia_pot_SBIP_optABC_over_angle_sep1e-3}
   }
  \vspace{-5pt}
  \subfigure[Surface separation $g_\text{bl}/R=10^{-1}$]{
    \includegraphics[width=0.45\textwidth]{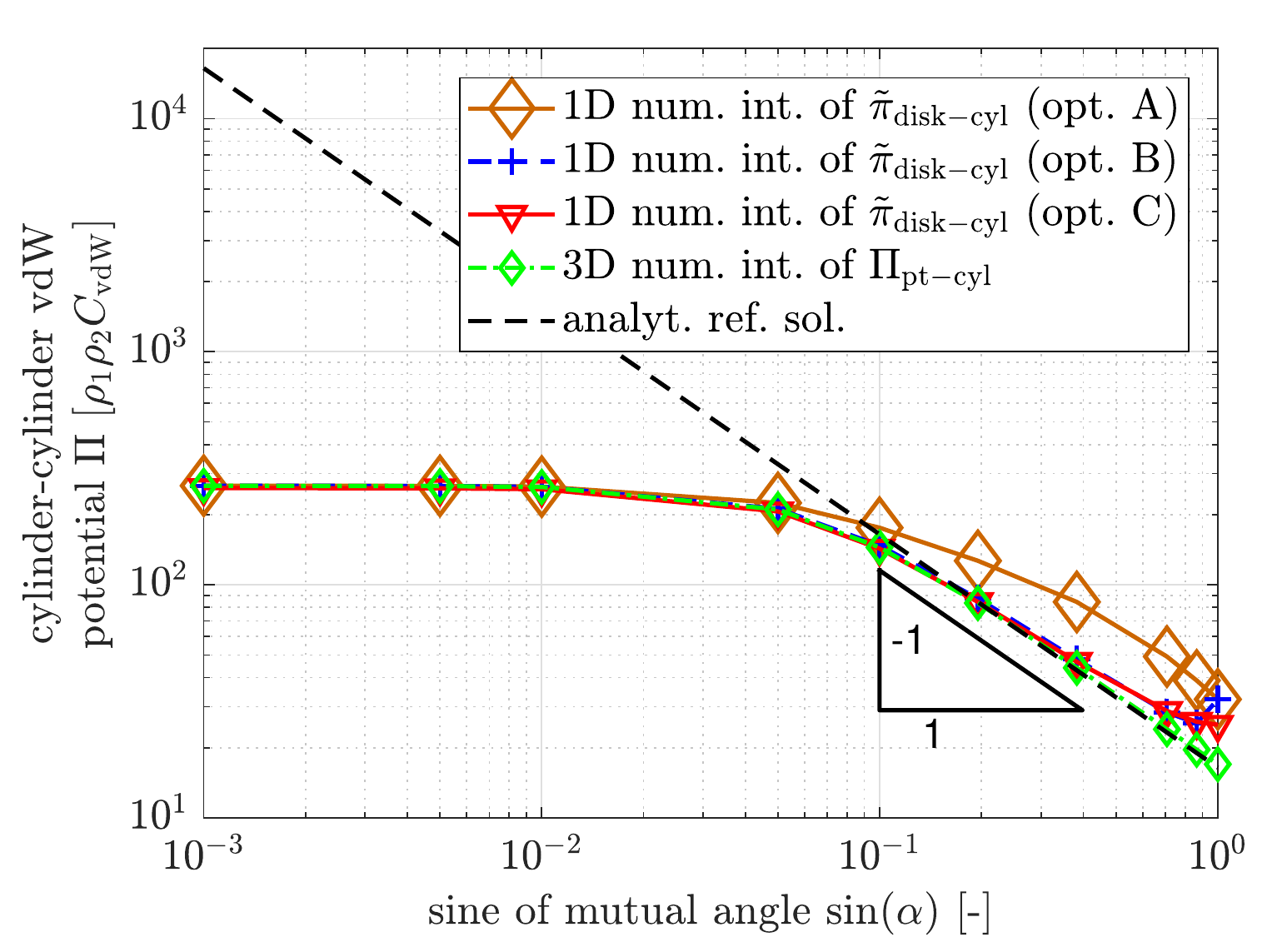}
    \label{fig::cyl-cyl_ia_pot_SBIP_optABC_over_angle_sep1e-1}
  }
   \caption{Interaction potential of two cylinders as a function of the dimensionless minimal surface separation~$g_\text{bl}/R$ at different mutual angles~$\alpha$ (first and second row) and as a function of~$\sin\alpha$ at different minimal surface separations~$g_\text{bl}/R$ (third row).
            Comparison of the options A (\eqref{eq::disk-cyl-pot_m_general} with~(\ref{eq::disk-cyl-pot_m_auxvariables_optA}); brown line with big diamonds), B (\eqref{eq::disk-cyl-pot_m_general} with~(\ref{eq::disk-cyl-pot_m_auxvariables_optB}); blue dashed line with pluses) and C (\eqref{eq::disk-cyl-pot_m}; red line with triangles) of the analytical expression for the disk-cylinder potential~$\tilde \pi_\text{6,disk-cyl}$ (used together with the SBIP approach\cite{GrillSBIP}).
            The numerical reference solution obtained via 3D Gaussian quadrature of the point-half space potential~$\Pi_\text{6,pt-hs}$ from \eqref{eq::pot_ia_vdW_point-cylinder} (green line with diamonds) and the analytical reference solutions summarized in~\secref{sec::introduction} (black dashed line) are plotted as reference.}
  \label{fig::cyl-cyl_ia_pot_SBIP_optABC_over_sep}
\end{figure}

$\,$\\
\figref{fig::cyl-cyl_ia_pot_SBIP_optABC_over_sep} specifically compares the different options A, B and C of the disk-cylinder interaction potential derived in~\secref{sec::disk-cyl-pot_m_derivation}.
Again, additional plots as function of the angle~$\alpha$ and in semi-logarithmic style are provided in \appref{sec::SBIP_verification_additional_plots} (cf.~\figref{fig::cyl-cyl_ia_pot_SBIP_optABC_over_angle_semilog}).
For parallel as well as perpendicular cylinders, option A and B are identical, as follows directly from the definition in \eqref{eq::disk-cyl-pot_m_general} and either~\eqref{eq::disk-cyl-pot_m_auxvariables_optA} or~\eqref{eq::disk-cyl-pot_m_auxvariables_optB}.
In all other cases, option B is closer to the numerical reference solution than option A and is thus the most accurate variant.
Most important, however, is the fact that both options A and B perfectly match the asymptotic solution for small separations for all mutual angles up to~$\alpha=\pi/2$ (see e.g.~\figref{fig::cyl-cyl_ia_pot_SBIP_optABC_over_sep_angle90} and \figref{fig::cyl-cyl_ia_pot_SBIP_optABC_over_angle_semilog}(a)), which has been identified as the most noticeable inaccuracy of the option C potential law above.
Therefore, future applications of the presented analytical disk-cylinder interaction potential laws with a strong focus on minimizing the approximation error and less restrictions with respect to the complexity of the equations probably want to use the option B expressions from Equations~(\ref{eq::disk-cyl-pot_m_general}) and (\ref{eq::disk-cyl-pot_m_auxvariables_optB}).
As outlined above, in the scope of this work the differences are considered small enough to use the significantly simpler option C expression from \eqref{eq::disk-cyl-pot_m} as the reduced interaction law within the SBIP approach from Ref.~\cite{GrillSBIP}.

\subsubsection{Intermediate conclusions}$\,$\\
The conclusions to take away from this important accuracy analysis are summarized as follows.
To begin with, the point-half space potential used as point-cylinder potential to compute the 3D numerical reference solution yields the correct asymptotic scaling behavior and allows to verify the two-cylinder potential in the range of intermediate separations where no analytical solution is known.
This is an important finding, because the same point-half space potential is used in the analytical derivations of the disk-cylinder potential~$\tilde \pi_\text{m,disk-cyl}$ of which all three considered options have been analyzed in this section.
As second important finding, all three investigated options show the correct asymptotic distance scaling, i.e. $\propto \! g^{-3/2}$ for parallel and $\propto  \! g^{-1}$ for perpendicular cylinders, as well as the theoretically predicted $(1\!/\!\sin\!\alpha)$-angle dependence in the decisive regime of small separations.
Despite the correct scaling behavior, the pleasantly simple option C (cf.~\eqref{eq::disk-cyl-pot_m}) shows a slight offset of the asymptotic small separation solution in the regime of large angles.
In contrast, options A (\eqref{eq::disk-cyl-pot_m_general} with~(\ref{eq::disk-cyl-pot_m_auxvariables_optA})) and B (\eqref{eq::disk-cyl-pot_m_general} with~(\ref{eq::disk-cyl-pot_m_auxvariables_optB})) ensure a very high accuracy in the absolute values of the asymptotic small separation solution for all mutual angles $\alpha$, which comes at the prize of an increased complexity of the expressions.
Thus, option C is considered as the optimal compromise between accuracy and complexity of the expression for the purposes of this work.
Taking into account the entire configuration space of separations and angles, option B shows the highest accuracy and is thus recommended for future applications with less restrictions in terms of the complexity of the expressions.

\subsection{Application examples: Simulations of adhesive, elastic nanofibers}
\label{sec::numerical_examples}
The simulation results shown in this section have first been presented as part of our recent article~\cite{GrillSBIP} introducing the SBIP approach as a novel beam-beam interaction model.
Our intention to outline a few examples at this point is twofold.
First, it illustrates our original motivation for the theoretical work presented in this contribution, as has been described in the introduction.
Second, it serves as a qualitative verification for more complex and general scenarios, where no reference solutions are available.
To this end, the presented analytical solution (option C, \eqref{eq::disk-cyl-pot_m}), embedded in the SBIP approach, has been implemented in C++ and integrated into the existing computational framework of the in-house research code BACI~\cite{BACI2020}.
More details on this framework can be found in Appendix C of our previous article~\cite{GrillSSIP}.
Note that the correctness of the implementation of the SBIP approach in general and \eqref{eq::disk-cyl-pot_m} as the disk-cylinder potential has been verified by means of a second, independent implementation in MATLAB~\cite{MATLAB2017b}, which has been used also for the accuracy analysis in \secref{sec::verif_disk-cyl-pot_twocylinders}.

\begin{figure}[htpb]%
  \centering
  \subfigure[Problem setup: undeformed configuration.]{
    \def\svgwidth{0.13\textwidth}
\begingroup%
  \makeatletter%
  \providecommand\color[2][]{%
    \errmessage{(Inkscape) Color is used for the text in Inkscape, but the package 'color.sty' is not loaded}%
    \renewcommand\color[2][]{}%
  }%
  \providecommand\transparent[1]{%
    \errmessage{(Inkscape) Transparency is used (non-zero) for the text in Inkscape, but the package 'transparent.sty' is not loaded}%
    \renewcommand\transparent[1]{}%
  }%
  \providecommand\rotatebox[2]{#2}%
  \newcommand*\fsize{\dimexpr\f@size pt\relax}%
  \newcommand*\lineheight[1]{\fontsize{\fsize}{#1\fsize}\selectfont}%
  \ifx\svgwidth\undefined%
    \setlength{\unitlength}{90bp}%
    \ifx\svgscale\undefined%
      \relax%
    \else%
      \setlength{\unitlength}{\unitlength * \real{\svgscale}}%
    \fi%
  \else%
    \setlength{\unitlength}{\svgwidth}%
  \fi%
  \global\let\svgwidth\undefined%
  \global\let\svgscale\undefined%
  \makeatother%
  \begin{picture}(1,2.5)%
    \lineheight{1}%
    \setlength\tabcolsep{0pt}%
    \put(0,0){\includegraphics[width=\unitlength,page=1]{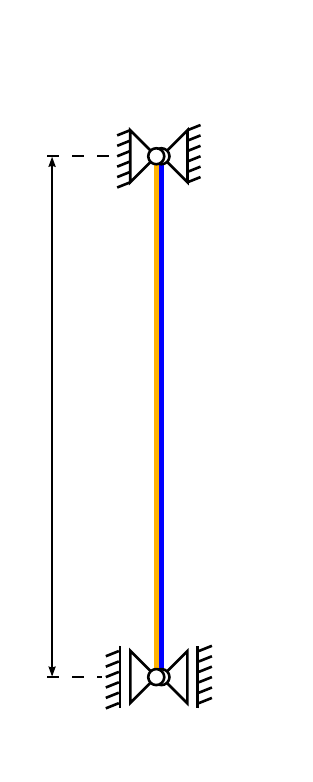}}%
    \put(0.72220523,0.16035767){\color[rgb]{0,0,0}\makebox(0,0)[lt]{\lineheight{0}\smash{\begin{tabular}[t]{l}$u_x, F_x^{br}$\end{tabular}}}}%
    \put(0.19677776,1.13176587){\color[rgb]{0,0,0}\makebox(0,0)[lt]{\lineheight{0}\smash{\begin{tabular}[t]{l}$l$\end{tabular}}}}%
    \put(0.71669108,2.17380371){\color[rgb]{0,0,0}\makebox(0,0)[lt]{\lineheight{0}\smash{\begin{tabular}[t]{l}$u_x, F_x^{tr}$\end{tabular}}}}%
    \put(0,0){\includegraphics[width=\unitlength,page=2]{num_ex_vdW_twoparallelbeams_pulloff_from_contact_problem_setup.pdf}}%
    \put(0.54009603,2.26407878){\color[rgb]{0,0,0}\makebox(0,0)[lt]{\lineheight{0}\smash{\begin{tabular}[t]{l}$y$\end{tabular}}}}%
    \put(0.81509603,2.03602702){\color[rgb]{0,0,0}\makebox(0,0)[lt]{\lineheight{0}\smash{\begin{tabular}[t]{l}$x$\end{tabular}}}}%
    \put(0,0){\includegraphics[width=\unitlength,page=3]{num_ex_vdW_twoparallelbeams_pulloff_from_contact_problem_setup.pdf}}%
  \end{picture}%
\endgroup%

    \label{fig::num_ex_vdW_twoparallelbeams_problem_setup}
  }
  \hspace{0.5cm}
  \subfigure[Quasi-static force-displacement curve. Force values to be interpreted as multiple of a reference point load that causes a deflection of~$l/4$ if applied at the fiber midpoint.]{
   \includegraphics[width=0.45\textwidth]{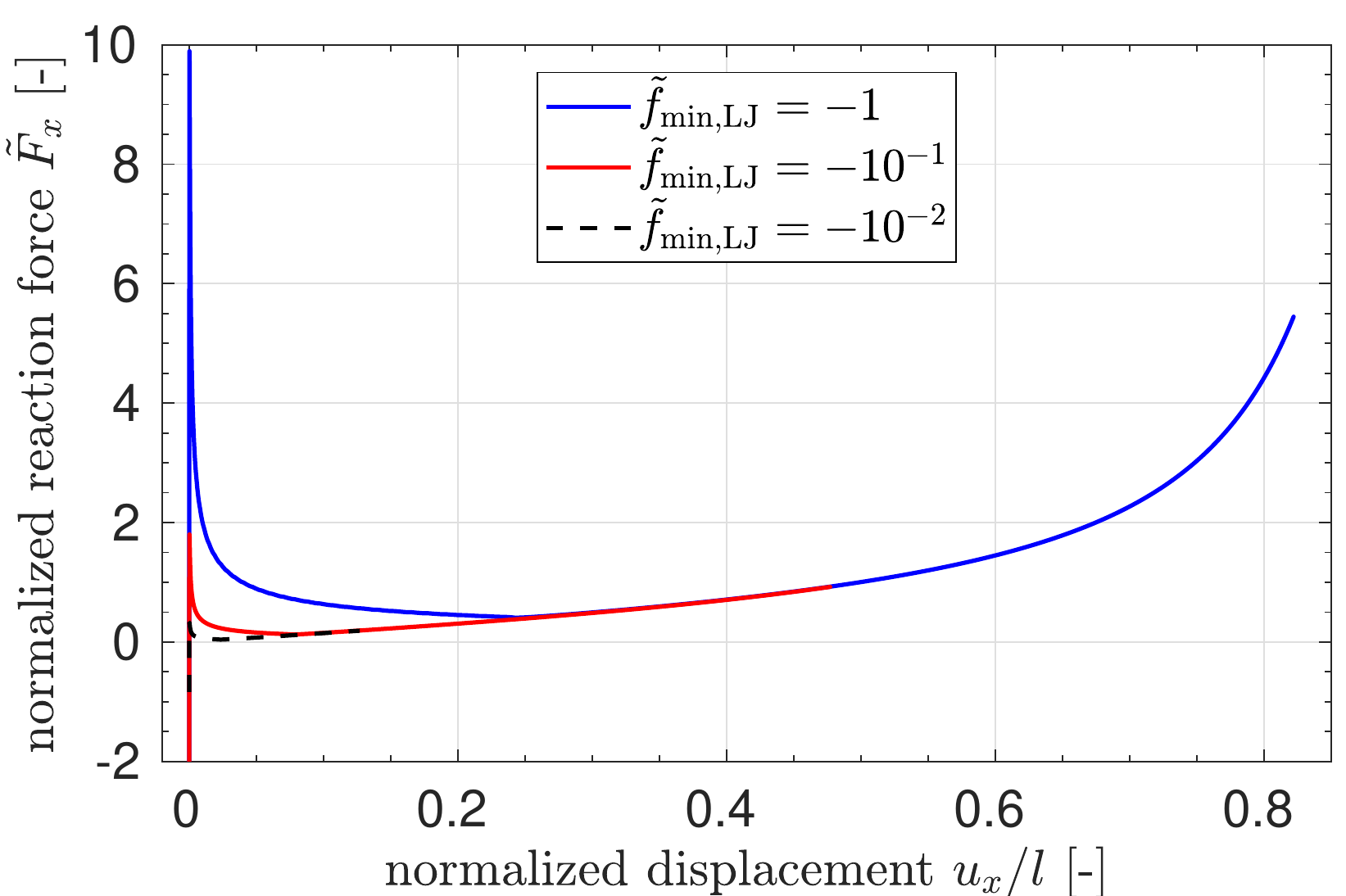}
   \label{fig::num_ex_vdW_twoparallelbeams_pulloff_force_over_displacement}
  }
  \hfill
  \subfigure[Detail view for small displacement values.]{
   \includegraphics[width=0.3\textwidth]{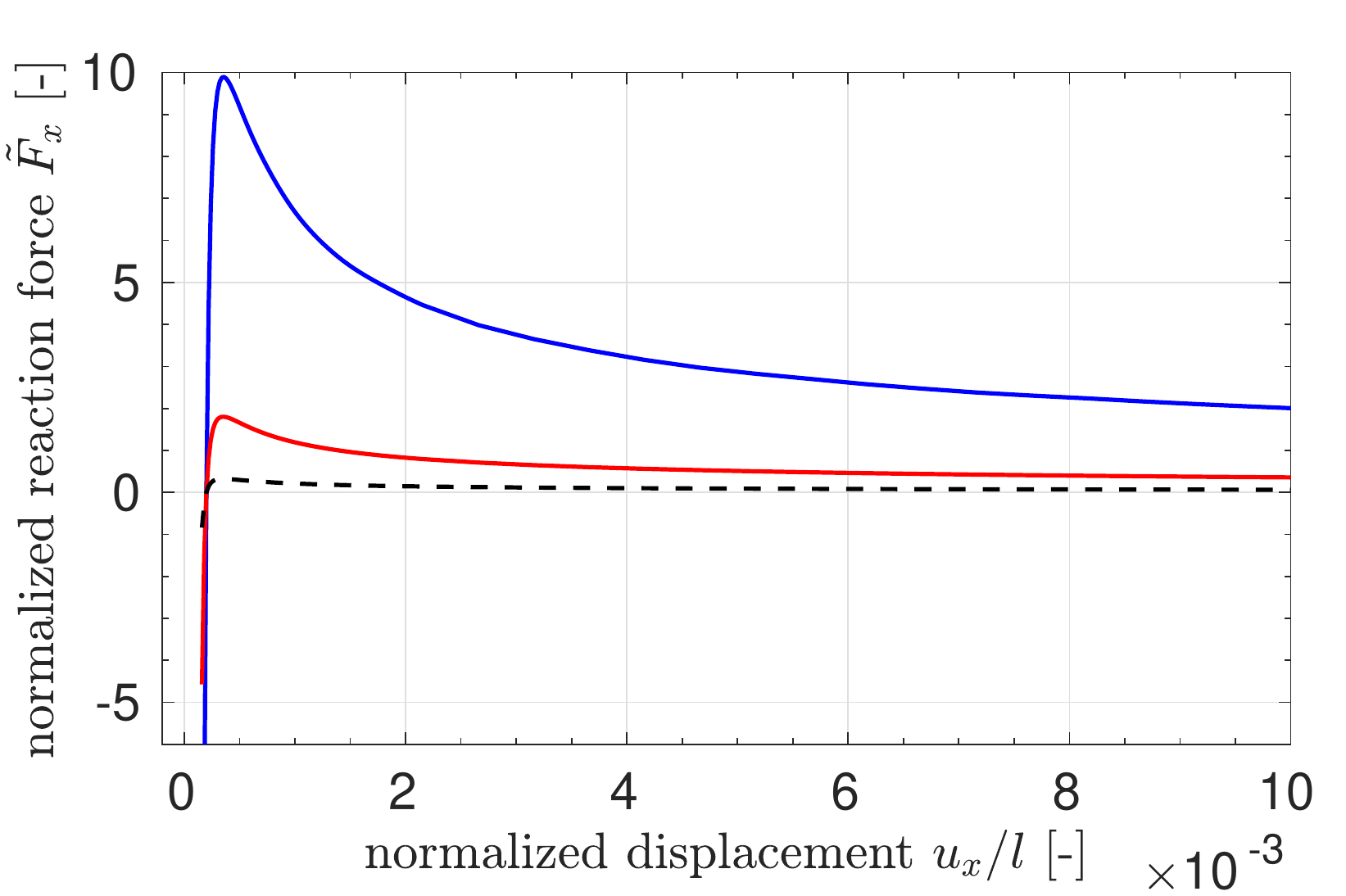}
   \label{fig::num_ex_vdW_twoparallelbeams_pulloff_force_over_displacement_zoom}
  }
  \caption{Numerical peeling experiment with two adhesive elastic fibers interacting via the LJ potential. Reprinted from our recent contribution \cite{GrillSBIP}.}
  \label{fig::num_ex_vdW_twoparallelbeams_peeling}
\end{figure}

\begin{figure}[htpb]%
  \centering
  \subfigure[]{
    \includegraphics[width=0.3\textwidth]{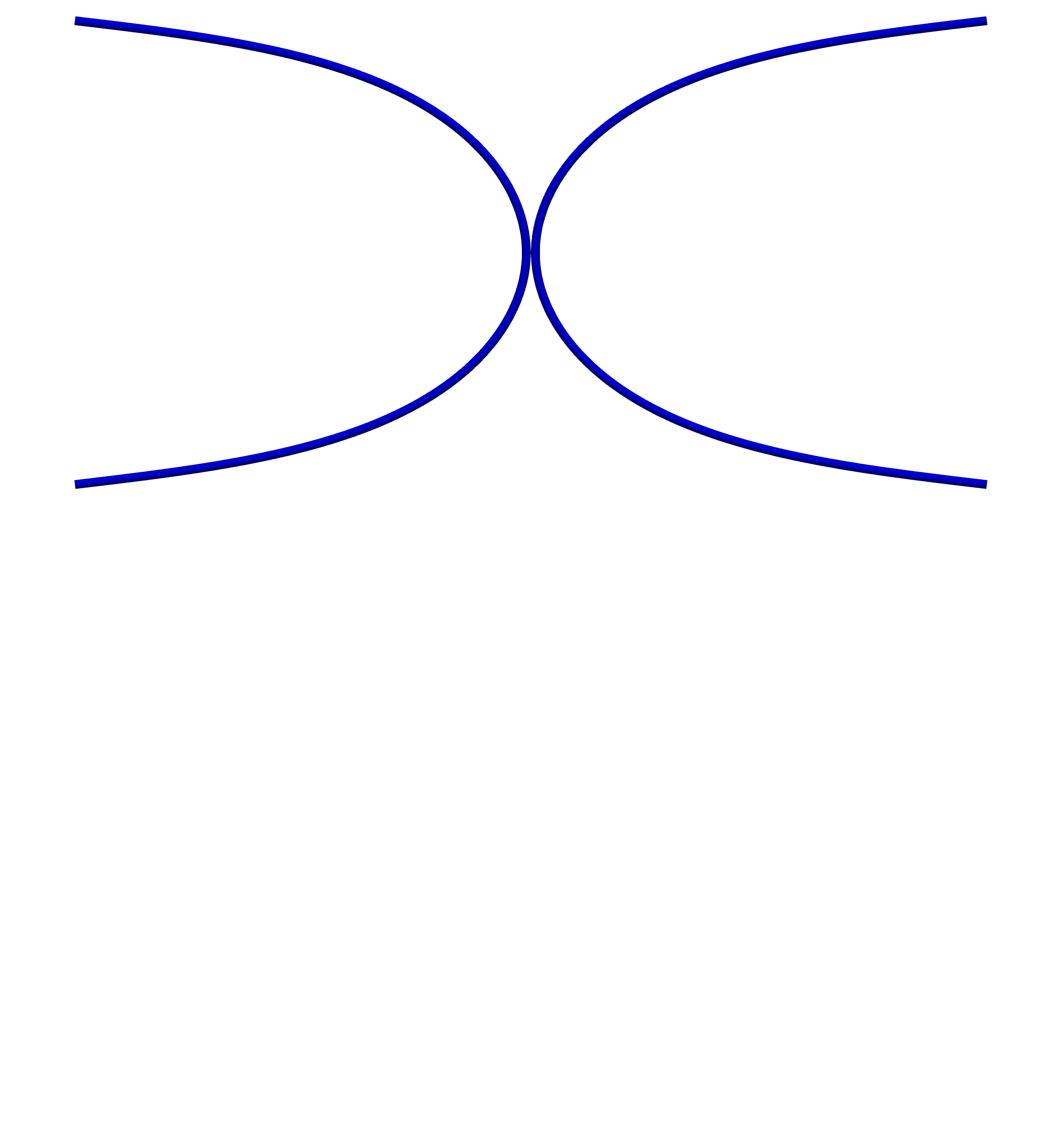}
    \label{fig::num_ex_vdW_twoparallelbeams_snapshot_comparison_finalstates_A}
  }
  \subfigure[]{
    \includegraphics[width=0.3\textwidth]{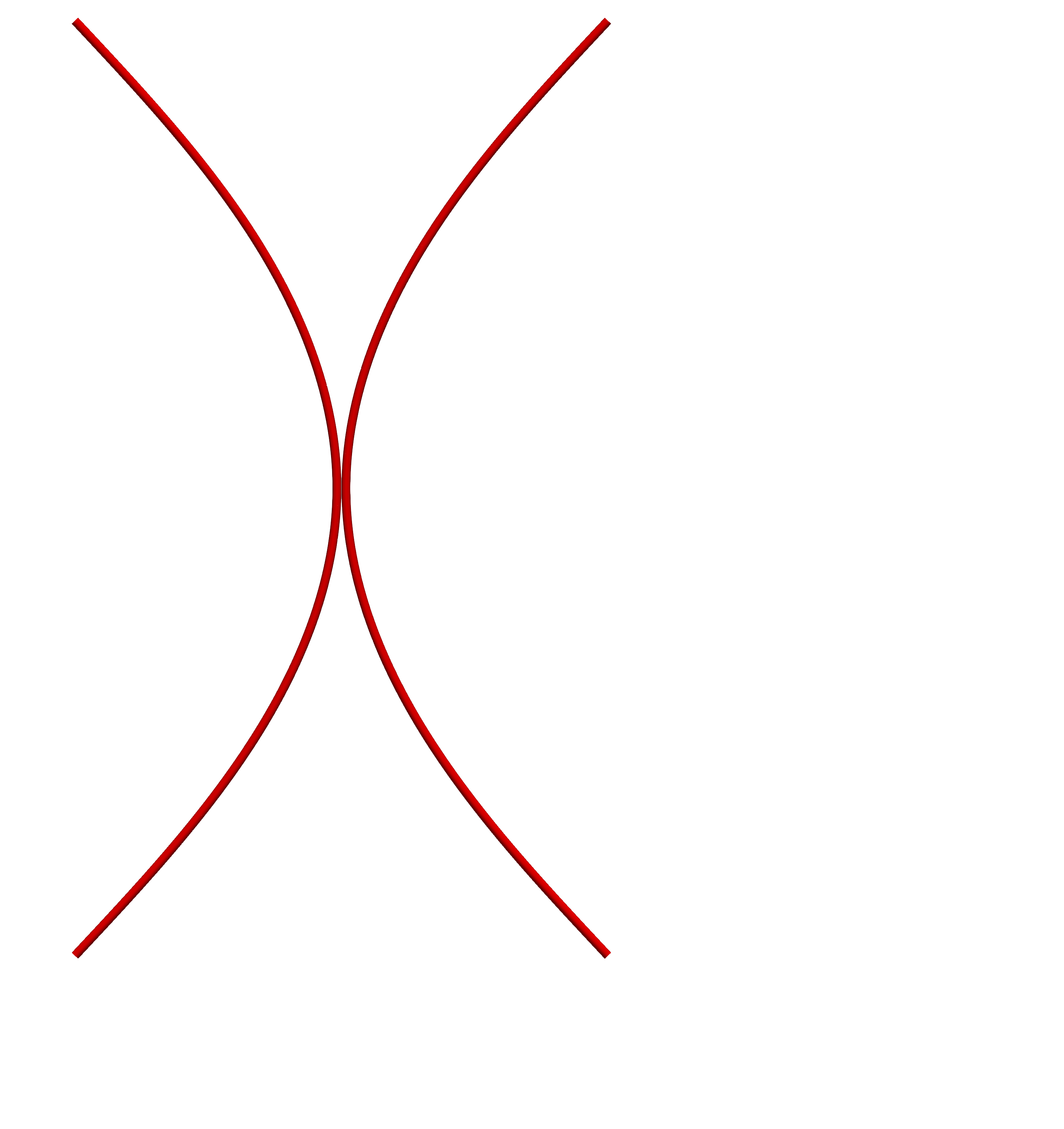}
    \label{fig::num_ex_vdW_twoparallelbeams_snapshot_comparison_finalstates_B}
  }
  \subfigure[]{
    \includegraphics[width=0.3\textwidth]{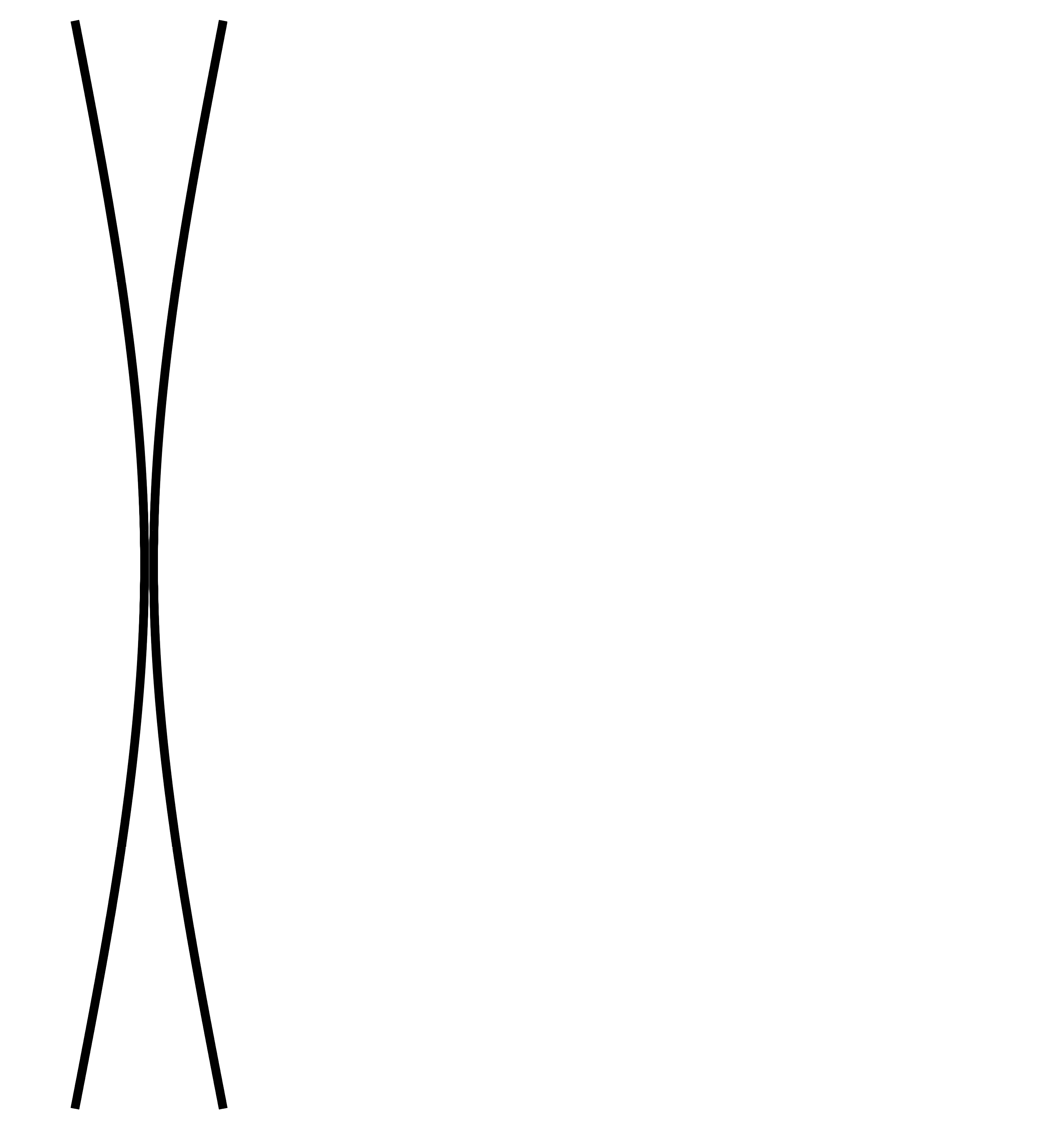}
    \label{fig::num_ex_vdW_twoparallelbeams_snapshot_comparison_finalstates_C}
  }
  \caption{Comparison of the final configurations before snapping free for three different levels in the strength of adhesion: from (a) strong to (c) weak adhesion.}
  \label{fig::num_ex_vdW_twoparallelbeams_peeling_snapshots_comparison_finalstates}
\end{figure}

\begin{figure}[htpb]%
  \centering
  \subfigure[]{
    \includegraphics[width=0.85\textwidth]{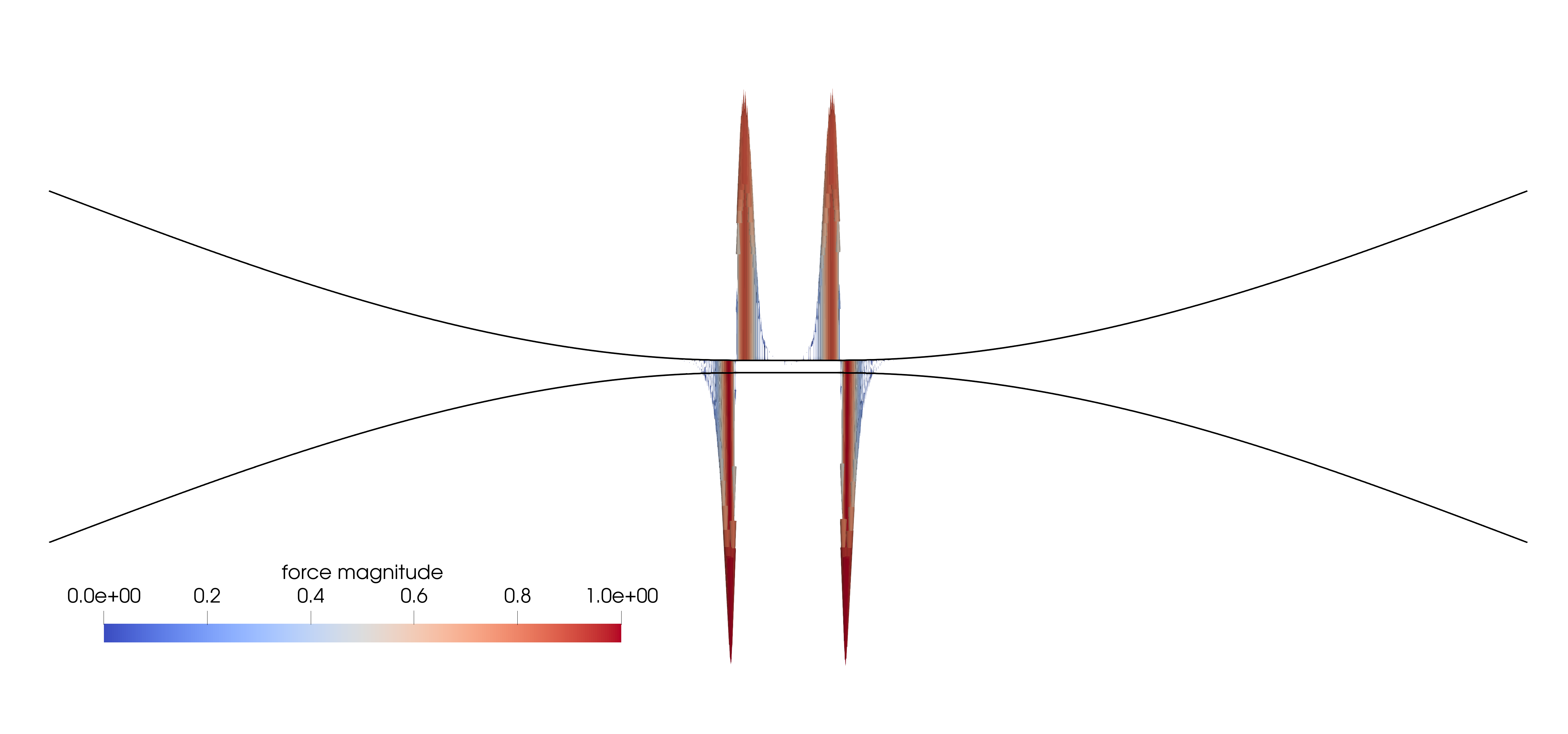}
    \label{fig::num_ex_vdW_twoparallelbeams_snapshot_forcedistribution_reldisp0_1}
  }
  \subfigure[]{
    \includegraphics[width=0.12\textwidth]{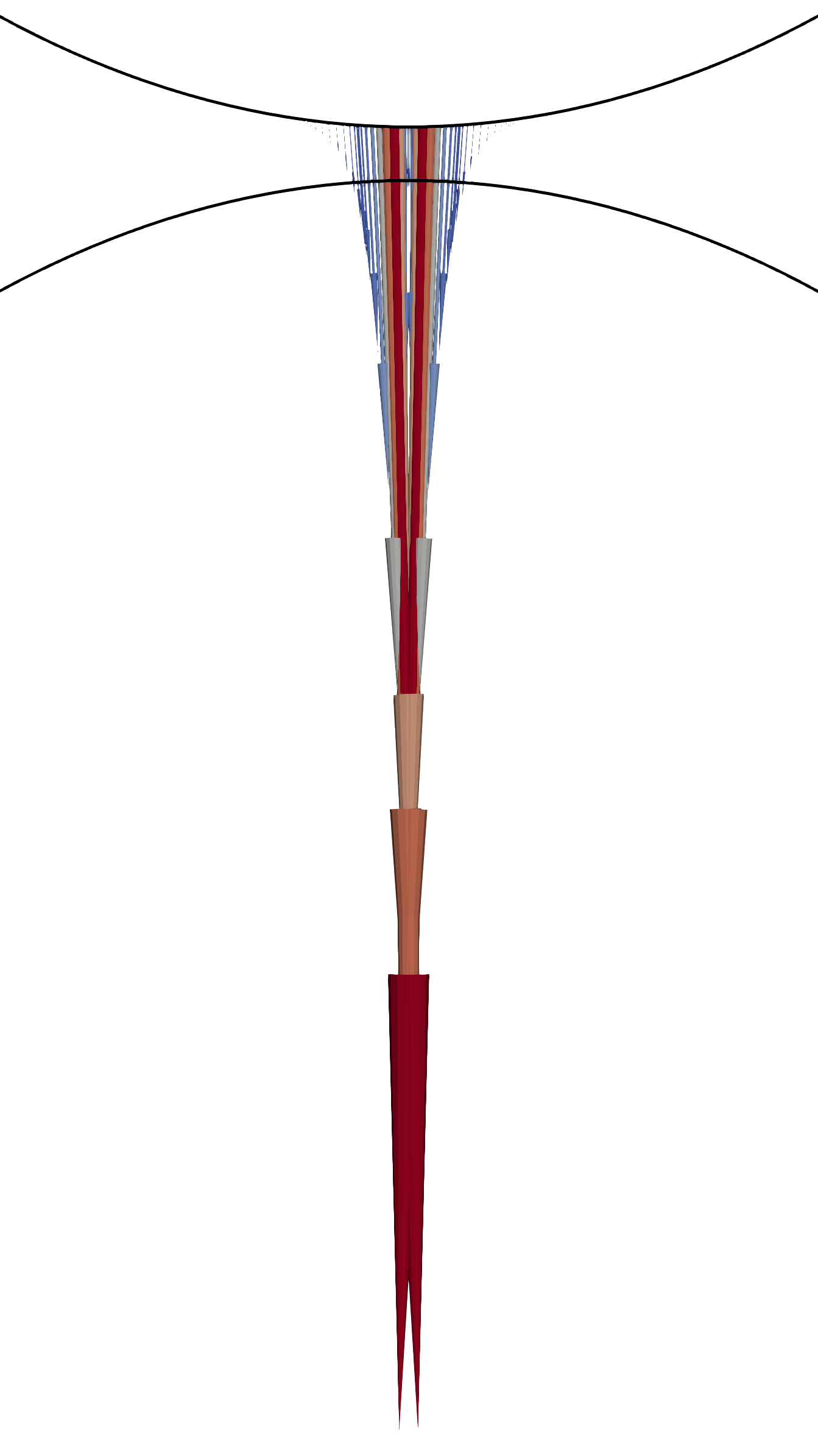}
    \label{fig::num_ex_vdW_twoparallelbeams_snapshot_forcedistribution_finalstate}
  }
  \caption{Detail study of the resulting line force distributions in two different displacement states obtained for the case of strong adhesion. For clarity, the fibers are depicted as their centerlines and forces are shown for the upper fiber only.}
  \label{fig::num_ex_vdW_twoparallelbeams_snapshot_forcedistribution}
\end{figure}

The first example mimics the peeling of two adhesive elastic fibers starting from contact along their entire length and ending as they snap free.
\figref{fig::num_ex_vdW_twoparallelbeams_peeling} shows the setup of this numerical experiment and the measured force-displacement curves for three different levels of the adhesion strength.
The resulting deformed shape of the fibers just before snapping free is compared in \figref{fig::num_ex_vdW_twoparallelbeams_peeling_snapshots_comparison_finalstates} and showcases the ability of this model to simulate even large deformations and changing mutual orientations of the fibers.
A detailed view of the line force distributions acting on the fibers as a result of the adhesive contact interaction is finally shown in \figref{fig::num_ex_vdW_twoparallelbeams_snapshot_forcedistribution}.
In the left snapshot, note especially the rapidly changing force magnitude and direction in the peeling zone and the net zero interaction forces in the middle part of the fibers, where adhesive and repulsive contact forces are in equilibrium.
Indeed, the surface-to-surface distance of the fibers in this region is equal to the theoretical prediction for two parallel cylinders of infinite length.
The snapshot on the right shows the interaction force distribution just before snapping free, which is fundamentally different in the sense that it is purely adhesive, without any repulsive component.
More detailed results and discussions can be found in the original publication~\cite{GrillSBIP} and the article, where this example was studied for the first time, using a previous beam-beam interaction model~\cite{GrillPeelingPulloff}.

\begin{figure}[htpb]%
  \centering
  \begin{minipage}{0.63\textwidth}
      \subfigure[Surfaces with 2x16x16 loops, pulled until just before snapping free.]{
        \includegraphics[width=\textwidth]{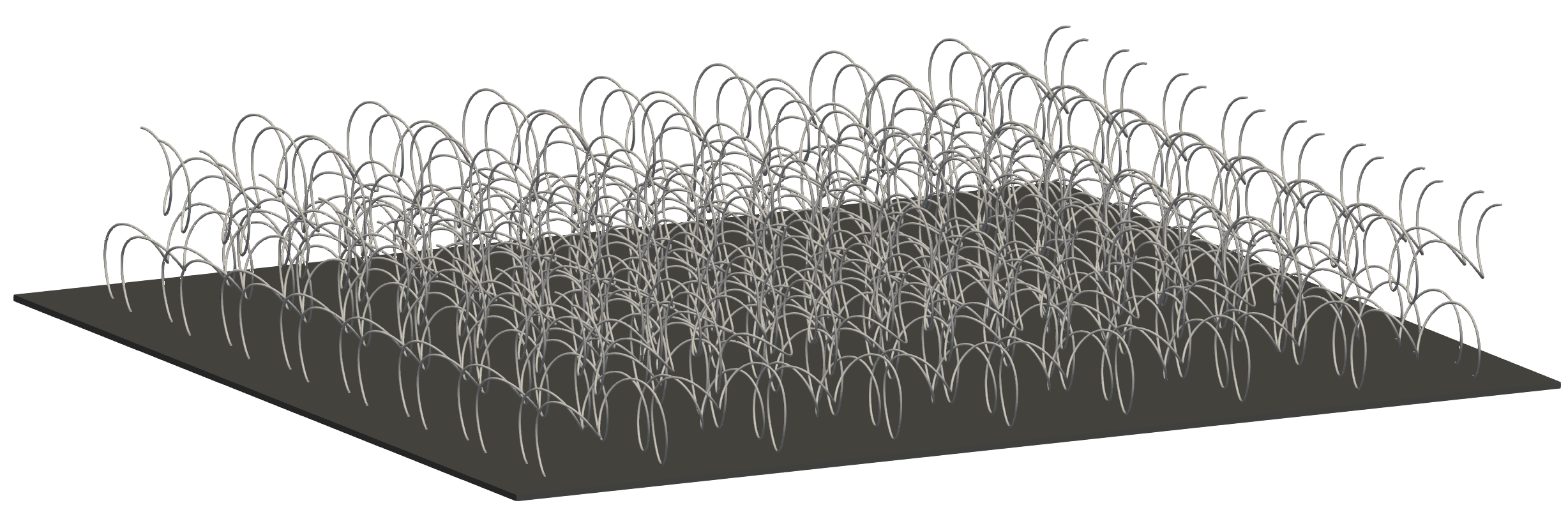}
        \label{fig::num_ex_adhesive_surfaces_simulation_snapshots_2x16x16loops_Pull_time4_25}
      }
      \subfigure[Surfaces with 2x8x8 loops, twisted by $75^\circ$.]{
        \vspace{0.5cm}
        \includegraphics[width=\textwidth]{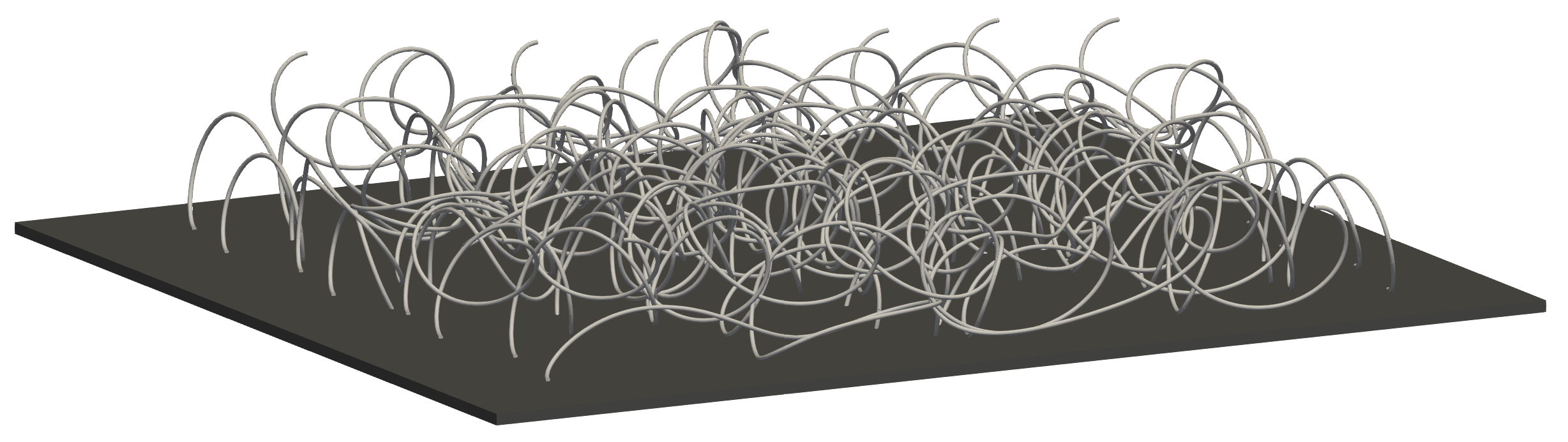}
        \label{fig::num_ex_adhesive_surfaces_simulation_snapshots_2x8x8_TwistAndPull_time4_step16969}
      }
  \end{minipage}
  \begin{minipage}{0.35\textwidth}
      \subfigure[Top view of (b).]{
        \includegraphics[width=\textwidth]{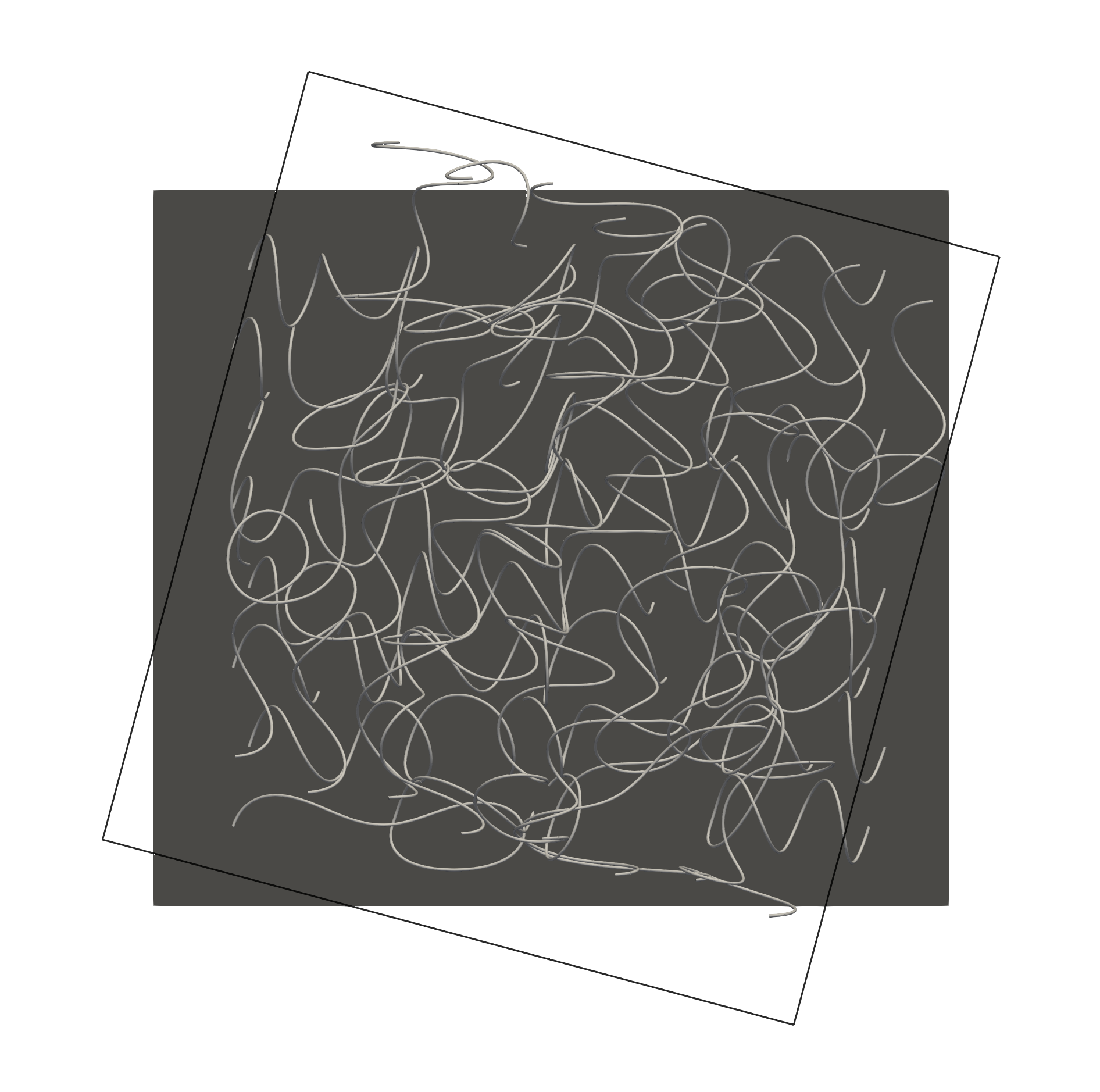}
        \label{fig::num_ex_adhesive_surfaces_simulation_snapshots_2x8x8_TwistAndPull_time4_step16969_Z-view}
      }
  \end{minipage}
  \caption{Selected simulation snapshots of two (rigid) surfaces grafted with helical, adhesive nanofibers that are being pulled or twisted.
  Top surface is hidden for better visibility of the fibers.
  Reprinted from our recent contribution \cite{GrillSBIP}.}
  \label{fig::num_ex_adhesive_surfaces_simulation_snapshots}
\end{figure}

The second example aims to showcase the ability to simulate large systems of practically relevant system sizes and high geometrical complexity.
It models the interaction of two (rigid) surfaces that are grafted with arrays of helical nanofibers (see \figref{fig::num_ex_adhesive_surfaces_simulation_snapshots}, where the top surface is hidden for better visibility of the fibers).
The shown simulation snapshots demonstrate a big variety both in the deformation state of the individual fibers and in the mutual orientation of interacting fibers.
Also the case of entangled fibers can be observed in the case where the surfaces are twisted by $75^\circ$ (bottom and right image).
It should be emphasized that simulations with such large system sizes and simulation times have only been made possible by using the analytical disk-cylinder potential law derived in this work.

\section{Conclusion and outlook}
\label{sec::conclusion_outlook}
In this article, analytical expressions for the resulting Lennard-Jones (LJ) interaction potential between a disk and a cylinder of infinite length have been derived that are valid for arbitrary mutual orientations in the decisive regime of small surface separations.
Based on the strategy of pairwise summation, a five-dimensional integral of the point-pair interaction potential over the area of the disk and the volume of the cylinder had to be solved.
Due to the absence of an exact analytical solution to this problem, we have proposed three different options A, B and C of the final expression, which vary in the approximations being made and thus in the complexity of the final expression and its accuracy.

All three investigated options show the correct asymptotic distance scaling, i.e. $\propto \! g^{-3/2}$ for parallel and $\propto  \! g^{-1}$ for perpendicular cylinders, as well as the theoretically predicted $(1\!/\!\sin\!\alpha)$-angle dependence in the decisive regime of small separations. Despite the correct scaling behavior, the pleasantly simple option C shows a slight offset of the asymptotic small separation solution in the regime of large angles. In contrast, options A (\eqref{eq::disk-cyl-pot_m_general} with~(\ref{eq::disk-cyl-pot_m_auxvariables_optA})) and B (\eqref{eq::disk-cyl-pot_m_general} with~(\ref{eq::disk-cyl-pot_m_auxvariables_optB})) ensure a very high accuracy in the absolute values of the asymptotic small separation solution for all mutual angles $\alpha$, which comes at the prize of an increased complexity of the expressions. Thus, option C is considered as the optimal compromise between accuracy and complexity of the expression for our purposes.
Taking into account the entire configuration space of separations and angles, option B shows the highest accuracy and is thus recommended for future applications with less restrictions in terms of the complexity of the expressions.
All the derived expressions are generic with respect to the exponent~$m$ of the inverse power law that is being used as the point-pair interaction potential law (with $m$>6), such that their application includes, but remains not limited to the most common case of adhesive van der Waals ($m=6$) and repulsive steric forces ($m=12$) of the LJ law.

Eventually, as we showed in a brief outlook to our current research work, the derived analytical disk-cylinder interaction potential laws may be used to formulate highly efficient computational models for the interaction of arbitrarily curved fibers, such that the disk represents the cross-section of the first and the cylinder a local approximation to the shape of the second fiber. Just to give a first impression of the possibilities, we showed a few snapshots of simulated scenarios including the peeling of deformable fibers and the interaction of nanofiber-grafted surfaces, underlining that the present work enables significant progress both in terms of accuracy and efficiency of such simulation models.
Regarding efficiency, the analytical closed-form expression replacing the otherwise required multi-dimensional numerical integration (of a numerically unfavorable inverse power law in the regime of small separations) leads to a boost in performance that allows to simulate significantly longer time scales and bigger systems such as the fiber-grafted surfaces with hundreds of fibers shown above.
In terms of accuracy, the derived expressions for the first time ensure the correct asymptotic scaling behavior for the limit of small separations and, as a result, allows accurate predictions for example of the maximal pull-off force when peeling two deformable fibers.

Other future applications for the derived analytical expressions need not remain limited to simulation models, but may well extend to theoretical work studying large and complex systems of adhesive components with circular cross-sections.
Moreover, the derivation of further analytical expressions  addressing e.g.~other types of interactions or cross-section shapes would be a promising extension of this work and further extend the possibilities of modeling approaches in this field.

\appendix

\section{Solutions for the point-cylinder interaction potential given an inverse power point-pair potential law with generic exponent}\label{sec::point-cyl-pot_m_derivation}
We aim to find an expression for the interaction potential~$\Pi_\text{m,pt-cyl}$ of a single point, i.\,e.~molecule, and an infinite cylinder of radius~$R$ and molecule density~$\rho$ via analytical pair-wise summation of the generic point-pair potential~$\Phi_\text{m}$:
\begin{align}
  \Pi_\text{m,pt-cyl} = \int \limits_{V_\text{cyl}} \rho \, \Phi_\text{m}(r) \dd V, \quad \text{with} \quad \Phi_\text{m}(r)=k_\text{m} \, r^{-m}, \quad m \geq 6.
\end{align}
Its solution shall serve as the basis for the disk-cylinder interaction potential~$\tilde \pi_{\text{m,disk-cyl}}$ as indicated in \eqref{eq::disk_cyl_pot_m_5Dint}.
To the best of the authors' knowledge, no exact (closed-form) analytical solution exists for this problem, however, as commonly applied in this context, good approximate analytical solutions for the dominating regime of small separations can be found by means of series expansion and truncation.
This solution approach will be exemplified in the remainder of this section, where we present two alternative solutions and finally compare their accuracy to choose the one to be used in the derivation of~$\tilde \pi_{\text{m,disk-cyl}}$ in \secref{sec::disk-cyl-pot_m_derivation}.

\subsection{Generalization of the solution by Montgomery et al.~for the case of vdW interaction}\label{sec::point-cyl-pot_Montgomery_generalization}
We start from the following expression for the van der Waals ($m=6$) interaction potential as obtained by Montgomery et al.~\cite{montgomery2000}:
\begin{align}\label{eq::pot_ia_vdW_point-cylinder_seriesform}
  \Pi_{\text{6,pt-cyl}}(g_{\text{pt-cyl}}) = \frac{1}{8} \pi^2 k_6 \rho \left( \frac{1}{g^3_{\text{pt-cyl}} } - \frac{1}{(g_{\text{pt-cyl}} + 2R)^3} + \text{H.O.T.} \right)
\end{align}
Here, $g_{\text{pt-cyl}}$ denotes the closest distance between the point and the cylinder surface.
Since we are interested in the limit of small separations~$g_{\text{pt-cyl}} / R \ll 1$, the second and all higher order terms will be substantially smaller as compared to the first term and we will restrict ourselves to this leading term, resulting in
\begin{align}\label{eq::pot_ia_vdW_point-cylinder}
  \Pi_{\text{6,pt-cyl}}(g_{\text{pt-cyl}}) = \frac{1}{8} \pi^2 k_6 \rho \, g^{-3}_{\text{pt-cyl}}.
\end{align}
All the steps of the derivation in~\cite{montgomery2000}, basically solving three nested integrals over the cylinder volume, can be generalized from the vdW case~$m=6$ to a generic exponent~$m \geq 6$.
Mainly due to the recursive nature of the following antiderivative (see e.\,g.~\cite[p.1020]{Bronshtein2003}) required for one of the integrals
\begin{align}
  \int X^{-(n+1)} \dd x &= \frac{x}{2na^2} \, X^{-n} + \frac{2n-1}{2na^2} \, \int X^{-n} \dd x \quad \text{with} \quad X=a^2+x^2\\
  \int X^{-1} \dd x &= \frac{x}{a} \, \arctan \left( \frac{x}{a} \right),
\end{align}
writing down the final expression for generic exponents~$m$ however is quite tedious.
For the later reference, at this point we therefore present the general form of the solution
\begin{align}\label{eq::point-cylinder-pot_m_variantMontgomery}
  \Pi_\text{m,pt-cyl} = K_\text{m,pt-cyl} \, g_\text{pt-cyl}^{-m+3}, \quad m \geq 6
\end{align}
and only provide the exact prefactors~$K_\text{m,pt-cyl}$ for the two cases of vdW ($m=6$) and repulsive part ($m=12$) of the LJ potential, which are actually applied in the numerical examples of this work:
\begin{align}
  K_\text{6,pt-cyl} = \frac{1}{8} \pi^2 k_6 \rho \qquad \text{and} \qquad K_\text{12,pt-cyl} = \frac{7}{256} \pi^2 k_{12} \rho
\end{align}
\subsection{Alternative solution obtained from the point-half space interaction}\label{sec::point-halfspace-pot_m_derivation}
As compared to the point-cylinder interaction scenario from the previous section, the geometry of an (infinite) half space is a much simpler integration domain and thus even an exact analytical solution can be found and stated in closed form also for a general inverse power law exponent~$m$ (see e.\,g.~\cite[p.210]{israel2011}):
\begin{align}
  \Pi_\text{m,pt-hs} = K_\text{m,pt-hs} \, g_\text{pt-hs}^{-m+3}, \quad \text{with} \quad K_\text{m,pt-hs} = \frac{2}{(m-2)(m-3)} \pi k_\text{m} \rho
\end{align}
This expression shall serve as an alternative approximate solution for the sought-after point-cylinder interaction potential
\begin{align}\label{eq::point-cylinder-pot_m_varianthalfspace}
  \Pi_\text{m,pt-cyl} \approx \Pi_\text{m,pt-hs}
\end{align}
and its approximation quality will be investigated in the following section.

\subsection{Investigation of the accuracy of both alternatives}\label{sec::point-cyl-pot_m_comparison}
This section aims to compare the accuracy of the variants from \eqref{eq::point-cylinder-pot_m_variantMontgomery} and \eqref{eq::point-cylinder-pot_m_varianthalfspace} presented above.
At first sight, the expression derived for the point-cylinder geometry (\eqref{eq::point-cylinder-pot_m_variantMontgomery}) appears to be the more natural choice.
However, as shown in more detail in~\secref{sec::verif_disk-cyl-pot_twocylinders}, the resulting cylinder-cylinder interaction potential~$\tilde \pi_\text{6,cyl-cyl}$ based on the approximate solution from \eqref{eq::point-cylinder-pot_m_variantMontgomery} deviates from the one obtained via analytical 6D integration stated in~\secref{sec::introduction}.
To be more precise, the solutions deviate by a constant scalar factor in the asymptotic behavior for very small separations, which turns out to be independent of the mutual angle of the cylinders.
Interestingly, exactly the same difference by a factor of~$3/4 \, \pi \approx 2.356$ has already been identified for the asymptotic case~$R \to \infty$, i.\,e., when comparing the result to the one for point-half space interaction in the original publication~\cite{montgomery2000}.
This initially motivated the investigation of the alternative solution for the point-cylinder interaction potential presented in~\secref{sec::point-halfspace-pot_m_derivation}.
Using this alternative solution, the resulting disk-cylinder interaction potential applied within in the SBIP approach indeed yields the asymptotically correct solution for the cylinder-cylinder interaction (cf.~ again~\secref{sec::disk-cyl-pot_m_derivation} and~\secref{sec::verif_approx_singlelengthspec} for the details), such that the difference in the results is tracked down to the underlying solution for~$K_\text{6,pt-cyl}$ stated above.
To this end, this assessment has been verified by means of a numerical reference solution obtained from 3D integration of the vdW point-pair potential over a cylinder volume in Maple~\cite{Maple2015}.
For the relevant regime of small separations~$g \ll R$, the numerical reference solution excellently agrees with the analytical solution obtained from the point-half space interaction in~\secref{sec::point-halfspace-pot_m_derivation}.
This rather counterintuitive result appears reasonable if we think of the two radii of curvature of the cylinder surface~$R_\text{c1} = R$ and~$ R_\text{c2} = \infty$, of which both are much greater than the surface separation in the considered regime~$g_\text{pt-cyl} \ll R$, such that the interacting point faces an almost flat surface.
Finally, in the limit~$g_\text{pt-cyl}/R \to 0$, this scenario coincides with the one of the point-half space interaction, which is the illustrative explanation why the corresponding analytical solution is the consistent one and therefore to be used in the derivation of~\secref{sec::disk-cyl-pot_m_derivation}.

\section{Supplementary verification plots}
\label{sec::SBIP_verification_additional_plots}
\begin{figure}[htb]%
  \centering
  \vspace{-5pt}
  \subfigure[Surface-to-surface separation $g_\text{bl}/R=10^{-3}$]{
    \includegraphics[width=0.45\textwidth]{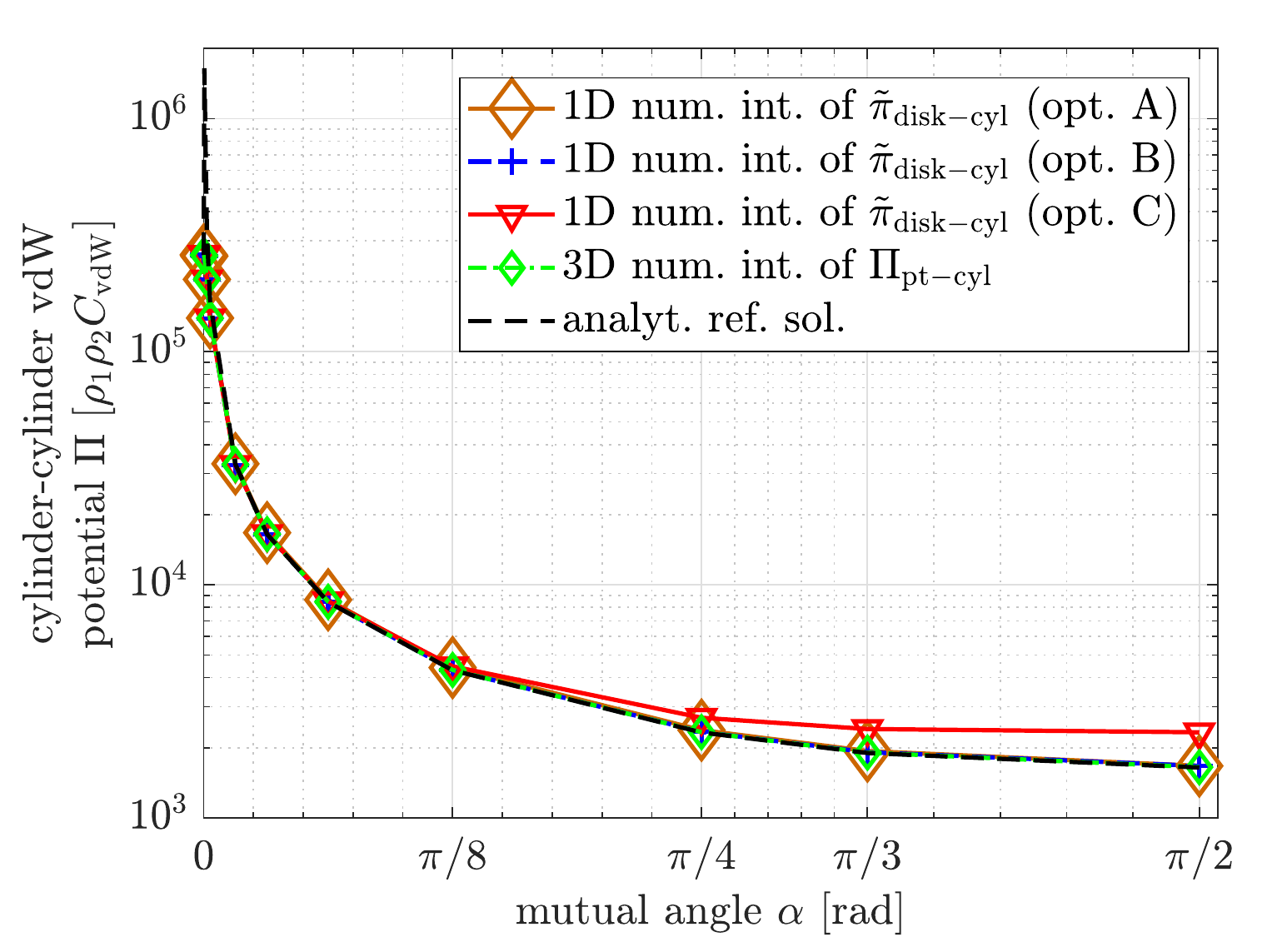}
    \label{fig::cyl-cyl_ia_pot_SBIP_optABC_over_angle_semilog_sep1e-3}
  }
  \vspace{-5pt}
  \subfigure[Surface-to-surface separation $g_\text{bl}/R=10^{-1}$]{
    \includegraphics[width=0.45\textwidth]{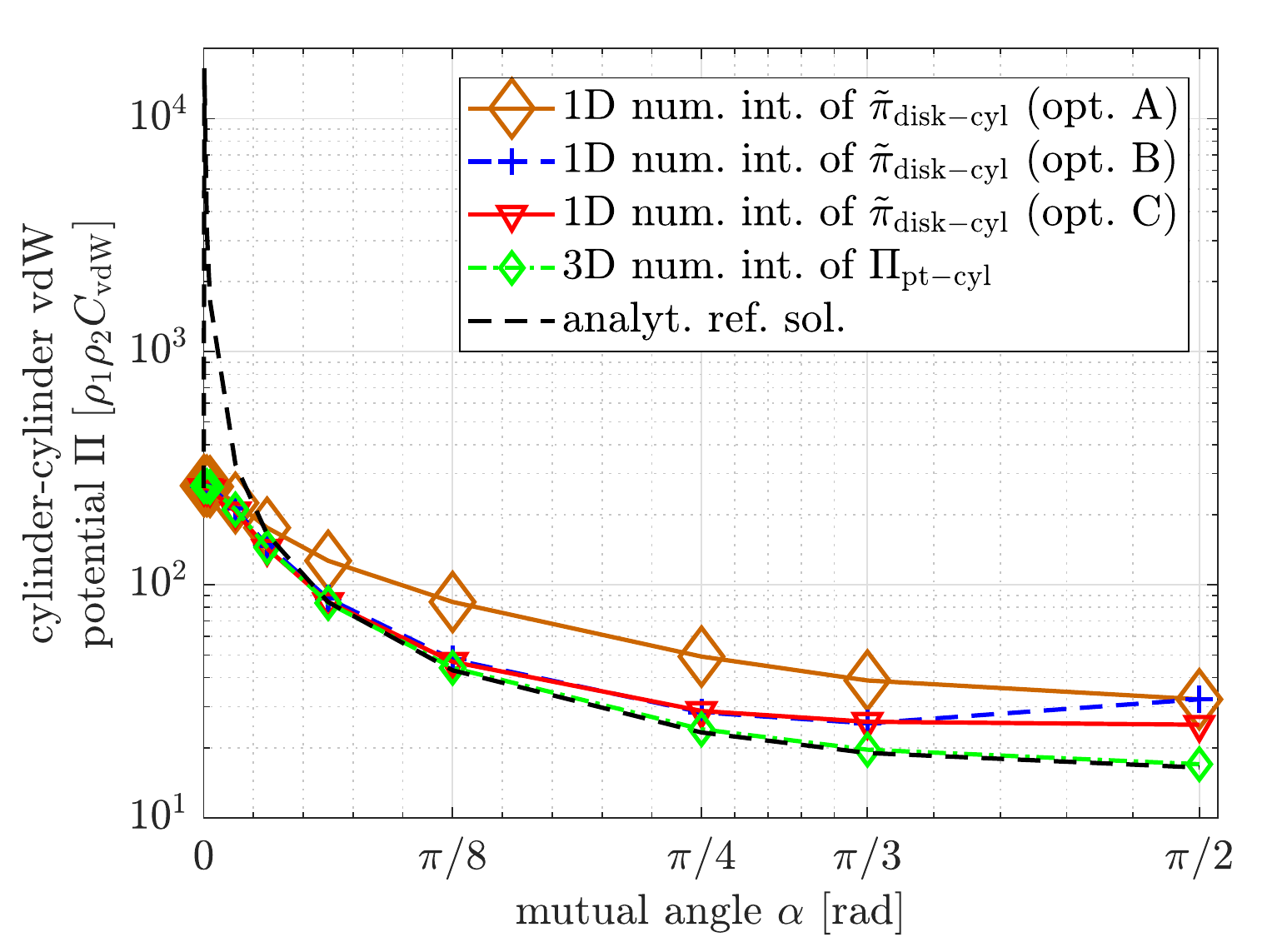}
    \label{fig::cyl-cyl_ia_pot_SBIP_optABC_over_angle_semilog_sep1e-1}
  }
  \caption{Interaction potential of two cylinders as a function of the mutual angle at different minimal surface separations~$g_\text{bl}/R$.
           Comparison of the options A (\eqref{eq::disk-cyl-pot_m_general} with~(\ref{eq::disk-cyl-pot_m_auxvariables_optA}); brown line with big diamonds), B (\eqref{eq::disk-cyl-pot_m_general} with~(\ref{eq::disk-cyl-pot_m_auxvariables_optB}); blue dashed line with pluses) and C (\eqref{eq::disk-cyl-pot_m}; red line with triangles) of the analytical expression for the disk-cylinder potential~$\tilde \pi_\text{6,disk-cyl}$ (used together with the SBIP approach\cite{GrillSBIP}).
           The numerical reference solution obtained via 3D Gaussian quadrature of the point-half space potential~$\Pi_\text{6,pt-hs}$ from \eqref{eq::pot_ia_vdW_point-cylinder} (green line with diamonds) and the analytical reference solutions summarized in~\secref{sec::introduction} (black dashed line) are plotted as reference.}
  \label{fig::cyl-cyl_ia_pot_SBIP_optABC_over_angle_semilog}
\end{figure}

\begin{figure}[htb]%
  \centering
   \vspace{-5pt}
   \subfigure[Surface-to-surface separation $g_\text{bl}/R=10^{-3}$]{
    \includegraphics[width=0.45\textwidth]{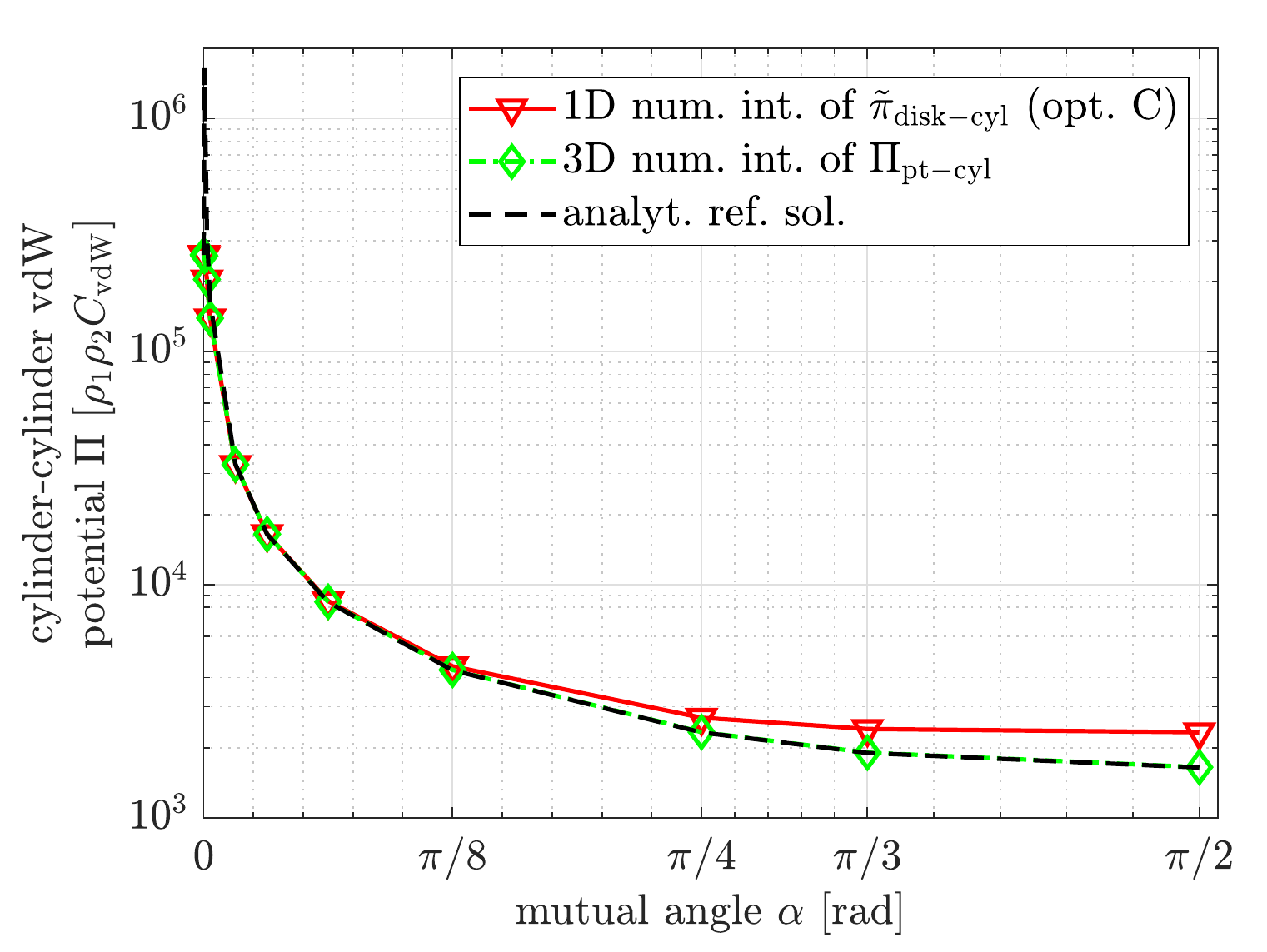}
    \label{fig::cyl-cyl_ia_pot_SBIP_optC_over_angle_semilog_sep1e-3}
   }
   \vspace{-5pt}
   \subfigure[Surface-to-surface separation $g_\text{bl}/R=10^{-2}$]{
    \includegraphics[width=0.45\textwidth]{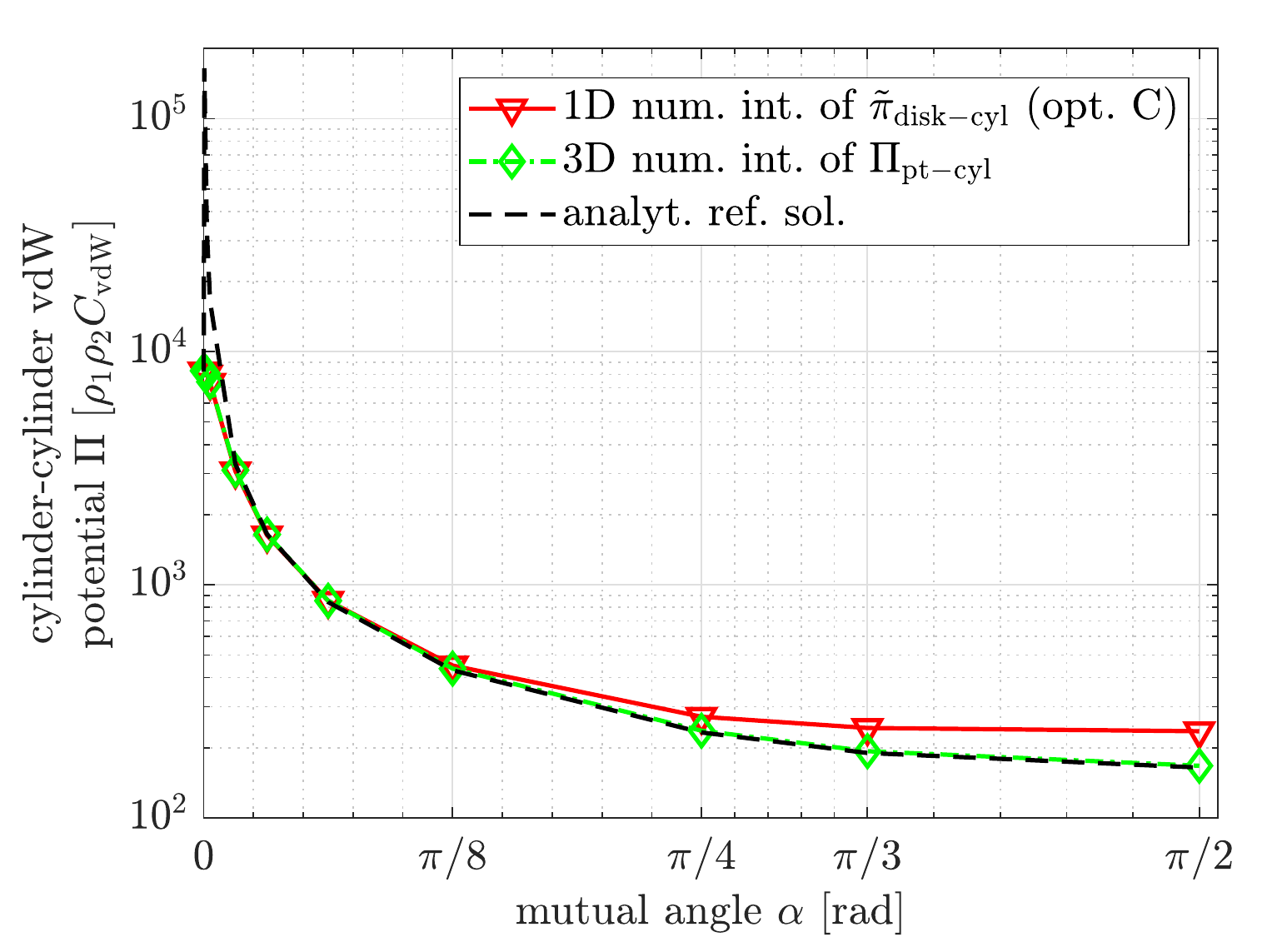}
    \label{fig::cyl-cyl_ia_pot_SBIP_optC_over_angle_semilog_sep1e-2}
   }
   \vspace{-5pt}
   \subfigure[Surface-to-surface separation $g_\text{bl}/R=10^{-1}$]{
    \includegraphics[width=0.45\textwidth]{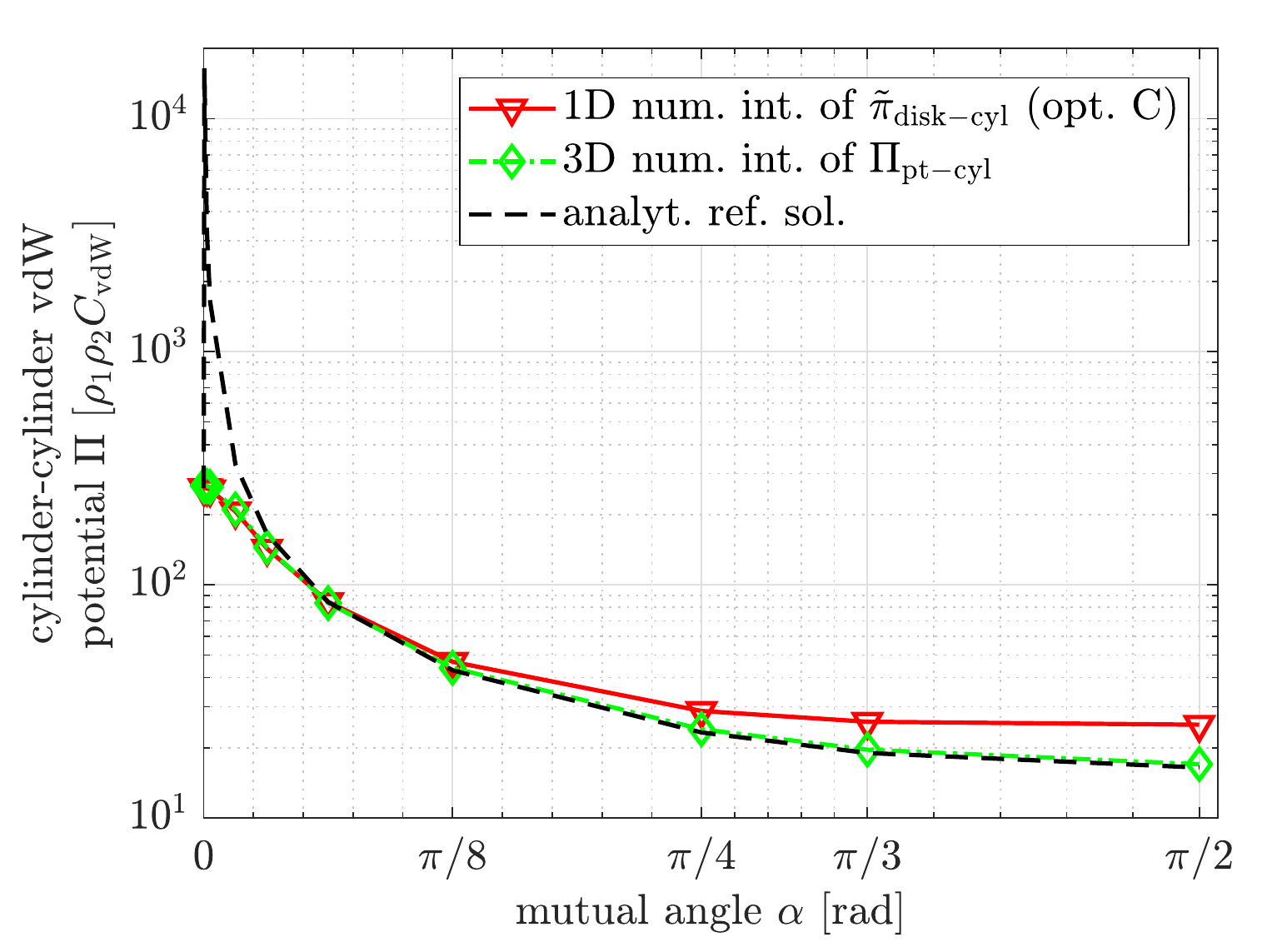}
    \label{fig::cyl-cyl_ia_pot_SBIP_optC_over_angle_semilog_sep1e-1}
   }
   \vspace{-5pt}
   \subfigure[Surface-to-surface separation $g_\text{bl}/R=1$]{
    \includegraphics[width=0.45\textwidth]{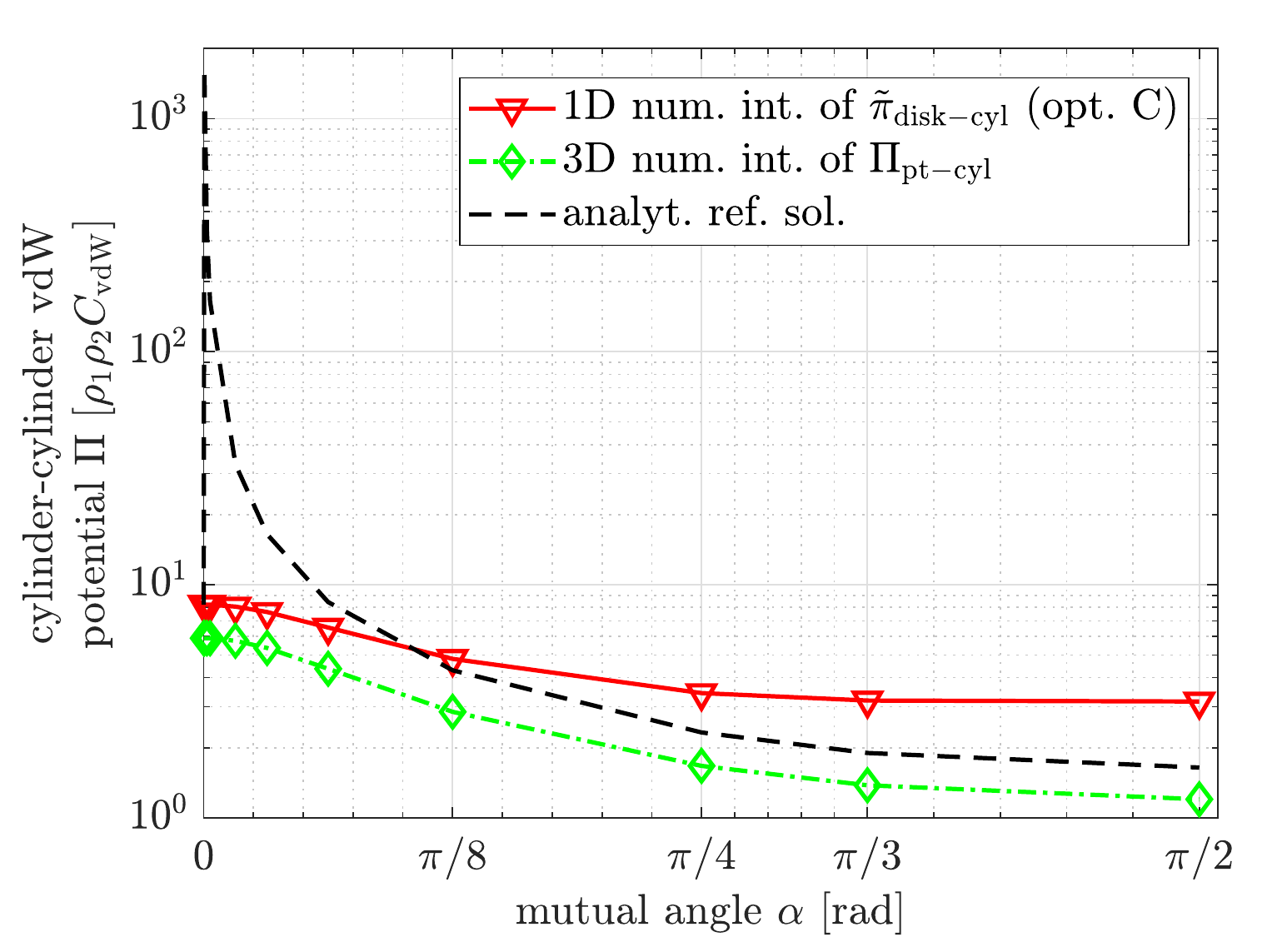}
    \label{fig::cyl-cyl_ia_pot_SBIP_optC_over_angle_semilog_sep1e0}
   }
   \caption{Interaction potential of two cylinders as a function of the enclosed angle at different smallest surface separations~$g_\text{bl}/R$. Verification of the analytical expression for the disk-cylinder potential~$\tilde \pi_\text{6,disk-cyl}$ from \eqref{eq::disk-cyl-pot_m} (used together with the SBIP approach\cite{GrillSBIP}; red line with triangles) by means of a numerical reference solution obtained via 3D Gaussian quadrature of the point-half space potential~$\Pi_\text{6,pt-hs}$ from \eqref{eq::pot_ia_vdW_point-cylinder} (green line with diamonds) and by means of analytical reference solutions summarized in~\secref{sec::introduction} (black dashed line).}
  \label{fig::cyl-cyl_ia_pot_SBIP_optC_over_angle_semilog}
\end{figure}

This appendix provides additional plots that analyze the accuracy of the disk-cylinder interaction potential law~$\tilde \pi_\text{m,disk-cyl}$ derived in \secref{sec::disk-cyl-pot_m_derivation} and its use within the general SBIP approach\cite{GrillSBIP}.
In particular, \figref{fig::cyl-cyl_ia_pot_SBIP_optC_over_angle_semilog} shows the vdW interaction potential of two cylinders as a function of the enclosed angle in a semi-logarithmic fashion and is thus closely related to \figref{fig::cyl-cyl_ia_pot_SBIP_optC_over_angle} using double-logarithmic plots to confirm the $1/\sin\alpha$-scaling.
Similarly, \figref{fig::cyl-cyl_ia_pot_SBIP_optABC_over_angle_semilog} shows the supplementary plots to \figref{fig::cyl-cyl_ia_pot_SBIP_optABC_over_angle_sep1e-3} -- \ref{fig::cyl-cyl_ia_pot_SBIP_optABC_over_angle_sep1e-1}.


\clearpage

\bibliography{analyt_disk_cyl_pot}

\end{document}